\documentclass[structabstract]{aa}
\usepackage{graphicx,txfonts,amssymb,epsfig,natbib}
\newcommand{\ARAA}{ARA\&A}
\newcommand{\AaA}{A\&A}

\newcommand{\ApJS}{Astrophysical Journal Supplement Series}
\newcommand{\AJ}{AJ}
\newcommand{\ApJ}{ApJ}
\newcommand{\MNRAS}{MNRAS}
\newcommand{\RvMP}{Reviews of Modern Physics}

\newcommand{\JChPh}{Journal of Chemical Physics}
\newcommand{\IAUS}{IAU Symposium}

\begin{document}
   \title{Excitation of H$_2$ in photodissociation regions as seen by Spitzer}

   \author{E. Habart\inst{1}, A. Abergel\inst{1}, F. Boulanger\inst{1}, C. Joblin\inst{2}, L. Verstraete \inst{1}, M. Compi\`egne \inst{1,3},  G. Pineau des For\^ets\inst{1} and J. Le Bourlot\inst{4}}

   \offprints{E. Habart. e-mail: emilie.habart@ias.u-psud.fr}

   \institute{
$^1$ Institut d'Astrophysique Spatiale (IAS), Universit\'e Paris-Sud, 91405 Orsay, France\\
$^2$ Universit\'{e} de Toulouse ;  UPS ;  CESR ;  9 avenue du colonel Roche, F-31028 Toulouse cedex 9, France and CNRS ; UMR5187 ; F-31028 Toulouse, France\\
$^3$ Canadian Institute for Theoretical Astrophysics, University of Toronto, 60 St. George Street, Toronto, ON M5S 3H8, Canada\\
$^4$ LUTH, Observatoire de Paris-Meudon, Universit\'e Paris 7, France
}

  \abstract
   {}
{ 
We present spectroscopic observations obtained with the infrared Spitzer Space Telescope, which provide 
insight into the H$_2$ physics and gas energetics in photodissociation Regions (PDRs) of low to moderate far-ultraviolet (FUV) fields and densities.
}
{ We analyze data on six well known Galactic PDRs (L1721, California, N7023E, Horsehead, rho Oph, N2023N),
sampling a poorly explored  range of excitation conditions  ($\chi \sim 5-10^3$),    
relevant to  the bulk of molecular clouds in galaxies. Spitzer observations of H$_2$ rotational lines
are complemented  with H$_2$ data, including ro-vibrational line measurements,  
obtained with  ground-based telescopes and ISO, to 
constrain the relative contributions of ultraviolet pumping and collisions to the H$_2$ excitation.
The data analysis  is supported by model calculations with the Meudon PDR code.
}
{ The observed  column densities  of rotationally excited H$_2$ are observed to be 
much higher than PDR model predictions. 
In the lowest excitation PDRs, the discrepancy between the model and the data is about one order of magnitude for rotational  levels $J \ge $3.  
We discuss whether  an enhancement in the H$_2$ formation rate or a  local increase in photoelectric heating, as proposed for brighter PDRs in former ISO studies, may improve the data-model comparison.
We find that an enhancement in the H$_2$ formation rates reduces the discrepancy, but the models still fall short of the data. }
 { This large disagreement suggests that our understanding of the formation and excitation of H$_2$ and/or 
   of PDRs energetics is still incomplete. We discuss several explanations, which could be further tested 
   using the Herschel Space Telescope.}


\authorrunning{Habart et al.}
\titlerunning{Excitation of H$_2$ in photodissociation regions}
\maketitle

%
%________________________________________________________________

\section{Introduction}
\label{introduction}

The bulk of interstellar matter is found in regions 
of low-to-moderate opacity to UV and visible light 
where stellar radiation governs the chemical and thermal state of the gas.
These photodissociation or photon-dominated regions \cite[PDRs, for a review see][]{hollenbach99}
are responsible for reprocessing much of the energy output from stars, reemitting this energy in the infrared-millimeter wavelengths, including a rich mixture of gas lines  (e.g., emission in fine-structure, rotational, and rovibrational lines).
They are privileged objects for studying the chemical and physical processes of the interstellar medium (ISM).
The motivation of this study is to test our understanding of the excitation of H$_2$ in a regime of space parameters that has been poorly studied.

Through its ability to observe pure rotational lines of H$_2$ towards a number of bright PDRs,
the Infrared Space Observatory (ISO) has given important information on the local microphysics of interstellar gas.
One conclusion  of these studies is that the H$_2$ line intensities and the gas temperature, 
derived from the first rotational levels of H$_2$, are higher than model calculations  \cite[]{bertoldi97,draine99a,thi99,kemper99,li2002,habart2003a}.
Changes in the description of heating and cooling processes or of the H$_2$ formation rate have been proposed to account
for this discrepancy  \cite[]{habart2004}.
%One explanation is that the H$_2$ formation rate is larger than usually  believed at high gas temperature moving the H$^0$/H$_2$ transition zone closer to the edge of the PDR where the gas is warmer \cite[]{habart2003a,habart2004}. 
%\cite{habart2004} have shown that in order to account for the observed H$_2$ line data in moderately excited PDRs an increase in the H$_2$ formation rate by a factor of 5 compared with the standard rate of 3$\times 10^{-17}$ cm$^3$ s$^{-1}$ derived by \cite{jura75} from UV absorption lines is required.
With H$_2$ UV absorption line observations performed by the Far Ultraviolet Spectroscopic Explorer (FUSE), 
the H$_2$ formation and excitation processes in diffuse and translucent clouds have also been reconsidered.
In particular, UV data show significant amounts of rotationally excited H$_2$,
for which UV photons could not be the unique heating source \cite[]{gry2002,nehme2008}.
%Radiative pumping is unable to explain the observed excitation of H$_2$ in its levels $J \ge 3$. 
The H$_2$ excitation in the diffuse interstellar medium may be tracing the dissipation
of interstellar turbulence in C-shocks \cite[]{flower98} or interstellar vortices \cite[]{falgarone2005}.

There has been very little exploration of the physics of PDRs with moderate FUV fields and densities, 
an intermediate regime between the diffuse clouds and the bright PDRs.
The sensitivity of the infrared spectrograph (IRS) onboard the Spitzer Space Telescope provides a unique opportunity to observe
H$_2$ pure rotational line emission in low-brightness sources. 
As part of the SPECPDR\footnote{See http://www.cesr.fr/$\sim$joblin/SPECPDR public/SPECPDR.html} program
dedicated to the study of PDRs with Spitzer,
we used this capability to study the H$_2$ line emission in regions with $\chi \sim 5-10^3$ times  the
Solar Neighborhood Far-UV interstellar radiation field \footnote{For a discussion on the definitions of the mean interstellar radiation field used in the literature on PDRs see Appendix B of \cite{allen2004} or Appendix C of \cite{lepetit2006}.} 
as given by \cite{draine78}.
Our sample consists of well-known PDRs, so that they are good test cases for models. 
These observations are analyzed in combination with previous ISO and ground-based data of rotational and rovibrational  H$_2$ lines when available, in order to help constrain the relative roles of ultraviolet pumping and collisions in establishing the level populations.

The paper is organized as follows. 
In Sect. \ref{sample}, we present our PDR sample. 
In Sect. \ref{observation}, we present the IRS spectrometer observations. 
In Sect. \ref{model_presentation}, we briefly describe the PDR model used to analyze the data, 
and in Sect. \ref{comparison_observation_model} we compare the observed H$_2$ line emission to the PDR model predictions.
In Sect. \ref{h2_diffus}, we compare our results with UV absorption measurements in diffuse clouds.
In Sect. \ref{origin}, we discuss several possibilities of explaining our results.
In Sect. \ref{galaxies}, in light of our PDR observations, we discuss H$_2$ line infrared emission measurements in galaxies and along Galactic lines of sight.
Our conclusions are summarized in Sect. \ref{conclusion}.

\section{Selected PDRs}
\label{sample}

In the framework of the SPECPDR program, we have observed several photodissociation regions - a total of eleven including nine galactic and two extra-galactic sources - with all the instruments onboard Spitzer.
In this paper, we select six of the nine galactic sources. 
They are close to the Sun ($d\sim$100-600 pc) with well-defined geometry and radiation field. Most of them are inclined with respect to the line of sight with nearly edge-on (shell-like) geometry.
They represent classic examples of the interaction between UV radiation and an interstellar molecular cloud.
They span a wide range of physical conditions; temperature of the exciting star: $T_{eff}$ = 15, 000 - 37, 000 K, FUV field: $\chi$ = 5 - 1000 in Draine units, proton gas densities: $n$ = $10^3$-$10^5$ cm$^{-3}$.
%Some of the other sources of the sample have specific UV radiation. %Ced 201
%\footnote{The other sources of the sample have specific UV radiation and/or metallicity conditions.}
A brief description of each source is given below.

\begin{table*}[htbp]
\caption{Sample of PDRs with distance, effective temperature of the exciting star, incident FUV radiation field strength, gas density, and geometry. }
\label{table_sample}
\begin{center}
\begin{tabular}{|l|l|l|l|l|l|l|} 
\hline 
Source   & Dist.$^a$ & Exciting star & FUV field$^b$ & Density$^c$ & Geometry$^d$ & Ref. \\
  & (pc) & (T$\rm_{eff}$, spec.type) & (Draine) & (cm$^{-3}$) &  &  \\
\hline 
L1721        &  134$_{-17}^{+23}$          &  20,000 K, B2V     & 4.5 [3-6]           & 3 10$^2$-3 10$^3$    & S-C           & (1)\\
California   &  543$_{-150}^{+334}$          &  37,0000, O7.5III      & 19 [7-37]            & 10$^3$-10$^4$		 & E-O     & (2)\\ 
N7023E       &  429$_{-90}^{+156}$          &  15,000 K, B5Ve     & 51 [27-82]             & 7 10$^3$		 & E-O     & (3)\\ 
Horsehead    &  352$_{-85}^{+166}$     &  33,000 K, O9.5V   & 60 [30-110]            & 10$^4$-10$^5$        & E-O             & (4)\\ 
rho Oph     &  136$_{-19}^{+26}$   &  22,000 K, B2V     & 250 [180-340]           & 1-5 10$^4$           & E-O,C           & (5)\\ 
N2023N       &  472$_{-173}^{+652}$  &  23,000 K, B1.5V   & 550 [100-1370]       & 10$^4$-10$^5$        & E-O,C           & (6)\\ 
\hline 
\end{tabular}
\end{center}
{\small 
$^a$~Estimated by Hipparcos \cite[]{perryman97} and \cite{hoogerwerf2000}.
%\cite{hoogerwerf2000} for orion
\noindent
$^b$~Incident FUV radiation field expressed in units $\chi$ of the \cite{draine78} average interstellar radiation field
with uncertainties in between brackets ([$\chi _{min}$, $\chi _{max}$]).
%For a discussion on the definitions of mean interstellar radiation field used in the literature on PDRs, see Appendix B of \cite{allen2004} or Appendix C of \cite{lepetit2006}.
\noindent
$^c$~Proton gas density $n\equiv n_{H^0}+2~n_{H_2}$ derived from different observational constraints.
\noindent
$^d$~Geometry codes : S-C : Spherical Cloud; E-O : Edge-On; C : Cavity. %; CG: Cometary Globule; F-O: Face-On.
\noindent
References : 
(1) \cite{habart2001a}; 
(2) \cite{boulanger88b};
(3) \cite{rapacioli2006,berne2007};
(4) \cite{abergel2003,teyssier2004,habart2005,pety2005}; 
(5) \cite{habart2003a,motte98};
(6) \cite{field98,compiegne2008} 
%(5) \cite{black87}; %for density 
%\cite{draine96,draine99a}; % for density and chi 
%\cite{field98};  % for density 
%\cite{wyrowski97a,wyrowski2000} % for density 
%(6) \cite{marconi98}; \cite{tielens85a}; \cite{tielens93}; \cite{tauber94}; \cite{wyrowski97}; \cite{herrmann97}; \cite{simon97}; \cite{hogerheijde95}; \cite{allers2005}.
}
\end{table*}

\par\bigskip\noindent
{\bf L1721}. At the low excitation end, our sample contains the Lynds Dark Nebula 1721 \cite[L1721,][]{lynds62}, a nearby isolated molecular cloud in the $\rho$ Ophiuchi region. 
This moderate opacity cloud is heated on one side by the star $\nu$ Sco and by a more isotropic interstellar radiation field created by the Upper Scorpius association \cite[e.g.,][]{habart2001a}.
The B2 star $\nu$ Sco is located at $\sim$130 pc from the Sun.
Visual extinction map and emissions of the dust and gas cooling lines ($[C^+]$ 158 $\mu$m, $[O^0]$ 63 $\mu$m) have been measured toward the L1721 cloud \cite[]{habart2001a}. 
The exciting radiation  field is estimated to be $\chi=$3 to 6 times the Draine radiation field.
% from the ratio of dust emission to gas column density.
%The density profile of the L1721 cloud is deduced from visual extinction data.
%ISO-LWS measurements of the major cooling lines $[C^+]$ 158$\mu$m and $[O^0]$ 63$\mu$m have been obtained across the L1721 cloud \cite[]{habart2001a}.

\par\bigskip\noindent
{\bf California}. The California Nebula (NGC 1499) is an emission nebula located in the constellation Perseus. 
The O7.5III star $\xi$ Per is the source of radiation of the California nebula.
This star is located at $\sim$540 pc.
Here, we focus on the filament situated at 45' from the star.
The intensity of the exciting radiation field after dilution and assuming that the distance of the exciting star to the PDR is equal to the distance projected onto the sky is estimated to be $\chi \sim$19 times the Draine field.
%Considering the uncertainties on the distance of the illuminated star, we find $\chi$=19$_{7}^{37}$.

\par\bigskip\noindent
{\bf N7023E}. NGC 7023 is a reflection nebula around the Herbig B5Ve illuminating star HD 200775.
This star is located at  $\sim$430 pc.
Three PDRs are visible around the star. Here, we focus on the NGC 7023 East located further east of the star (170'').
Spitzer spectroscopic dust studies of the East PDR was performed by \cite[]{berne2007}.
%According the authors, 
The exciting radiation  field is estimated to be 87 times the Habing field, i.e. $\chi \sim$51 times the Draine field.
%NGC 7023-E B5e, 15 000 K, 87 (Habing), nH=7e3,T=340, Rapacioli et al. (2006)

\par\bigskip\noindent
{\bf Horsehead}. The Horsehead nebula (B33) emerges from the edge of the L1630 molecular complex.
The $\sigma$ Ori star, a O9.5V system, is the main heating source in this region.
This star  is located at  $\sim$350 pc.
We focus here on the western edge of the nebula situated at 0.5$^{\circ}$ from $\sigma$ Ori.
Emissions of the gas and dust associated with the western edge have been mapped from the visible to millimeter wavelengths by various
authors \cite[e.g.,][]{pound2003,teyssier2004,pety2004,habart2005,ward-thompson2006,philipp2006,goicoechea2006,pety2007}.
%\cite[e.g., ]{abergel2003,pound2003,teyssier2004,pety2004,habart2005,hily-blant2005,ward-thompson2006,philipp2006,goicoechea2006,pety2007}.
This PDR is illuminated and viewed nearly edge-on.
The intensity of the exciting radiation field after dilution and assuming that the distance of the exciting star to the PDR is equal to the distance projected onto the sky is $\chi \sim$60 times the Draine field.

\par\bigskip\noindent
{\bf Rho Oph}. The  Ophiuchi molecular cloud is a nearby star-forming region.
The western part of this cloud is bounded by a bright filament structure, forming an edge-on (shell-like) PDR, illuminated by the B2 HD147889 star \cite[]{abergel99}. 
The HD147889 star is located at $\sim$130 pc.
We focus here on the western edge situated at $\sim$600'' from the illuminated  star.
%The B2 star HD147889, which appears to be the center of a more-or-less spherical cavity, is in this region the main heating source \cite[]{abergel99,liseau99}.
ISO spectroscopic dust and gas (H$_2$) studies of this PDR  were performed \cite[]{habart2003a}.
The exciting radiation  field is estimated to be $\chi \sim$250 times the Draine field.

\par\bigskip\noindent
{\bf N2023N}. NGC 2023 is a reflection nebula excited by a B1.5V star (HD37903), which is embedded in the L1630 cloud. % (e.g. Gatley et al. 1987). 
The HD37903 star is located at $\sim$470 pc.
In this work, we focus on the filament situated at 150'' north of the star.
Spitzer spectroscopic dust studies of the N2023 north PDR was performed by \cite{compiegne2008}.
The intensity of the exciting radiation field after dilution and assuming that the distance of the exciting star to the PDR is equal to the distance projected onto the sky is estimated to be $\chi \sim$550 times the Draine field.
According to the study of \cite{compiegne2008}, there is a visual extinction of $A_V$ equal to 1.25 between the star and the filament.
The intensity of the exciting radiation field after dilution and extinction is 120 times the Habing field, i.e. $\chi \sim$70 times the Draine field.

\par\bigskip\noindent
In Table \ref{table_sample}, we summarize the distances, radiation sources, gas densities and geometries for each region as determined from the literature. The distances have been estimated by Hipparcos \cite[]{perryman97} and other measurements \cite[]{hoogerwerf2000}.
As given above for most of our sources, the radiation field $\chi$ factor is determined from the expected FUV luminosity of the exciting star and assuming that the distance of the exciting star to the PDR is equal to the distance projected onto the sky. 
This is an upper limit in principle. 
Uncertainties in $\chi$ due to the errors in the distance of the exciting star from the Sun are given, so, estimations of $\chi$  based on observations of dust and gas lines emissions are also reported.
The gas density for each region is also given in Table \ref{table_sample}.
The density inferred from various atomic/molecular species shows a relatively large dispersion. %(typically $10^3$-$10^4$ or $10^4$-$10^5$cm$^{-3}$). 
This dispersion could result from density gradients from the outer hot layer to the inner molecular cold layer \cite[see, for example,][]{walmsley2000,habart2003a,habart2005}. 
However, what is needed for our study is the density in the H$_2$-emitting region.
% while the other molecular density tracers could reflect the density of the cold gas in the cloud. %For HH and rho Oph : nh=1e4 in the H$_2$ zone.

\section{Observations}
\label{observation}

\begin{figure*}[htbp]
\leavevmode
\begin{minipage}[c]{9cm} 
%\centerline{ \psfig{file=/Users/emiliehabart/PDR/PDR_plot/all_objets/figure_article_spectrum_average_L1721.ps,width=9cm,height=6cm,angle=0} }
\centerline{ \psfig{file=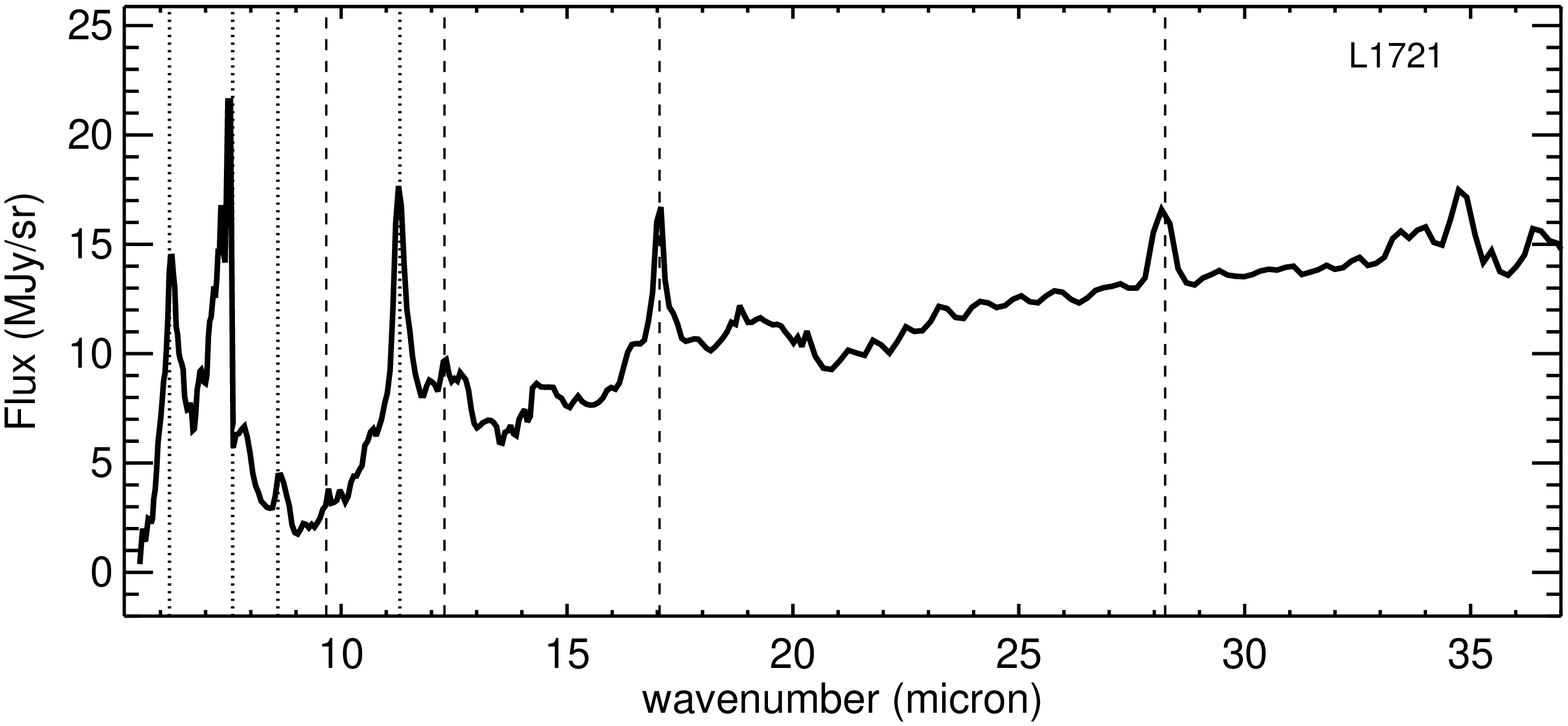,width=9cm,height=6cm,angle=0} }
\end{minipage}
\begin{minipage}[c]{9cm} 
%\centerline{ \psfig{file=/Users/emiliehabart/PDR/PDR_plot/all_objets/figure_article_spectrum_average_California.ps,width=9cm,height=6cm,angle=0} }
\centerline{ \psfig{file=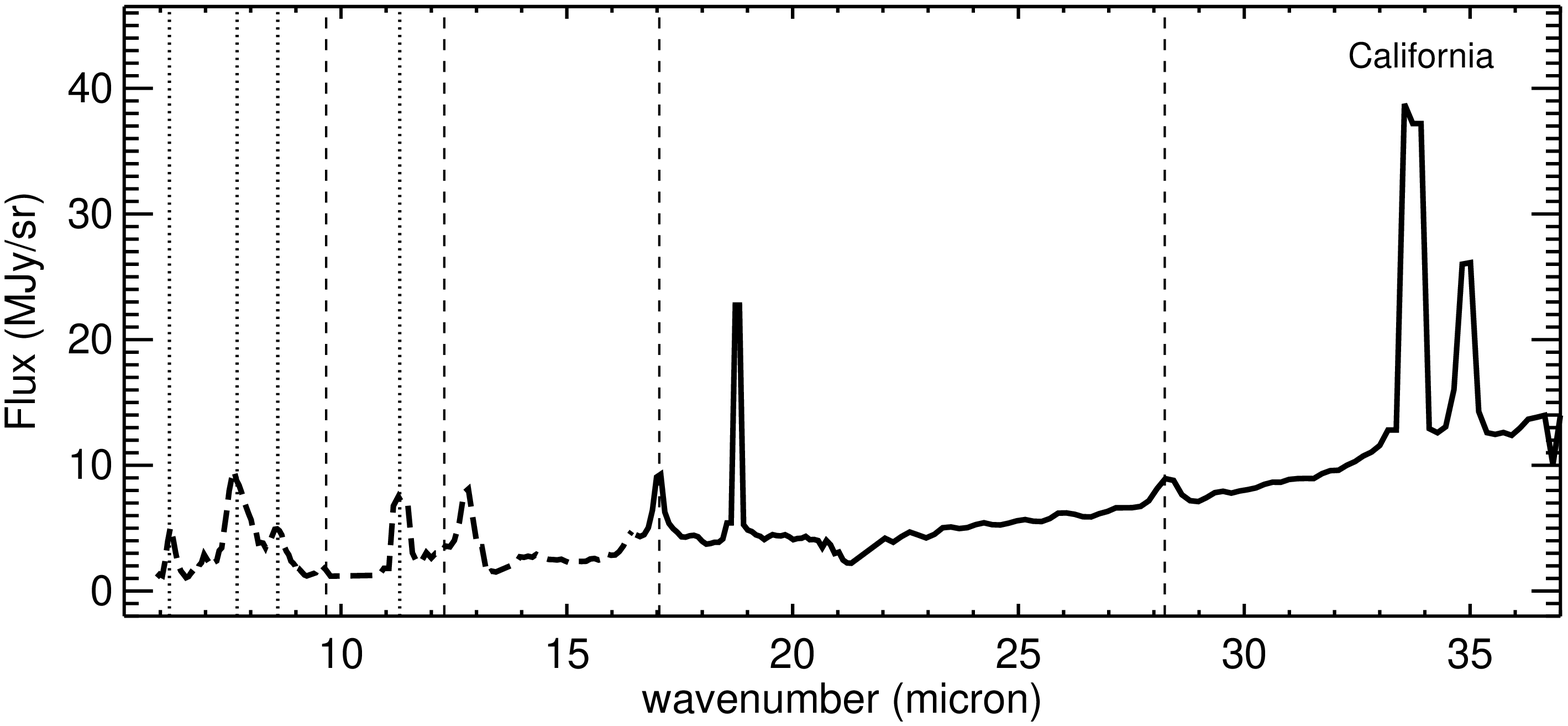,width=9cm,height=6cm,angle=0} }
\end{minipage}
\begin{minipage}[c]{9cm} 
%\centerline{ \psfig{file=/Users/emiliehabart/PDR/PDR_plot/all_objets/figure_article_spectrum_average_N7023E.ps,width=9cm,height=6cm,angle=0} }
\centerline{ \psfig{file=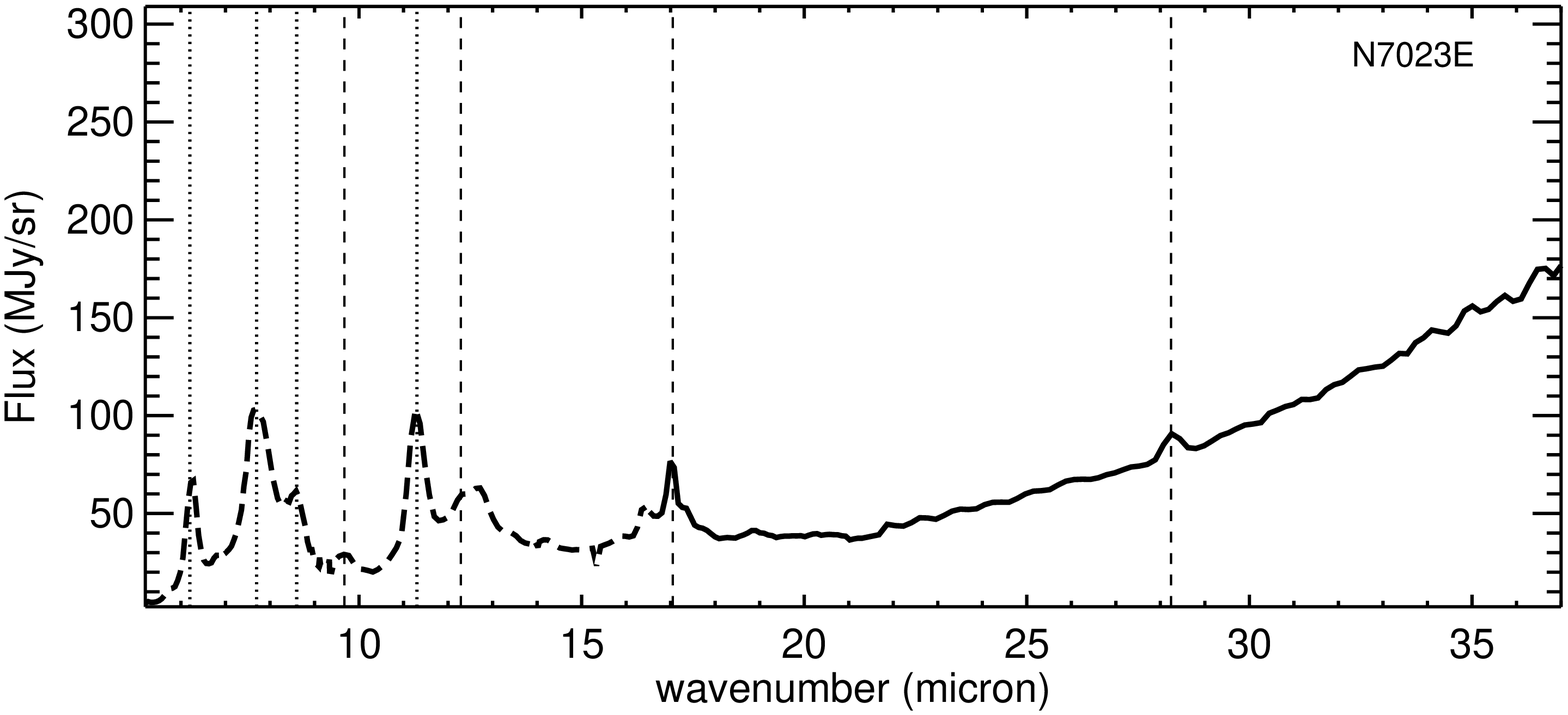,width=9cm,height=6cm,angle=0} }
\end{minipage}
\begin{minipage}[c]{9cm} 
%\centerline{ \psfig{file=/Users/emiliehabart/PDR/PDR_plot/all_objets/figure_article_spectrum_average_Horsehead.ps,width=9cm,height=6cm,angle=0} }
\centerline{ \psfig{file=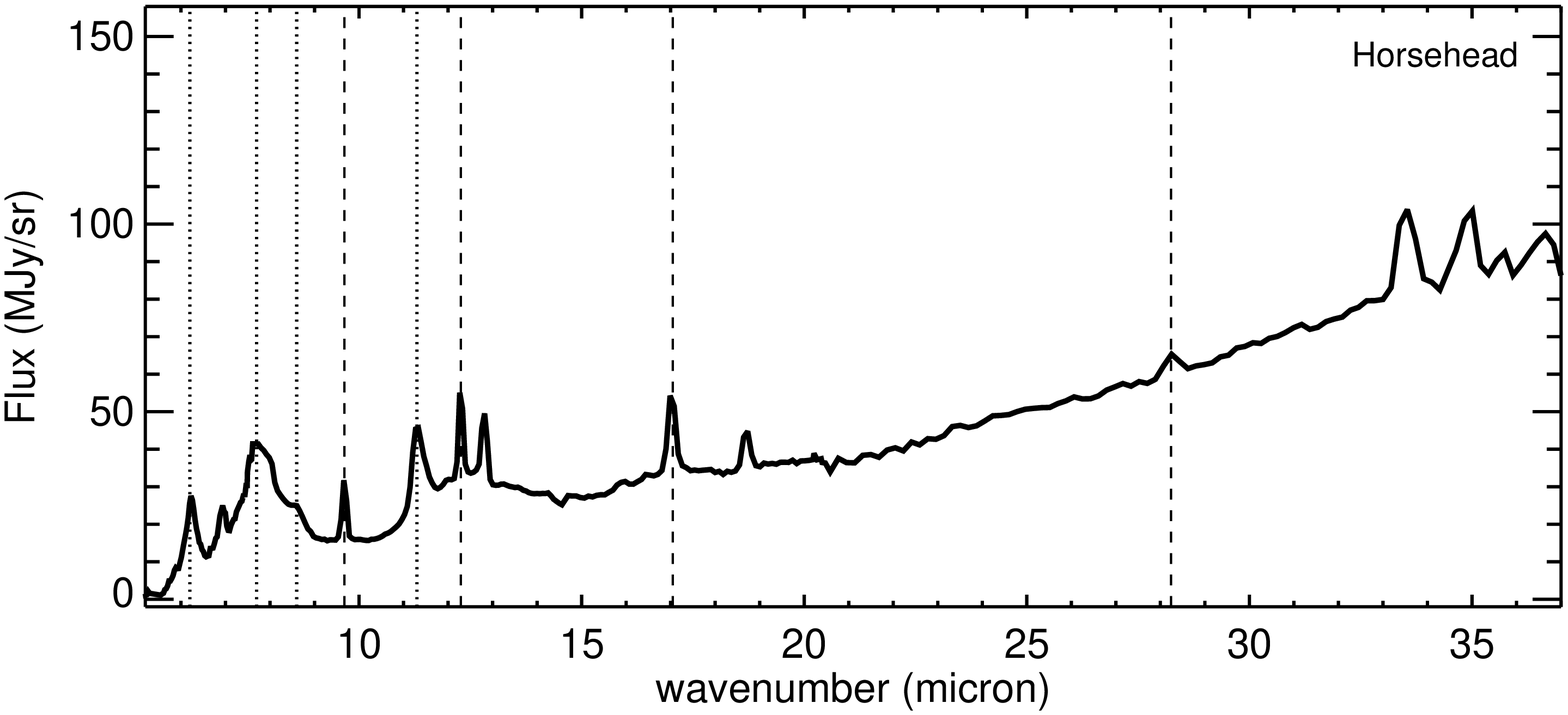,width=9cm,height=6cm,angle=0} }
\end{minipage}
\begin{minipage}[c]{9cm} 
%\centerline{ \psfig{file=/Users/emiliehabart/PDR/PDR_plot/all_objets/figure_article_spectrum_average_rho_Oph.ps,width=9cm,height=6cm,angle=0} }
\centerline{ \psfig{file=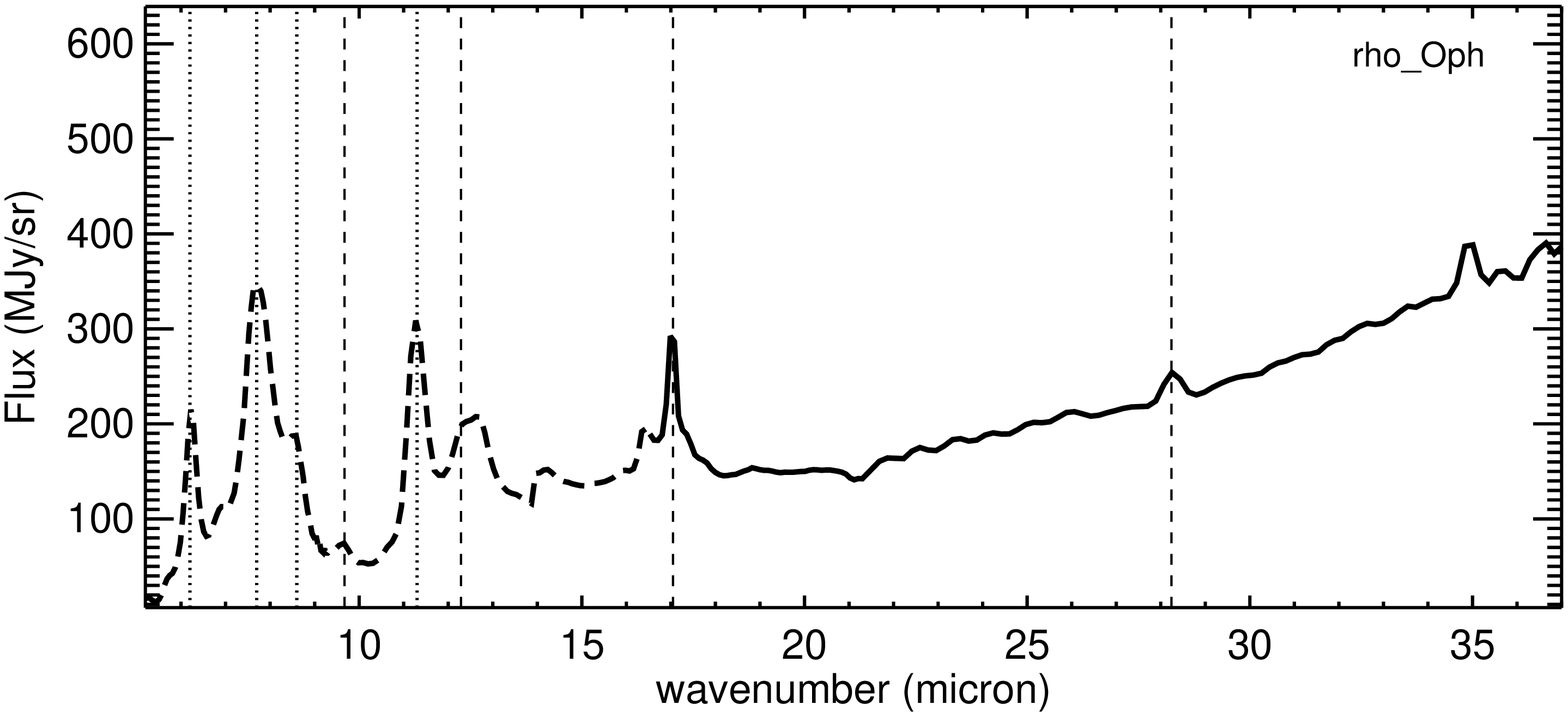,width=9cm,height=6cm,angle=0} }
\end{minipage}
\begin{minipage}[c]{9cm} 
%\centerline{ \psfig{file=/Users/emiliehabart/PDR/PDR_plot/all_objets/figure_article_spectrum_average_N2023N.ps,width=9cm,height=6cm,angle=0} }
\centerline{ \psfig{file=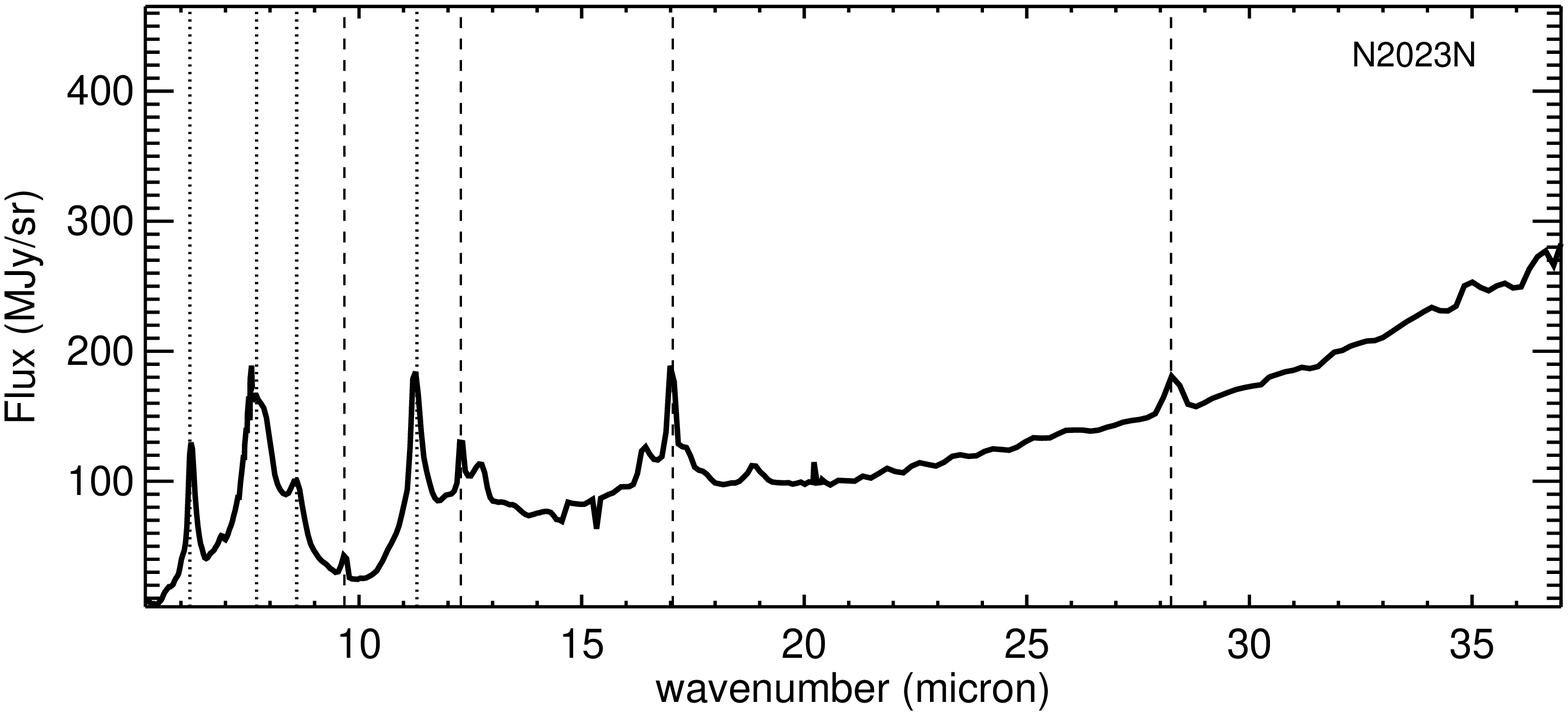,width=9cm,height=6cm,angle=0} }
\end{minipage}
%-------------------------------------
\caption{\em Average IRS spectra obtained with the low spectral resolution mode. 
The average areas used are marked  in Fig. \ref{fig_h2_map_obs}.
For L1721 and California, all pixels are averaged in order to increase the signal-to-noise ratio, since these extended PDRs are fainter and do not present any detected spatial variation  within the field observed with IRS.
In the case of California, N7023E and rho Oph, only the LL IRS submodule data are available.
These three objects have already been observed with the CVF of ISOCAM. 
We present here the combined ISO-CVF (dashed lines) and IRS-LL (solid lines) spectra.
The IRS wavelength coverage allows us to detect several strong H$_2$ pure rotational lines from 0-0 S(0)-S(3) at 28.2, 17.03, 12.29, and 9.66 $\mu$m, the aromatic band features at 6.2, 7.7, 8.6, 11.3 $\mu$m, the dust mid-IR continuum emission 
and the fine structure lines of ionized gas [NeII] at 12.8 $\mu$m, [SIII] at 18.7 and 33.4 $\mu$m and [SiII] at 34.9 $\mu$m.
The dashed vertical lines delineate the H$_2$ 0-0 S(0)-S(3) lines position.
The dotted vertical lines delineate the aromatic band feature position.}
\label{fig_spectrum_low_resolution}
\end{figure*}

\begin{figure*}[htbp]
\leavevmode
\begin{minipage}[c]{6cm} 
%\centerline{ \psfig{file=/Users/emiliehabart/spitzer/data/IRS_High_Resolution_Laurent/IRS_HR_H2_LINE_S2_L1721.ps,width=6cm,height=6cm,angle=0} }
\centerline{ \psfig{file=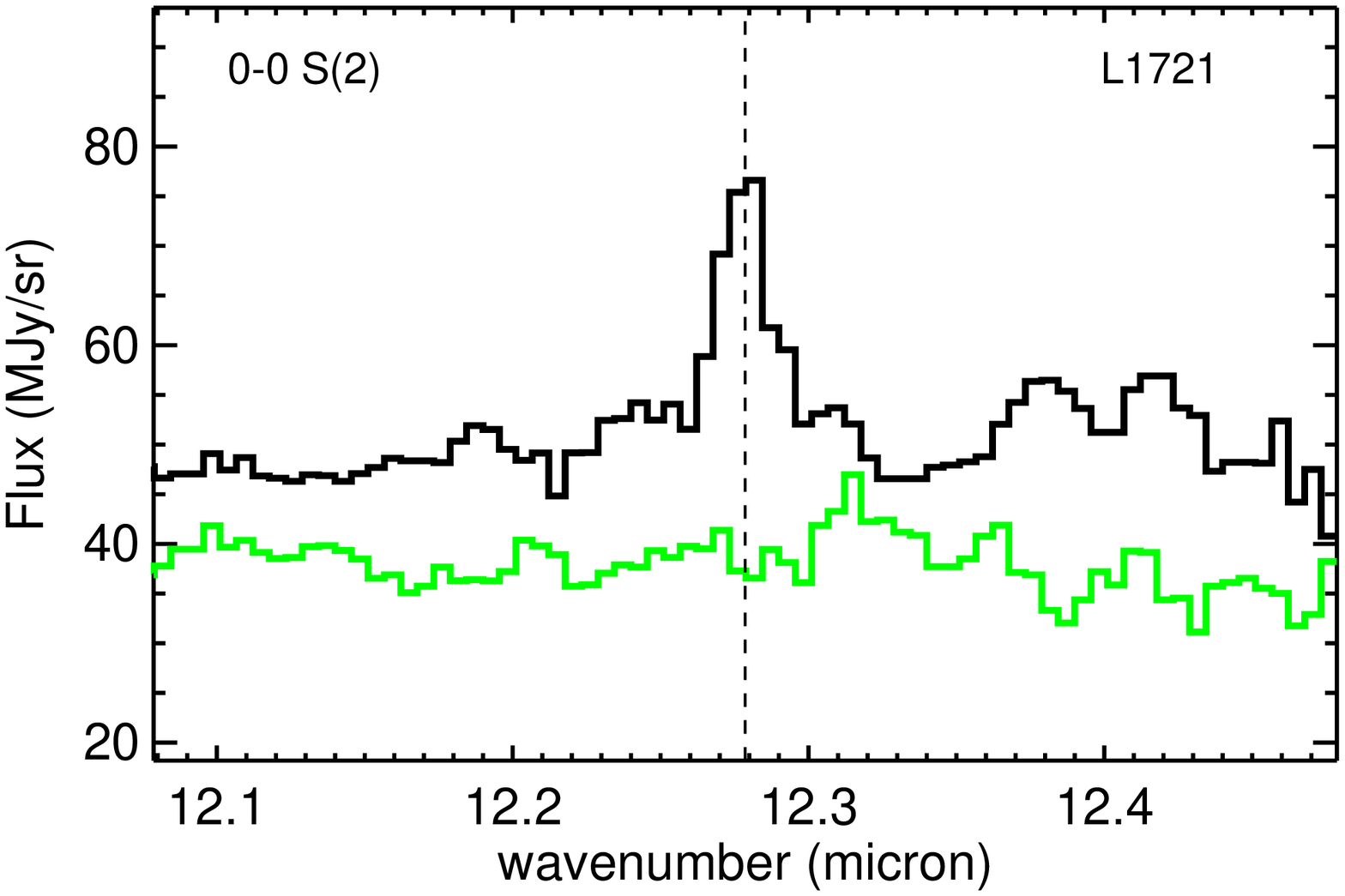,width=6cm,height=6cm,angle=0} }
\end{minipage}
\begin{minipage}[c]{6cm} 
%\centerline{ \psfig{file=/Users/emiliehabart/spitzer/data/IRS_High_Resolution_Laurent/IRS_HR_H2_LINE_S1_L1721.ps,width=6cm,height=6cm,angle=0} }
\centerline{ \psfig{file=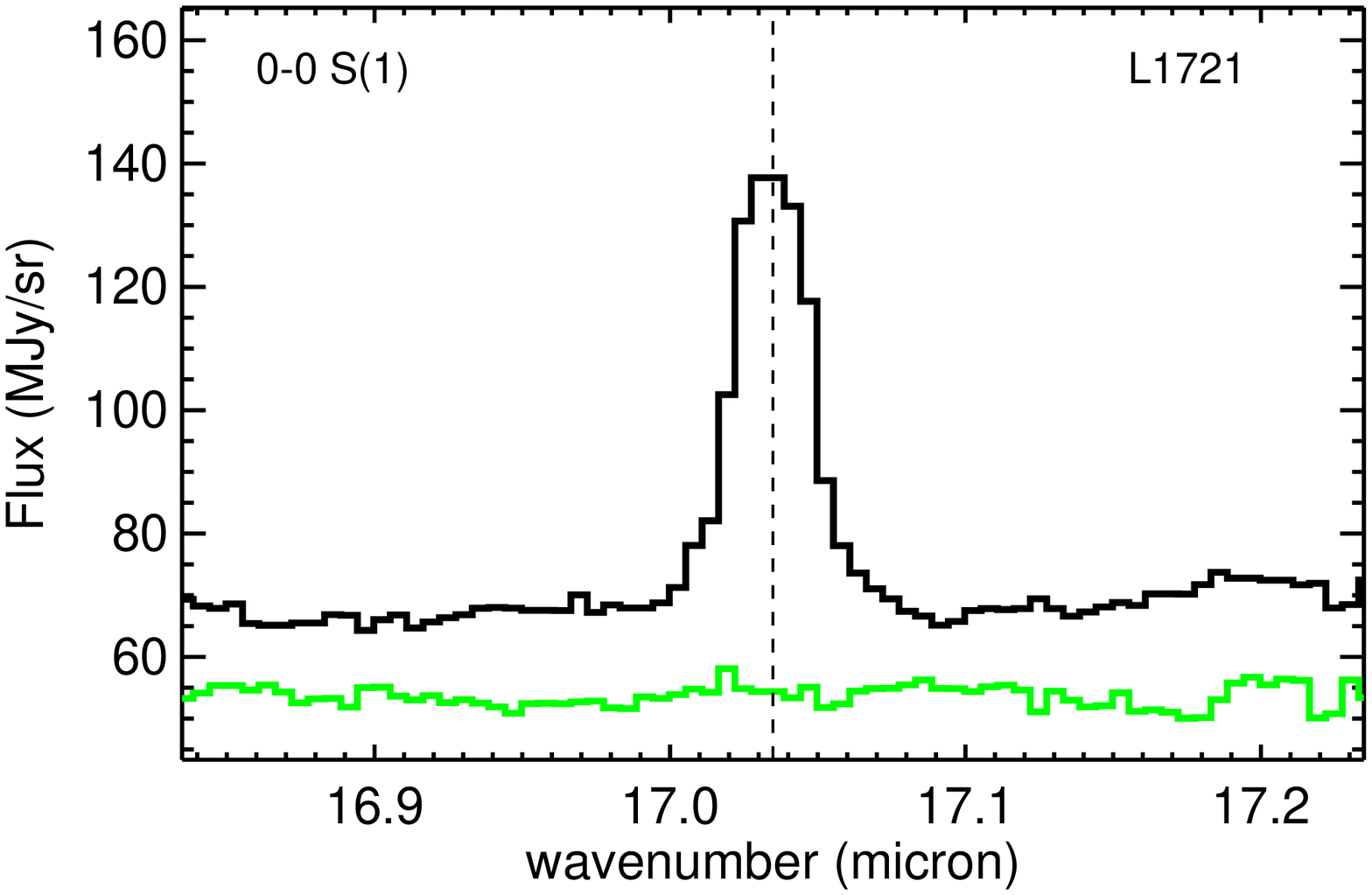,width=6cm,height=6cm,angle=0} }
\end{minipage}
\begin{minipage}[c]{6cm} 
%\centerline{ \psfig{file=/Users/emiliehabart/spitzer/data/IRS_High_Resolution_Laurent/IRS_HR_H2_LINE_S0_L1721.ps,width=6cm,height=6cm,angle=0} }
\centerline{ \psfig{file=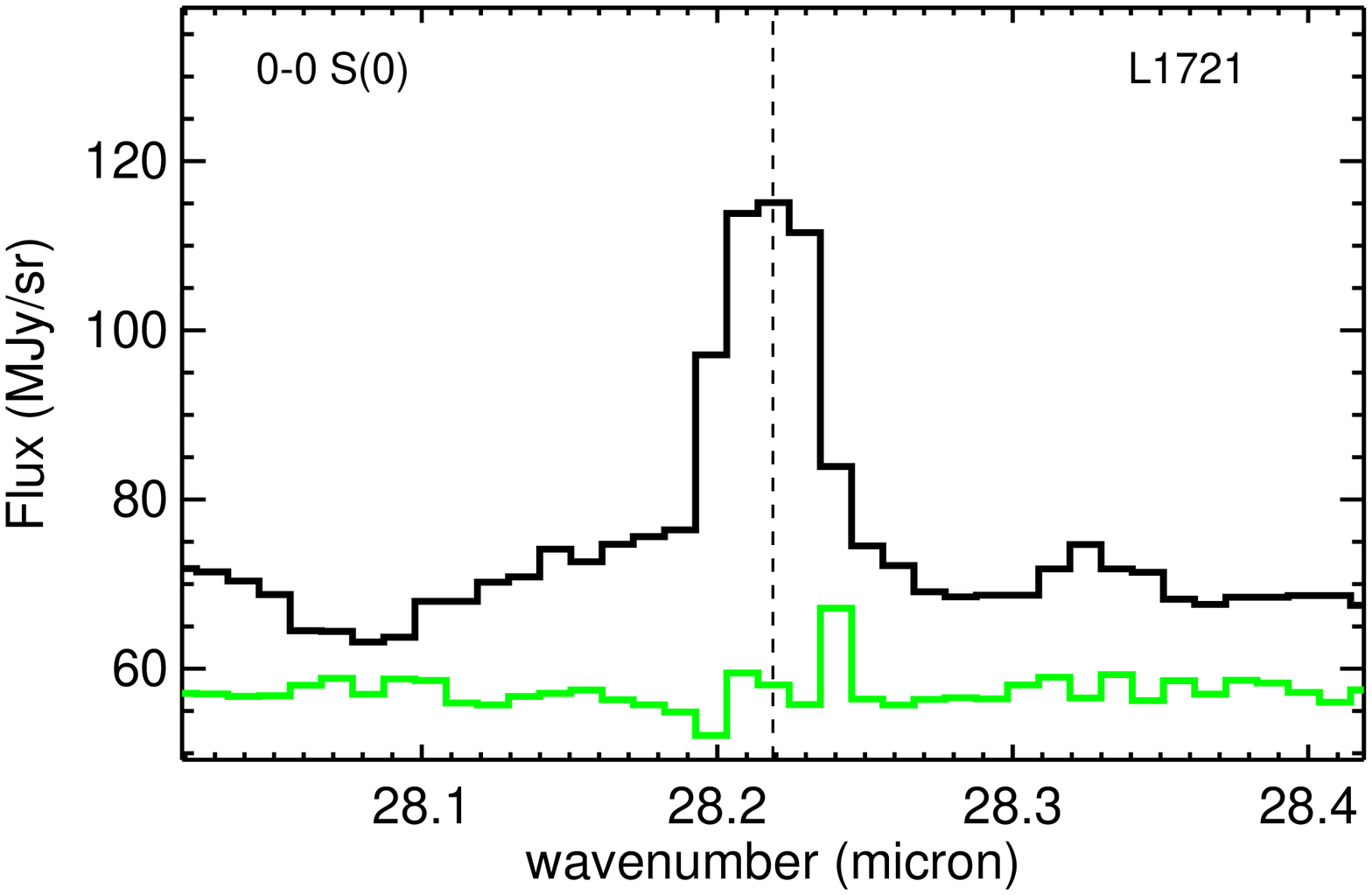,width=6cm,height=6cm,angle=0} }
\end{minipage}
\begin{minipage}[c]{6cm} 
%\centerline{ \psfig{file=/Users/emiliehabart/spitzer/data/IRS_High_Resolution_Laurent/IRS_HR_H2_LINE_S2_California.ps,width=6cm,height=6cm,angle=0} }
\centerline{ \psfig{file=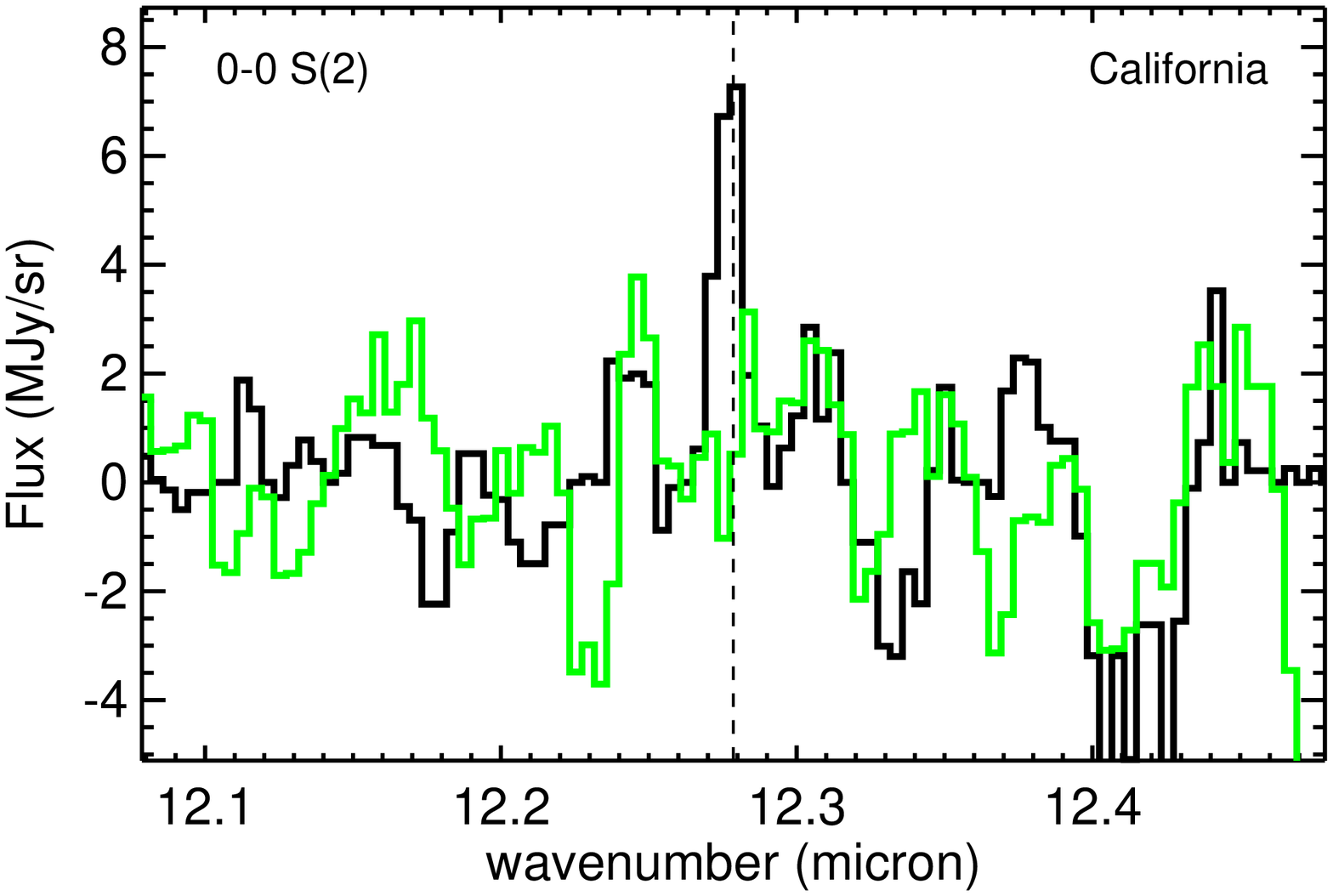,width=6cm,height=6cm,angle=0} }
\end{minipage}
\begin{minipage}[c]{6cm} 
%\centerline{ \psfig{file=/Users/emiliehabart/spitzer/data/IRS_High_Resolution_Laurent/IRS_HR_H2_LINE_S1_California.ps,width=6cm,height=6cm,angle=0} }
\centerline{ \psfig{file=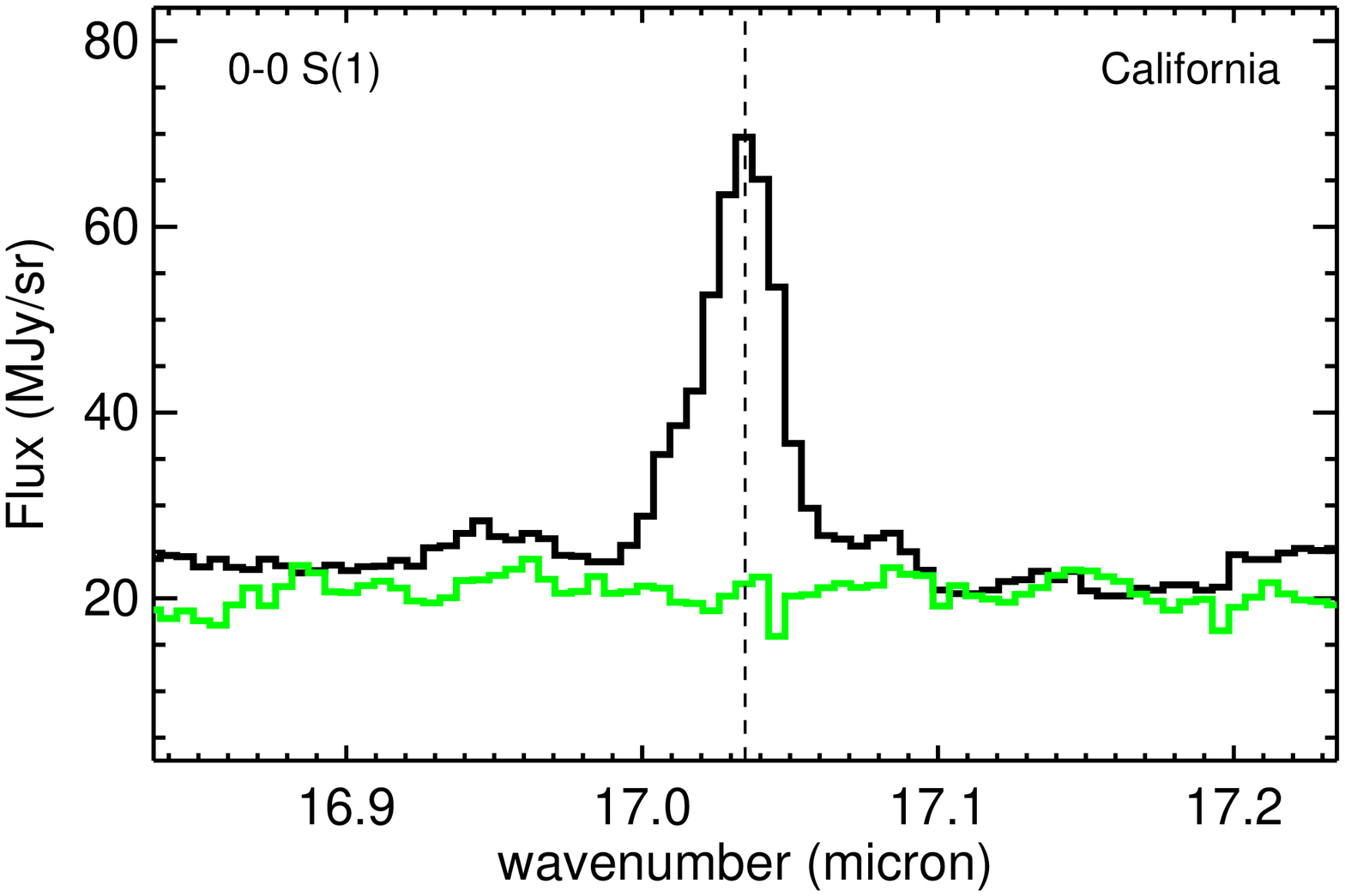,width=6cm,height=6cm,angle=0} }
\end{minipage}
\begin{minipage}[c]{6cm} 
%\centerline{ \psfig{file=/Users/emiliehabart/spitzer/data/IRS_High_Resolution_Laurent/IRS_HR_H2_LINE_S0_California.ps,width=6cm,height=6cm,angle=0} }
\centerline{ \psfig{file=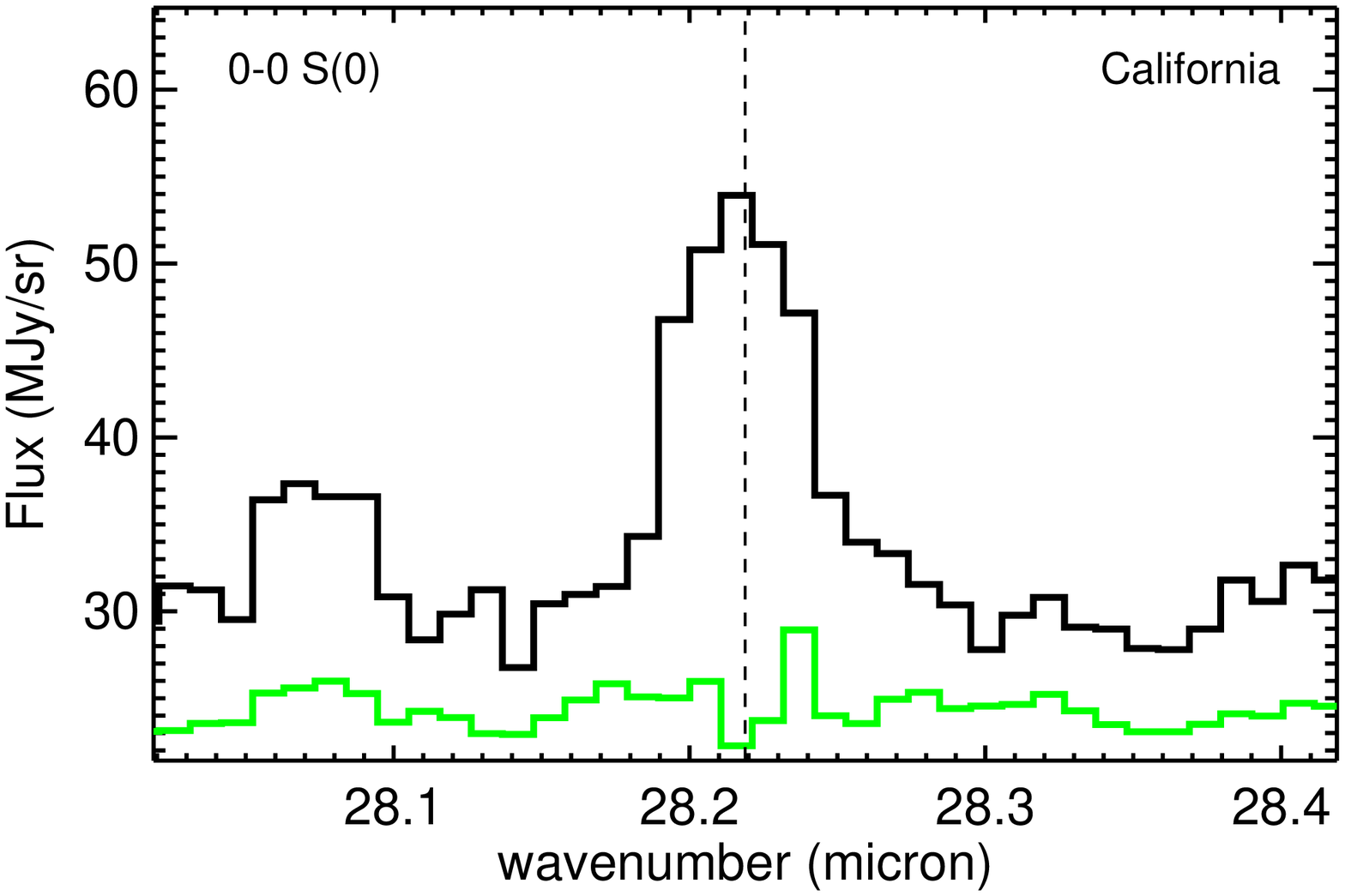,width=6cm,height=6cm,angle=0} }
\end{minipage}
\begin{minipage}[c]{6cm} 
%\centerline{ \psfig{file=/Users/emiliehabart/spitzer/data/IRS_High_Resolution_Laurent/IRS_HR_H2_LINE_S2_N7023E.ps,width=6cm,height=6cm,angle=0} }
\centerline{ \psfig{file=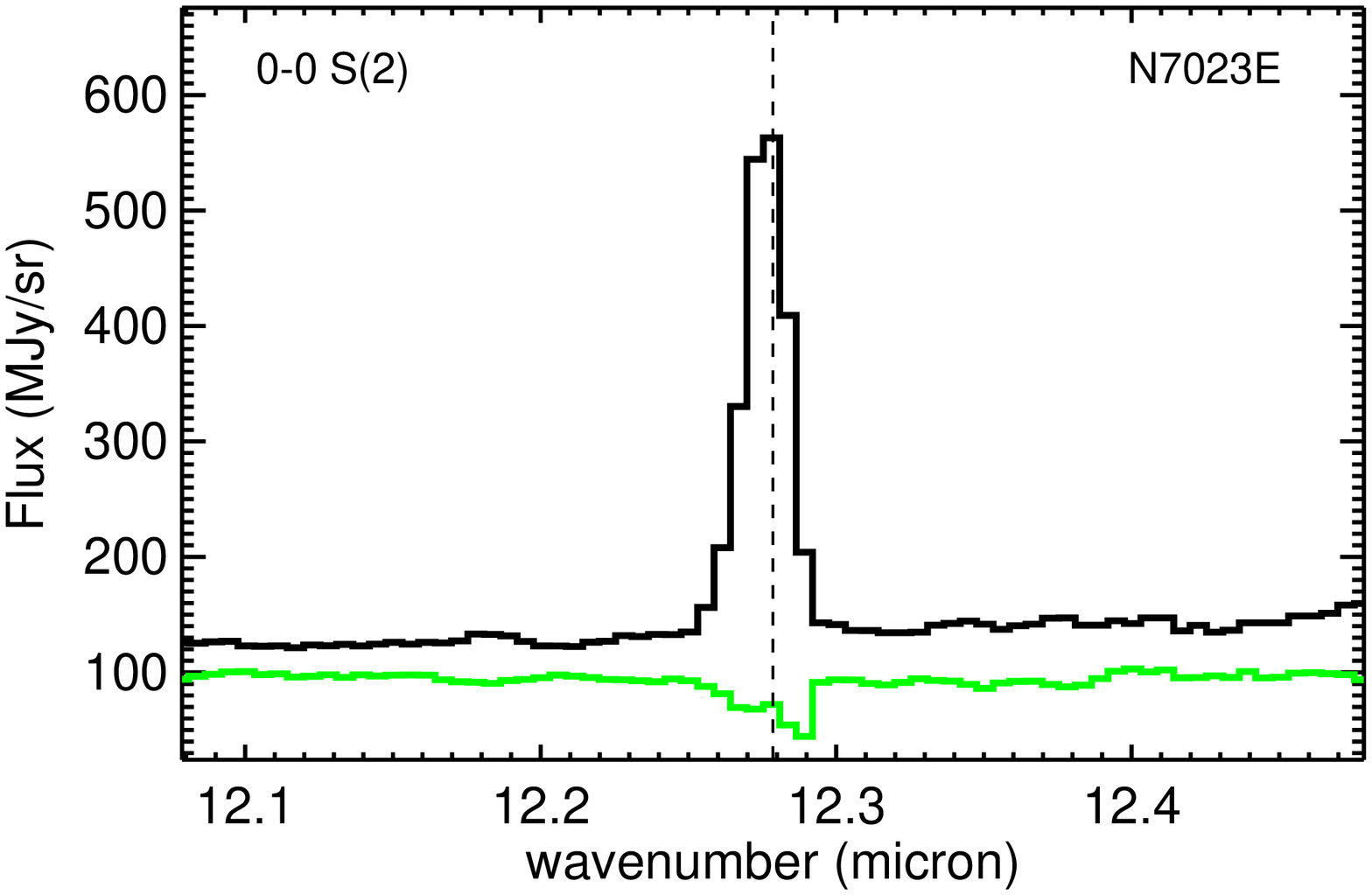,width=6cm,height=6cm,angle=0} }
\end{minipage}
\begin{minipage}[c]{6cm} 
%\centerline{ \psfig{file=/Users/emiliehabart/spitzer/data/IRS_High_Resolution_Laurent/IRS_HR_H2_LINE_S1_N7023E.ps,width=6cm,height=6cm,angle=0} }
\centerline{ \psfig{file=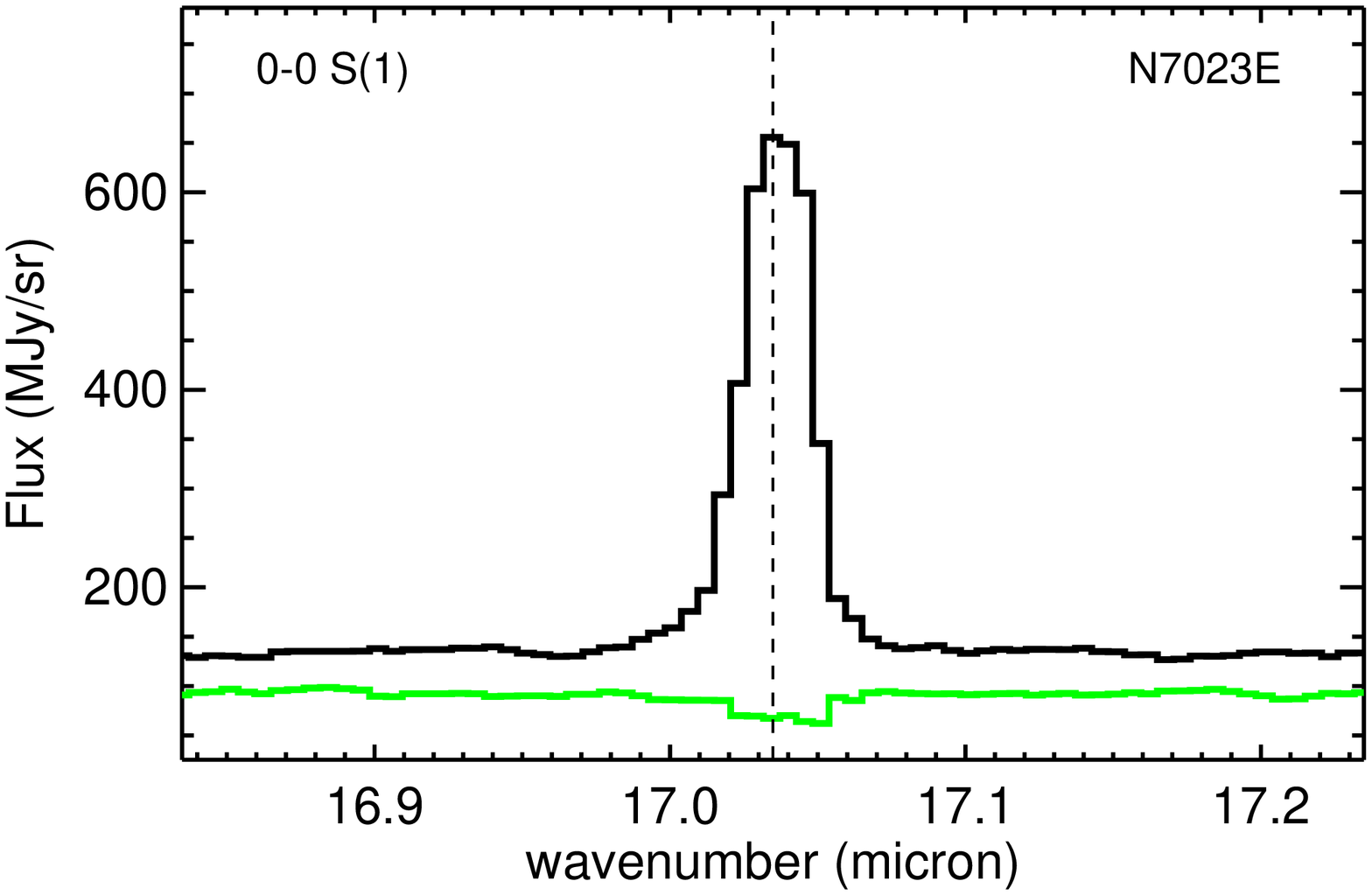,width=6cm,height=6cm,angle=0} }
\end{minipage}
\begin{minipage}[c]{6cm} 
%\centerline{ \psfig{file=/Users/emiliehabart/spitzer/data/IRS_High_Resolution_Laurent/IRS_HR_H2_LINE_S0_N7023E.ps,width=6cm,height=6cm,angle=0} }
\centerline{ \psfig{file=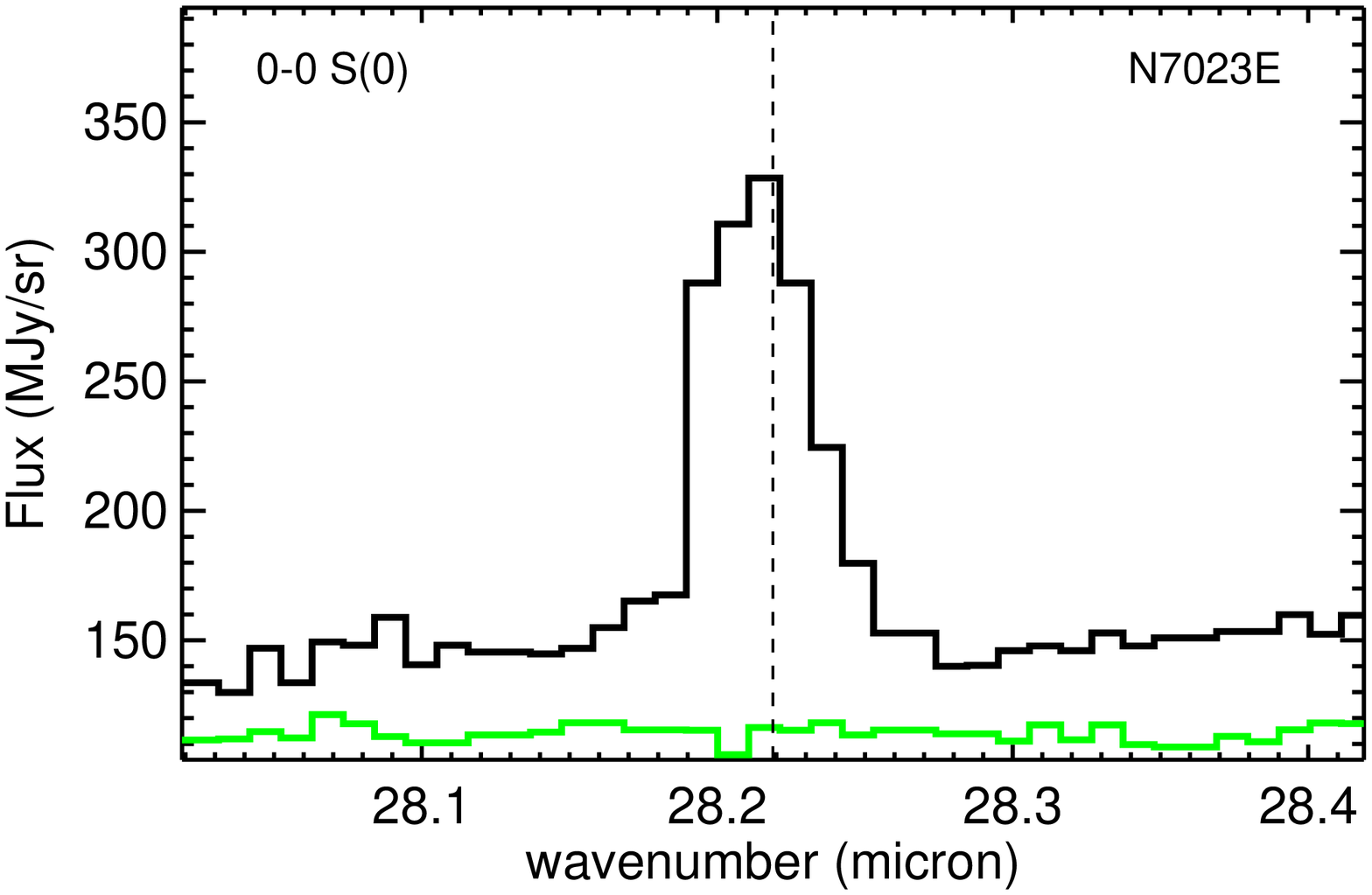,width=6cm,height=6cm,angle=0} }
\end{minipage}
%----------------------------------------------------
\caption{\em 
Average IRS spectra obtained with the high spectral resolution mode for three objects of our sample (L1721, California, and N7023E). 
In each panel, the line spectrum is shown with the local rms noise (for clarity, an arbitrary offset value has been added to the noise). 
The black and green lines represent the line spectrum and the noise, respectively.
%The local rms noise has been estimated from (UP-DOWN)/2: UP is the line scan where the wavelength increases with time and DOWN is the line scan where the line decreases with time. 
With increasing wavelength, the panels show H$_2$ pure rotational lines at  28.2, 17.03, and 12.29 $\mu$m.
}
\label{fig_spectrum_high_resolution}
\end{figure*}

%----------------------------------------------------
\begin{figure*}[htbp]
\leavevmode
\begin{minipage}[c]{9cm} 
%\centerline{ \psfig{file=/Users/emiliehabart/PDR/PDR_plot/all_objets/figure_article_carte_h2_17mic_convolved_N7023E.ps,width=8.0cm,angle=0} }
\centerline{ \psfig{file=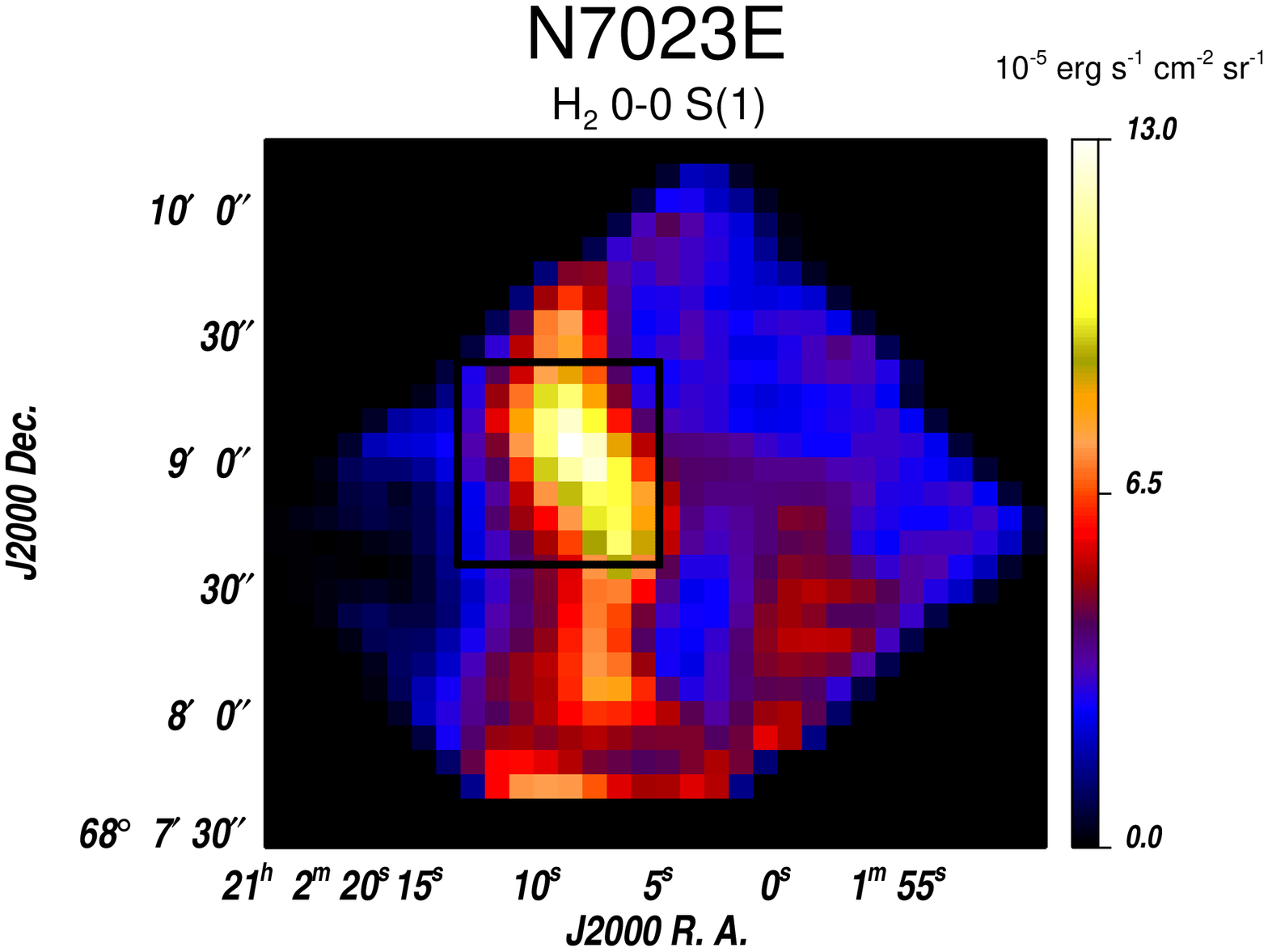,width=8.0cm,angle=0} }
\end{minipage}
\begin{minipage}[c]{9cm} 
%\centerline{ \psfig{file=/Users/emiliehabart/PDR/PDR_plot/all_objets/figure_article_carte_h2_17mic_convolved_Horsehead.ps,width=8.0cm,angle=0} }
\centerline{ \psfig{file=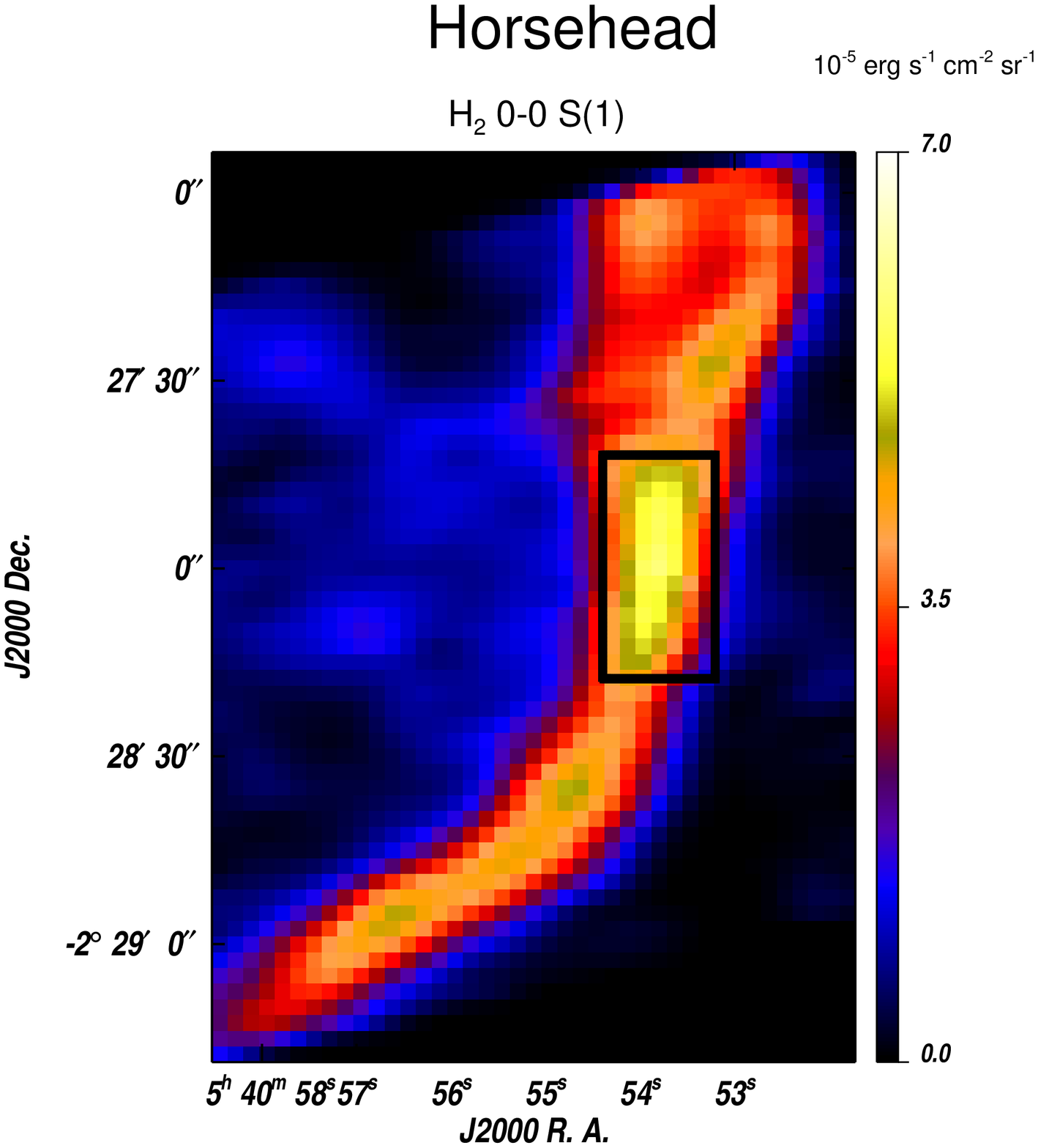,width=8.0cm,angle=0} }
\end{minipage}
\begin{minipage}[c]{9cm} 
%\centerline{ \psfig{file=/Users/emiliehabart/PDR/PDR_plot/all_objets/figure_article_carte_h2_17mic_convolved_rho_Oph.ps,width=8.0cm,angle=0} }
\centerline{ \psfig{file=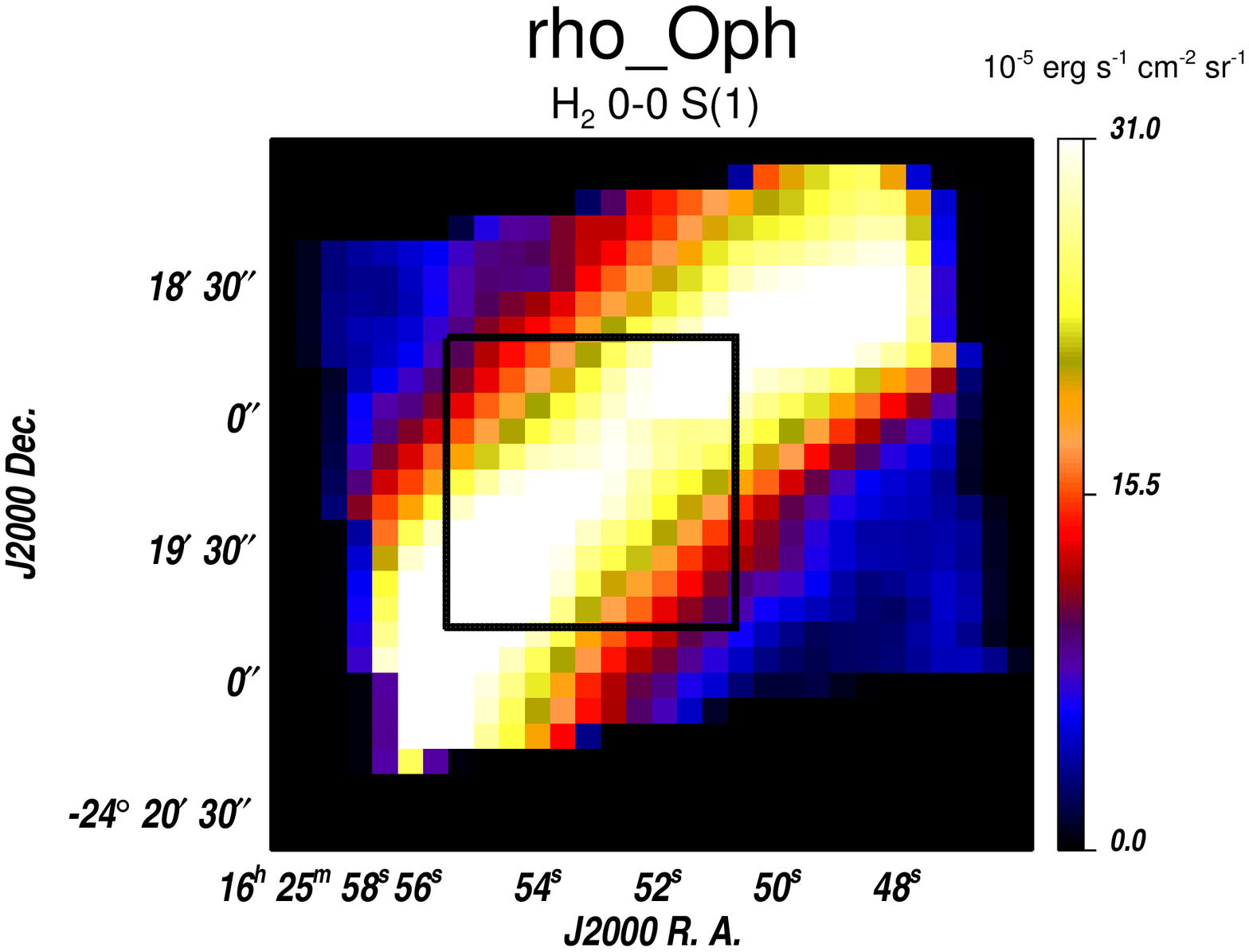,width=8.0cm,angle=0} }
\end{minipage}
\begin{minipage}[c]{9cm} 
%\centerline{ \psfig{file=/Users/emiliehabart/PDR/PDR_plot/all_objets/figure_article_carte_h2_17mic_convolved_N2023N.ps,width=8.0cm,angle=0} }
\centerline{ \psfig{file=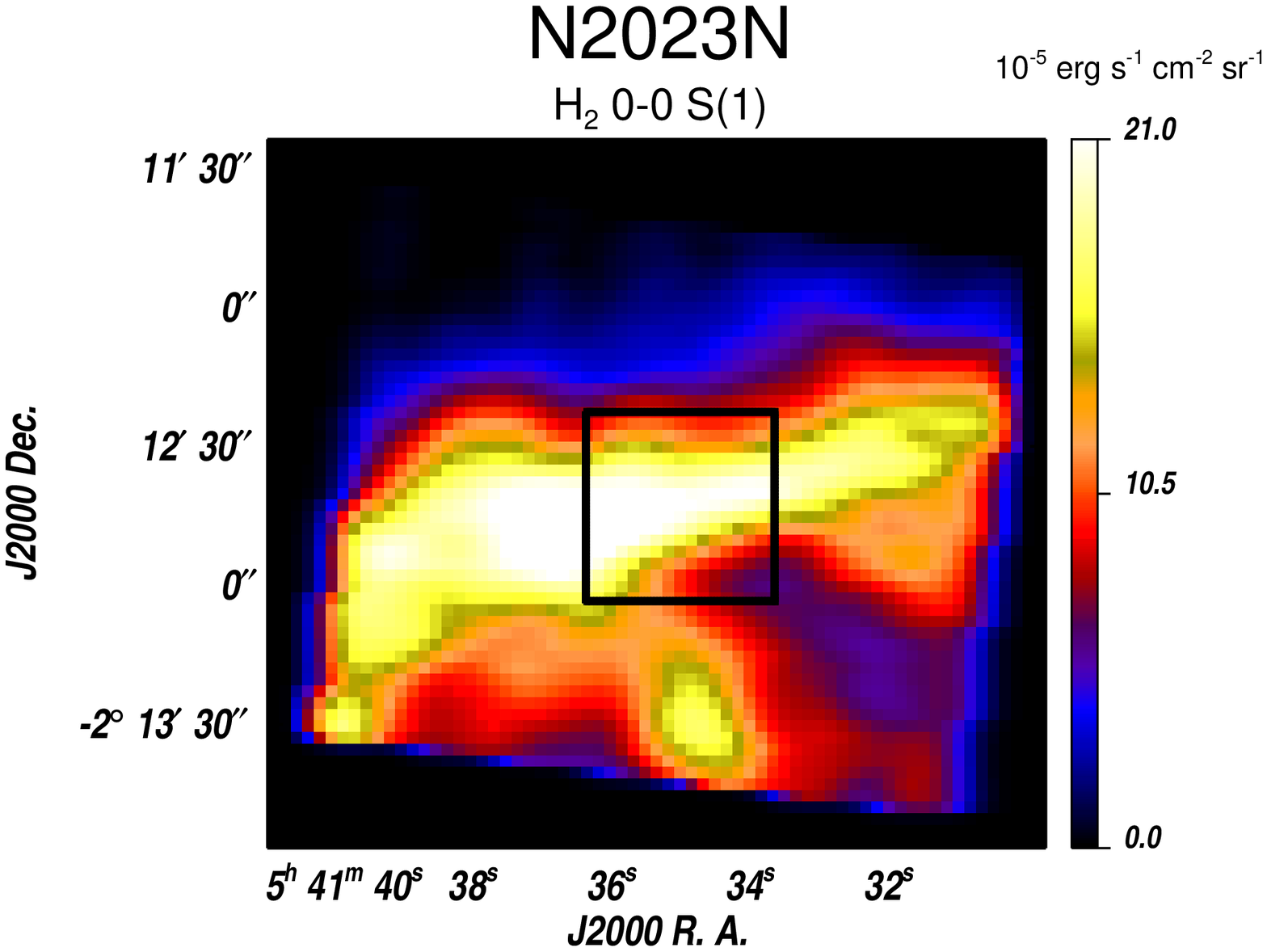,width=8.0cm,angle=0} }
\end{minipage}
%----------------------------------------------------
\caption{\em 
Map of the H$_2$ 0-0 S(1) emission line obtained with the IRS low spectral resolution mode.
The black boxes show the region where the spectra of Fig. \ref{fig_spectrum_low_resolution} have been averaged.
Figures of map in the other H$_2$ 0-0 S(0), S(2), and S(3) emission lines are shown in the Appendix.}
\label{fig_h2_map_obs}
\end{figure*}

\begin{table*}[htbp]

\caption{
H$_2$ observed line intensities, populations in each upper level, rotational excitation temperatures, ortho-to-para ratio, and column densities.
}
\label{table_obs}
\begin{center}

\begin{tabular}{|l|l|l|l|l|l|l|} 
\hline 
Source  & $I^a_{0-0~S(0)}$ & $I^a_{0-0~S(1)}$  & $I^a_{0-0~S(2)}$ & $I^a_{0-0~S(3)}$ & $I^a_{1-0~S(1)}$ & Ref.\\ 
        & (erg s$^{-1}$ cm$^{-2}$ sr$^{-1}$) & (erg s$^{-1}$ cm$^{-2}$ sr$^{-1}$) & (erg s$^{-1}$ cm$^{-2}$ sr$^{-1}$) & (erg s$^{-1}$ cm$^{-2}$ sr$^{-1}$) & (erg s$^{-1}$ cm$^{-2}$ sr$^{-1}$) & \\
\hline 
     L1721 &   6.9e-06[1.5e-06] &   2.3e-05[1.1e-06] &   9.1e-06[2.6e-06] &   $<$1.0e-05 &   - & (1) \\
California &   4.6e-06[7.0e-07] &   1.3e-05[1.3e-06] &   6.4e-06[3.2e-06] &   5.5e-06[2.7e-06] &   - & (1) \\
    N7023E &   3.0e-05[1.3e-06] &   1.5e-04[1.1e-05] &   1.5e-04[2.0e-05] &   9.2e-05[4.6e-05] &   - & (1) \\
 Horsehead &   1.2e-05[2.8e-06] &   4.3e-05[9.4e-06] &   5.4e-05[1.3e-05] &   5.9e-05[1.3e-05] &   7.5e-06[1.5e-06] & (1,2)\\
   rho Oph &   6.6e-05[1.4e-05] &   2.5e-04[5.5e-05] &   2.4e-04[4.8e-05] &   8.6e-05[8.6e-06] &   2.3e-05[5.7e-06] & (1,2)\\
    N2023N &   5.5e-05[1.2e-05] &   1.5e-04[3.2e-05] &   1.1e-04[4.0e-05] &   6.2e-05[3.1e-05] &   1.1e-05[2.2e-06] & (1,2)\\
\hline 
\end{tabular}

\par\bigskip
\begin{tabular}{|l|l|l|l|l|l|l|} 
\hline 
Source  & $N^b_{(0,2)}$ & $N^b_{(0,3)}$ & $N^b_{(0,4)}$ & $N^b_{(0,5)}$ & $N^b_{(1,3)}$ & $\frac{N_{(0,J\ge 4)}}{N_{(0,J<4)}}$ \\ 
  & (cm$^{-2}$) & (cm$^{-2}$) & (cm$^{-2}$) & (cm$^{-2}$) & (cm$^{-2}$)  &\\
\hline 
   L1721 &   4.2e+19[9.4e+18] &   5.2e+18[2.5e+17] &   2.6e+17[7.3e+16] &   $<$6.2e+16 &  - &      0.007[0.002] \\
California &   2.8e+19[4.2e+18] &   3.0e+18[2.9e+17] &   1.8e+17[9.1e+16] &   3.4e+16[1.7e+16] &  - &      0.007[0.004] \\
    N7023E &   1.8e+20[8.1e+18] &   3.4e+19[2.6e+18] &   4.2e+18[5.7e+17] &   5.7e+17[2.8e+17] &  - &      0.022[0.003] \\
 Horsehead &   7.1e+19[1.7e+19] &   9.8e+18[2.1e+18] &   1.5e+18[3.8e+17] &   3.7e+17[8.3e+16] &   1.0e+15[2.0e+14] &      0.023[0.008] \\
   rho Oph &   4.0e+20[8.6e+19] &   5.8e+19[1.2e+19] &   6.9e+18[1.3e+18] &   5.4e+17[5.4e+16] &   3.0e+15[7.7e+14] &      0.016[0.005] \\
    N2023N &   3.3e+20[7.3e+19] &   3.5e+19[7.2e+18] &   3.0e+18[1.1e+18] &   3.8e+17[1.9e+17] &   1.9e+15[3.7e+14] &      0.009[0.004] \\
\hline 
\end{tabular}

\par\bigskip
\begin{tabular}{|l| l|l|l|l| l|l|} 
\hline 
Source   & $T^c_{42}$ & $T^c_{53}$ & $T^c_{64}$ & $T^c_{75}$ & OTP$^d_{53}$ & OTP$^d_{75}$    \\
         & (K) & (K) & (K) & (K)    &   &   \\
\hline 
     L1721 &      206.[ 19.] &      $<$305. &   - &   - &        - &          - \\
California &      208.[ 25.] &      301.[ 38.] &   - &   - &        2.3 &          - \\
    N7023E &      269.[ 11.] &      327.[ 43.] &   - &   - &        1.3 &          - \\
 Horsehead &      264.[ 30.] &      398.[ 49.] &      458.[ 92.] &      747.[208.] &        1.5 &        2.7 \\
   rho Oph &      252.[ 22.] &      290.[ 18.] &      327.[ 20.] &      543.[ 48.] &        1.1 &        2.6 \\
    N2023N &      221.[ 26.] &      300.[ 45.] &   - &   - &        1.6 &          - \\
\hline 
\end{tabular}

\par\bigskip
\begin{tabular}{|l| l|l|l|l| } 
\hline 
Source  & N(H$_2$) $^e_{42}$  & N(H$_2$) $^e_{53}$ & N(H$_2$) $^e_{64}$ & N(H$_2$) $^e_{75}$   \\
        & (cm$^{-2}$)  &(cm$^{-2}$)&(cm$^{-2}$)&(cm$^{-2}$)\\
\hline 
 
     L1721 &      4.9e+20 &    5.1e+19 &       - &       - \\
California &      3.2e+20 &    3.0e+19 &       - &       - \\
    N7023E &      1.5e+21 &    2.9e+20 &       - &       - \\
 Horsehead &      6.1e+20 &    5.7e+19 &    7.5e+19 &    6.0e+18 \\
   rho Oph &      3.6e+21 &    6.4e+20 &    1.1e+21 &    2.3e+19 \\
    N2023N &      3.5e+21 &    3.5e+20 &       - &       - \\

\hline 
\end{tabular}

\end{center}
{\small
$^a$~Intensities at the emission peak with uncertainty (1$\sigma$) between brackets. 
%H$_2$ line intensities have been convolved to $\sim$10'' beam or multiplied by a beam factor.
%Line fluxes and error bars are obtained by fitting a Gaussian and a linear baseline to the profiles. 
%In the case of Oph W, the measured fluxes in the 0-0 S(0), S(1) lines were very similar to that obtained with ISO-SWS \cite[]{habart2003a}. The 0-0 S(J$\ge$2) intensity line was estimated from ISO data.
%In the case of Oph W, we used the Spitzer and ISO data where ISO-SWS and ISO-CVF data were available, only the Long Low ($\lambda\sim 14-38\mu$m, $\lambda/\Delta\lambda\sim$64-128) IRS sub-modules were used. 
%For the 0-0 S(2), S(3) lines, we used the ISO-SWS measurement. 
%In the case of Oph W, the 0-0 S(J$\ge$2) intensity line was estimated from ISO data.
%In the case of N2023S and the Orion Bar, we report the H$_2$ pure rotational lines measurements made with ISO-SWS.
%For the Orion, the measurements reported are similar to that recently inferred from ground-based spectrometer by Allers et al. (2005). 
%, Allers et al. 2005
%H$_2$ line intensities have not been corrected for dust attenuation.
\noindent
$^b$~The level populations $N_{(v_u,J)}$ are derived under the assumption that the line is optically thin.
%The have been estimated from the observed line intensities, corrected for dust extinction (see text for more details).  
%The last column gives the population ratio between the high $J=4,5$ and low $J=2,3$ levels.
\noindent
$^c$~Excitation rotational temperature, $T_{J_2J_1}$, derived from the ratio of the $N_{(0,J_2)}$ and $N_{(0,J_1)}$ column densities.  
%determined from 0-0 S(1)/S(0), 0-0 S(2)/S(0), 0-0 S(3)/S(1) and 0-0 S(4)/S(2) line intensity ratios assuming optically thin, thermalized emission.   
%For the 0-0 S(1)/S(0)  line intensity ratio, we assumed an ortho-to-para ratio of 3 (left values) and 1 (right values).  
\noindent
$^d$~Fit of the ortho-to-para (OTP) ratio of the  $J_1-J_2$ levels and assuming thermalized emission at the temperature $T_{J_2J_1}$.  
% For L1721, not enough H$_2$ lines have been measured to do a proper fit.  
%;!!!!!!!!!!!!ORTHO-TO-PARA RATIO
%;Given a rotational temperature compute of ortho/para to have a best fit of N(Ju) 
%;N(Ju)~(2*Ju+1)*(2I+1)*exp(-Tu/Trot)
%;Then 
%;ortho/para~(2I[ortho]+1)/(2I[para]+1)
%;ortho/para~N(Ju impaire)/N(Ju paire)/((2*Ju impaire+1)/(2*Ju paire+1))/exp(-(Tu impaire-Tu paire)/Trot)
%Fit of the ortho-to-para (OTP) ratio of the  $J=5-7$ levels and assuming thermalized emission at the temperature $T_{75}$:O/P=3
\noindent
$^{e}$~The total H$_2$ column density inferred from the  $N_{(0,J_1)}$ column densities
assuming optically thin, thermalized emission at the temperature $T_{J_2J_1}$. 
\noindent
Ref. (1) Spitzer Space Telescope; 
(2) ESO 3.6-m telescope at La Silla \cite[]{habart2003a}
and New Technology Telescope with the SOFI instrument \cite[]{habart2005,compiegne2008}.
%(3) data obtained from  ESO ground based telescopes \cite{habart2003a};
%Note that in the case of Oph W, the 0-0 S(2), S(3) lines intensity were estimated from ISO data.
%(4) data obtained with the SOFI instrument at the ESO New Technology Telescope.
%(5) ISO-SWS and ground data from \cite{draine99a} and \cite{field98,burton98};
%(5) ISO-SWS and ground data from \cite{moutou99} and \cite{field98}; %, \cite{burton98};
%(6) ISO-SWS and ground data from Bertoldi (private communication) and \cite{vanderwerf96}. 
%The measurements reported here for the pure rotational lines intensities are similar to that recently inferred from ground-based spectrometer by \cite{allers2005}. 
}
\end{table*}

\subsection{IRS spectrometer data}

The selected PDRs were observed using the IRS spectrometer in the low ($\lambda/\Delta\lambda\sim$60-127) and/or high ($\lambda/\Delta\lambda\sim$600) spectral resolution  mode. 
In the low spectral resolution mode, spectral line maps were obtained using the ``spectral mapping''  mode
with a number and size of steps suitable for each PDRs
in order to cover the infrared emitting region. 
Projection onto the same spatial grid (2.5''/pixel) and convolution into the same beam (Gaussian with a full width high maximum of 10.7'') were performed to compare the different line maps obtained.
In the high spectral resolution mode, several individual sky positions or smaller spectral line maps were obtained.
%The positions and areas observed with the different modules are show in Fig. \ref{fig_chart} for each object.
We used the S17 version of the data delivered by the Spitzer Science Center. 
For details of our observations and data reduction method see \cite{joblin2005} and \cite{compiegne2007}. 
% AA  Il ne faut pas mettre berne qui utilise CUBISM pour reduire les donnedes

We present in Fig. \ref{fig_spectrum_low_resolution} the averaged IRS spectra obtained with the low spectral resolution mode in the short-low (SL, 5.2 - 14.5 $\mu$m) and long-low (LL, 14 - 38 $\mu$m) modules. 
The spectra are averaged in a box marked in Fig. \ref{fig_h2_map_obs} showing maps in the H$_2$ 0-0 S(1) line emission at 17.03 $\mu$m. The averaged area centered at the emission peak covers the emission zone observed by both the IRS SL and LL submodules.
%FOR N2023 SMALL BOX (NOT ALONGATED) AS MAP S(2), S(3) SMALL WITH SL
%FOR RHO OPH SMALL BOX (NOT ALONGATED) AS ISO-CVF and ISO-SWS DATA (NO SL DATA) 
%The averaged area covered the emission zone and are centered at the emission line peak.
For L1721 and California, all pixels are averaged to increase the signal-to-noise ratio, 
since these extended PDRs are fainter and do not present any detected spatial variation 
within the field observed with IRS.
In the case of California, N7023E and rho Oph, only the LL IRS submodule is available.
These three objects have already been observed with the Circular Variable Filter (CVF, $\lambda/\Delta\lambda\sim$45, 5.15-16.5 $\mu$m) of ISOCAM.
We present the combined ISO-CVF and IRS-LL spectra in Fig. \ref{fig_spectrum_low_resolution}.

The IRS wavelength coverage allowed us to detect several strong H$_2$ pure rotational lines from 0-0 S(0) to S(3) at 28.2, 17.03, 12.29, and 9.66 $\mu$m, the aromatic band features at 6.2, 7.7, 8.6, 11.3 $\mu$m, the dust mid-IR continuum emission 
and the fine structure lines of ionized gas [NeII] at 12.8 $\mu$m, [SIII] at 18.7 and 33.4 $\mu$m and [SiII] at 34.9 $\mu$m.
The rotational lines 0-0 S(4), S(5) at 8.03 and 6.91 $\mu$m, blended with the strong aromatic feature at 7.7 $\mu$m and possibly the ionized argon line  [ArII] at 6.98 $\mu$m, are marginally detected in some sources. 
%Detection of  0-0 S(4), S(5) at 8.03 and 6.91 $\mu$m : Horsehead : yes; N2023N : TBD ; L1721 : too faint ; California, N7023E, & rho Oph : not with CVF
In the CVF spectra of the California, N7023E and rho Oph, the rotational line 0-0 S(2) at 12.29 $\mu$m appears blended with the [NeII] line at 12.8 $\mu$m because of the CVF lower spectral resolution.
To estimate the flux in the 0-0 S(2) line for these objects, we used the high spectral resolution IRS data and/or previous ISO observations (see below).
In the case of the faint L1721 PDR, the 0-0 S(2) and S(3) lines are marginally detected in the IRS low spectral resolution data. We used the high spectral resolution data to estimate the flux in the 0-0 S(2) line (as described below). 
For the 0-0 S(3) line, an upper limit was derived from the low spectral resolution data.

For L1721, California and N7023E, we present in Fig. \ref{fig_spectrum_high_resolution} the spectra obtained with the high spectral resolution mode in the short-high (SH, 9.9-19.6 $\mu$m) and long-high (LH, 18.7-37.2 $\mu$m) modules.
The data are averaged in the module slits.
The wavelength coverage allows us to detect the 0-0 S(0), S(1) and S(2) lines at 28.2, 17.03, and 12.29 $\mu$m.
Both the S(1) and S(0) lines, which are visible in the low spectral resolution data, are clearly detected in the high spectral resolution data.
We find good agreement between the low and high spectral resolution measurements.
The S(2) line is clearly better detected in the high spectral resolution data.
 Nevertheless, for one object (California), the S(2) line is marginally detected with only 2$\sigma$ detection in amplitude.

When available, we complement and compare our measurements with ISO short wavelength spectrometer (SWS, $\lambda/\Delta\lambda\sim$1500-2500) data.
For one object of our sample (rho Oph),  a series of pure rotational lines (0-0 S(0) through S(5)) 
have been detected using SWS at several positions towards the infrared emitting region \cite[]{habart2003a}. The SWS positions are localized in the IRS-LL field (see Fig. 1 in \cite{habart2003a} and Fig. \ref{fig_h2_map_obs} in this paper).
%Only the IRS-LL sub-module is available for rho Oph.
%For the rotational S(0) and S(1) lines detected with IRS-LL (only this sub-module were available for rho Oph), we compare the line flux measured by IRS and SWS.
For the first rotational S(0) and S(1) lines detected with both IRS/Spitzer and SWS/ISO, 
convolution into the same beam (SWS beam $\sim$20'') were performed in order to compare the data.
We find good agreement between the IRS and SWS data with differences on the lines flux $\le$30\%.
To estimate the flux in the other rotational lines S(2)-S(5), we spatially averaged the SWS positions data situated over the box marked in  Fig. \ref{fig_h2_map_obs}.
%In order to compare the SWS and IRS measurements, we compute for each SWS position the flux of the S(0) and S(1) lines measured with IRS-LL and averaged in the SWS beam ($\sim$20''). We find a good agreement between the IRS and SWS measurements.
%Note : four SWS positions. 3 of them in the IRS-LL field. position 2 in the center of the area - position 1 et 3 at the extremite.

In Table \ref{table_obs}, we report the observed intensities of the first four H$_2$ pure rotational lines. %derived from the averaged spectra.
Line fluxes and error bars were generally obtained by integrating the average spectra within the lines. 
Typical error bars are about 10-30\%.
In Table \ref{table_obs}, we also report the intensities for the H$_2$ 1-0 S(1) rovibrational line
 obtained from high spatial resolution ground based observations when available. 
The 1-0 S(1) data have been spatially averaged over the box marked in  Fig. \ref{fig_h2_map_obs}.

\subsection{Column densities, temperatures, and ortho-to-para ratio}
\label{column_densities_temperatures}

The intensity $I_{ul}$ of a transition between two levels $u$ (upper) and $l$ (lower)
can be converted into the population $N_u$ of the upper level, under the assumption that the line is optically thin so 
that $I_{ul} = N_u A_{ul} h\nu/4\pi$. 
The level populations $N_u$ derived for each transition are given in Table \ref{table_obs}.
We did not correct for extinction for the mid-IR H$_2$ rotational line intensities, 
as the effect of reddening is within our error bars for $A_V$=5-15,
which is the typical range of values for our PDRs (see references in Table \ref{table_sample}).
It is only at shorter wavelengths that dust extinction will significantly attenuate the emission.
For the 1-0 S(1) line at 2.12 $\mu$m, we correct for dust attenuation using a factor $A_K/A_V=0.14$.
% and $A_V$ derived from visual extinction maps, sub-mm dust emission or CO observations (see references in Table \ref{table_sample}).

Based upon the $N_u$ column densities, we present the excitation diagrams in Fig. \ref{fig_diagram_mod_obs}.
In Table \ref{table_obs}, we report  the rotational excitation temperatures, $T_{J_2J_1}$, 
derived from the ratio of the $N(0,J_2)$ and $N(0,J_1)$ column densities. Because the ortho-to-para ratio could show departures from equilibrium, the quantity $T_{J_2J_1}$ is estimated when $J_2$ and $J_1$ are either both odd or both even. 
In Table \ref{table_obs}, we give the rotational excitation temperatures $T_{42}$, $T_{53}$, $T_{64}$, and $T_{75}$.
%The error bars on the rotational temperatures are about 10-20\%.
%In the excitation diagrams, the population of the upper level of the transition, $N_u$, is plotted against the energy of the upper level, $E_u$. when the lowest levels are thermalized by collisions, their populations in a log scale should be proportional to 1/$T_{gas}$.
%For each object of our sample, the derived excitation diagram is similar over the H$_2$ lines emitting region. 
We find that the rotational diagrams exhibit a noticeable curvature.
A single temperature cannot describe the full set of observed line intensities
and the inequalities $T_{42} <T_{53} < T_{64} < T_{75}$ are a manifestation of the curvature apparent in the rotational diagrams.
The v=0 rotational distributions are characterized by an excitation temperature between $\sim$200 and 700~K. 
In order to fit the  rotational diagrams, a combination of at least two H$_2$ gas components is required:
one cool/warm ($\sim$100-300K) and another warm/hot ($\sim$300-700K) with much lower column densities (about a few percent of the main cool/warm component). 
%A very small column density of hot H$_2$ is sufficient to produce the intensities observed for the higher J lines. 
%Note that the inferred relative populations $N_u$ of high ($J=4,5$) to low ($J=2,3$) rotational levels are remarkably comparable in our PDRs in spite of their different characteristics.

In Table \ref{table_obs}, we also report a fit of the ortho-to-para ratio (OTP) for the $J_1$ to $J_2$ levels with $J_1=3$, $J_2=5$ and
 $J_1=5$, $J_2=7$ and assuming thermalized emission at the temperature  $T_{J_2J_1}$.
Unfortunately, not enough H$_2$ lines have been measured for L1721 to do a proper fit of the OTP ratio.
For California and N7023E, the OTP ratios derived are about $\sim$1-2 for the $J_1=3$ to $J_2=5$ levels.
For the rotational levels $J_1=5$ to $J_2=7$, not enough H$_2$ lines have been measured in the CVF spectra to fit the OTP ratio.
In the moderate excited PDRs, such as Horsehead and rho Oph, the OTP ratios derived are about $\sim$1 for the $J_1=3$ to $J_2=5$ levels, while about $\sim$3 for the $J_1=5$ to $J_2=7$ levels.
The local thermodynamic equilibrium ratio OTP$_{\rm LTE}$ is equal to 3 at the relevant temperatures ($T>$200K).
%In the low/moderate excited regions, the OTP ratio derived are lower than the local thermodynamic ratio (OTP$_{\rm LTE}\sim$ 3) at the temperatures of relevance.
The non-equilibrium behavior has already been noted in the ISO observations of rho Oph \cite[]{habart2003a} and other PDRs \cite[e.g.,][]{moutou99,fuente99}.
The H$_2$ column densities inferred from the measured line intensities and assuming thermalized emission at the temperatures derived are also reported in Table \ref{table_obs}.

In spite of their simple appearance, any interpretation of these diagrams has to take several factors into account. 
First, the gas temperature could vary very rapidly through the PDR layer,
and it is likely that with our spatial resolution ($\sim$10'' or $\sim$0.02 pc at 400 pc) the temperature structure is not spatially resolved.
%Moreover, projection effects along the line of sight can be important. 
%Temperatures of different gas components coming from different zones can thus appear together.
Second, only the lowest rotational levels are collisionally excited, and this holds up to different $J$ for low- and high-FUV field and gas density PDRs.
Third, the excited rotational populations can be affected by UV pumping.
An analysis using PDR models is thus necessary.
%In the following, we present the results of PDR model calculations for the range of FUV radiation field and densities values prevailing in our sample.

\section{PDR model}
\label{model_presentation}

\subsection{Description}

To analyze the H$_2$ emission observations, we used an updated version of the Meudon PDR code described in \cite{lepetit2006}.
The model considers a stationary plane-parallel slab of gas and dust illuminated by an ultraviolet radiation field coming from one or both sides of the cloud. It solves the chemical and thermal balances iteratively at each point of the cloud.

Here, we present a grid of isochoric models with radiation field coming from one side.
The adopted densities are $n_{\rm H}=10^{3}$, 5 $10^{3}$, $10^{4}$, 5 $10^{4}$, and $10^{5}$ $\rm{cm}^{-3}$ in the range of values prevailing in our PDR sample (see Table \ref{table_sample}). 
For the incident radiation field, we use  the Draine UV radiation field scaled by the factor $\chi$.
The adopted scaling factors of the Draine radiation field at the cloud edge are $\chi=0.5, 5, 50, 500, 5000$ in the range of values prevailing in our PDR sample (see Table \ref{table_sample}).  
%We also compute models with blackbody or Kurucz model spectrum with the suitable stellar parameters given in Table \ref{table_sample}. %we have found in our modeling that this is not very important for the present studies. 

For the gas-phase abundance, our assumed abundances are consistent with estimates for the average abundances in the interstellar diffuse and translucent medium as derived from UV absorption measurements \cite[]{savage96,meyer97,meyer98}.
For the grain properties, the size distribution is assumed to be a power-law function following the MRN law; the UV extinction curve is taken as equal to a standard galactic expression; the grain absorption cross sections is taken from \cite{draine84,laor93}.
These values are determined for spherical  particles of graphite and silicates for a sample of radii greater than 1 nm. Values corresponding to radii smaller than 50 nm are considered to mimic polycyclic aromatic hydrocarbon (PAH) properties.
The heating rate due to the photoelectric effect on small dust grains (including PAHs) is derived from the formalism of \cite{bakes94} with some significant upgrades \cite[see][]{lepetit2006}.
A summary of the most important model input parameters are given in Table~\ref{table_mod}.

\begin{table*}[htbp]
\caption{PDR model parameters}
\label{table_mod}
\begin{center}
\begin{tabular}{|l|l|l|} 
\hline
Parameter  & Value & Comment \\
\hline 
%$A_v$ & 20 & total visual extinction \\
%cos($\theta$) & 0.15 & viewing angle \\
$\zeta$ & 5 10$^{-17}$ s$^{-1}$ & cosmic ray ionization\\
$\rm v_{turb}$ & 2 km s$^{-1}$ & turbulent velocity\\
Extinction     &   standard galactic &  parameterization of \cite{fitzpatrick90}\\
$R_V$ &  3.10 &  Av / E(B-V) \\
$\omega$ &  4.20e-01 & dust albedo\\
$g$ &  6.00e-01 & dust anisotropy factor\\
$G$& 1.00e-02 & Dust to gas mass ratio\\
$\rho$ & 3 g/cm3 &   Dust density (g/cm3)\\
$\alpha$ & 3.50e+00 & MRN dust size distribution index  \\
$a_{min}$ & 1.00e-07 cm & dust minimum radius \\
$a_{max}$ & 1.50e-05 cm & dust maximum radius \\
$R_f$ & $1.5~10^{-16}$ cm$^3$ s$^{-1}$ & H$_2$ formation rate \cite[]{habart2004}\\
$q_{H-H_2}$ &  & H + H$_2$ collision rate \cite[]{lebourlot99}\\
He/H & 0.1 & \\ %Gas abundance\\
O/H  &  3.2 10$^{-4}$   & Gas abundance \cite[]{meyer98}\\ % \\
C/H  &  1.3 10$^{-4}$   & Gas abundance \cite[]{savage96}\\ %Gas abundance\\
N/H  &  7.5 10$^{-5}$   & Gas abundance \cite[]{meyer97}\\ %Gas abundance \\
%S/H  &  1.85 10$^{-5}$   &\\ %Gas abundance \\
%Fe/H &  1.5 10$^{-8}$   & Gas abundance\\
\hline 
\end{tabular}
{\small }
\end{center}
%For the collisional excitation and de-excitation of H$_2$ with H$^0$,  He and H$_2$, we adopt the inelastic rates of le Bourlot et al. 1999 and references therein.
\end{table*}

\subsection{H$_2$ specifications}

For the collisional excitation and de-excitation of H$_2$, we adopt here the quantum  mechanical calculations given by \cite{lebourlot99}.  
It must be underlined that the cross sections for collisional excitation and deexcitation of H$_2$ are still not completly known. 
The quantum  mechanical calculations adopted here contain only rate coefficients for the first fifty levels ($T<$20,000 K) and do not include reactive transitions. 
However, the populations of the ($\rm v$=0, $J<10$) levels, where collision data exist, should be only secondarily affected by the missing collision data \cite[]{shaw2005,kaufman2006}.
Only for high excited levels (e.g., the 1-0 S(1) line) and high densities and UV fields (e.g., Orion Bar for which the population of the excited states is determined either by collisions or by collisionally modified FUV pumping) will the rate uncertainties affect the  emission line intensities.

Although there is a consensus that H$_2$ forms on the surface of dust grains  \cite[]{gould63,hollenbach71,jura75,duley84}, the mechanism has not yet been fully understood.
This is partly because we are ignorant of interstellar grain properties and we do not understand surface reactions in the interstellar context.
Here, we adopt a simple empirical prescription using an H$_2$ formation rate as derived from astronomical observations.
In the ISM studies, the commonly adopted value of the formation rate $R_f^0$ is 3 $10^{-17}$ cm$^3$ s$^{-1}$
inferred from UV absorption lines  from diffuse clouds \cite[]{jura75,gry2002}.
However, studies based on ISO observations suggest that a higher rate of about five times this standard value is required to account for H$_2$ emission in moderately excited PDRs \cite[such as rho Oph, S140, and IC 63,][]{habart2004}.
Such a high H$_2$ formation rate in warm gas could be explained by a formation mechanism involving a chemisorbed H atom, analogous to the mechanism presented by \cite{cazaux2004}. 
Based on this, we assume a single constant value of $R_f = 5 \times R^0_f$.
This assumption is discussed in Sect. \ref{origin}.

Another parameter is the internal energy of the nascent H$_2$.
The energy released by the nascent molecule (4.5 eV) is distributed between the grain excitation, internal energy, and kinetic energy of released molecules. The branching ratios are not known and probably depend on conditions in the cloud and the nature of the grain itself.
Using standard option based on an equipartition argument, we assume that the internal energy (1.5 eV) is distributed with a Boltzmann law among all the energy levels.
This assumption is discussed below.

\subsection{H$_2$ excitation}
\label{model_prediction}

The excited states of H$_2$ are populated by inelastic collisions with gas phase species, UV pumping or the formation process. 
%The population distribution of H$_2$ levels is a function of the gas density and temperature and the UV flux.
The first pure rotational lines (e.g., 0-0 S(0) and S(1)) essentially result from collisional excitation, since their upper states ($\rm v=0$, $J=$2 and 3) are relatively low lying and their critical densities are low even at low temperatures \cite[$n_{crit} < 10^4$ cm$^{-3}$ for $T\ge 100$ K,][]{lebourlot99}.
%The population distribution of the low rotational $\rm v=0-0, J\le 3$ H$_2$ levels (i.e., 0-0 S(0)-S(1)) is essentially in LTE 
These line intensities depend mainly on the gas temperature at the photodissociation front.
In contrast, the high rotational and ro-vibrational lines  (e.g., 1-0 S(1)) result essentially from the decay of electronically excited states that are pumped through the absorption of FUV photons. 
The line intensities do not reflect the gas temperature but the fraction of the FUV photon flux pumping H$_2$. 
UV pumping could also contribute significantly to the excitation of the pure rotational 0-0 S(2)-S(5) lines, since their upper states ($\rm v=0$, $J=$4-7) are relatively high and their critical densities are high even at moderate temperatures \cite[$n_{crit} \ge 10^4$ cm$^{-3}$ for $T \le 500$ K,][]{lebourlot99}. %($n_{crit}\ge 10^4$ cm$^{-3}$ for $T \le 500$ K).
For a given line, the relative importance of UV pumping and collisions depends on the PDR physical conditions.
In our low $n$ and $\chi$ regions (such as L1721, California with $n\sim 10^3$ cm$^{-3}$, $\chi \sim$5-20), 
the rotational  $J \ge 3$ levels are  mainly controlled by ultraviolet pumping rather than collisions.
For moderate $n$ and $\chi$  regions (such as N7023E, Horsehead with $n\sim 10^4-10^5$ cm$^{-3}$, $\chi \sim$50),  collisions dominate the excitation for the $J\le 3-4$ levels, while UV pumping contributes mostly to the excitation of the $J \ge 4-5$ levels.
For higher $n$ and $\chi$  regions (such as rho Oph, N2023N with $n\sim 10^4-10^5$ cm$^{-3}$, $\chi \sim$250-550), collisions dominate the excitation until the $J = 5-6$ levels. 
%On the contrary, in the high-$n$ and -$\chi$  regions (such as N2023S, Orion Bar) inelastic collisions will dominate.

The internal energy of the nascent H$_2$ can also specifically affect the level populations.
%Here, we used the standard assumption based on the equipartition argument (see Sect. \ref{model}) but the internal energy distribution of the newly formed H$_2$ on grain is not known.
However, for an equilibrium between photodissociation and reformation, the ratio of the formation to UV pumping rates is only about $\sim$15/85.
Of all the UV photons absorbed by H$_2$ only 10 to 15 \% lead to dissociation.
Thus, unless the level distribution of newly formed H$_2$ is strongly concentrated toward a small
number of high energy levels, %(which is not the case for the rotational $J<10$ levels),
%(for the sharing out the energy formation of 4.5 eV), 
the H$_2$ formation excitation will not specifically affect the H$_2$ spectrum.
%Moreover, for the low pure rotational levels inelastic collisions are dominant.

\section{Comparison between PDR model and observations}
\label{comparison_observation_model}

\begin{figure*}[htbp]
\leavevmode
\begin{minipage}[c]{9cm} 
%\centerline{ \psfig{file=/Users/emiliehabart/PDR/PDR_plot/all_objets/figure_diagramme_obs_mod_L1721.ps ,width=9cm,height=7.5cm,angle=0} }
\centerline{ \psfig{file=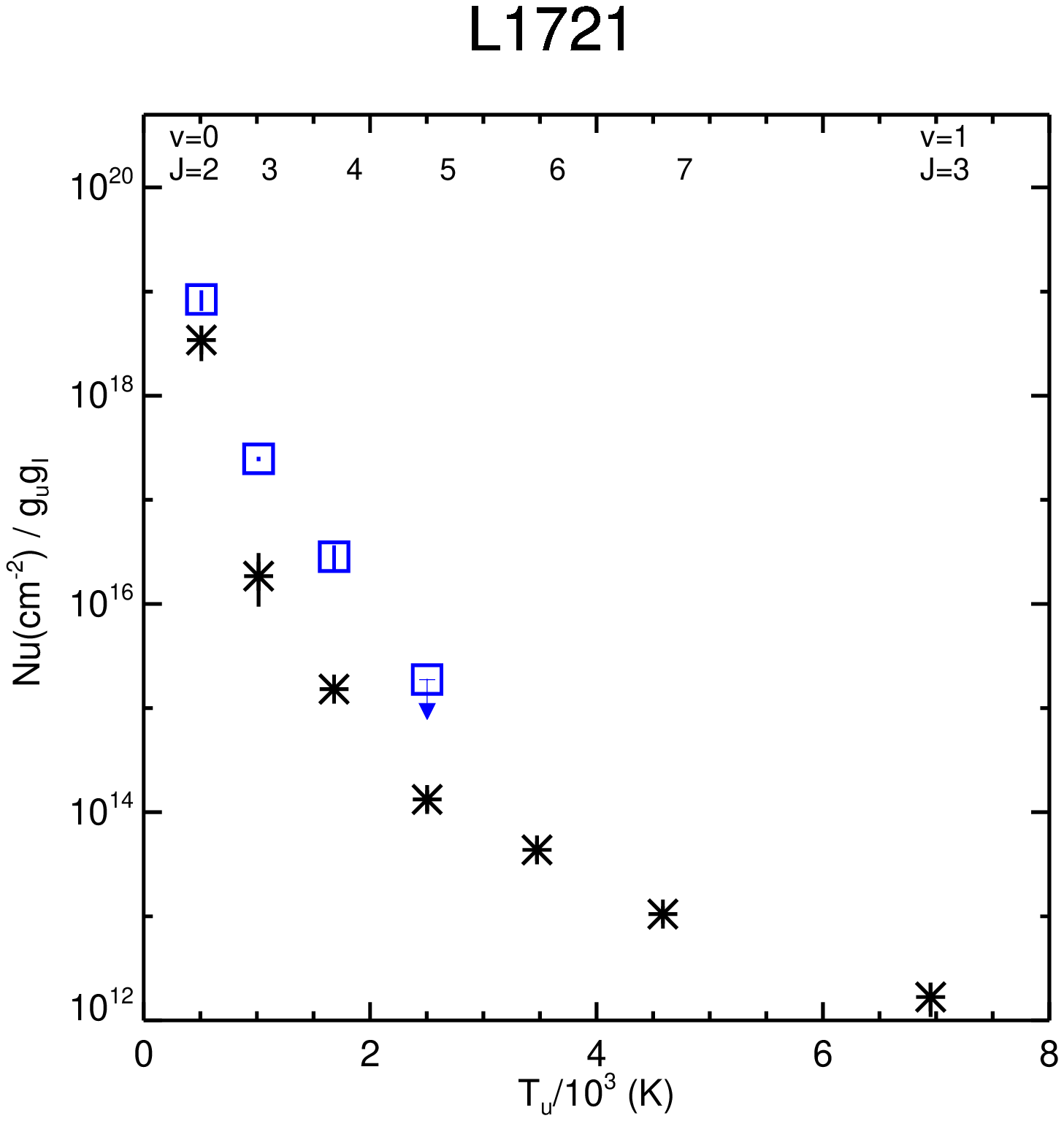 ,width=9cm,height=7.5cm,angle=0} }
\end{minipage} 
\begin{minipage}[c]{9cm} 
%\centerline{ \psfig{file=/Users/emiliehabart/PDR/PDR_plot/all_objets/figure_diagramme_obs_mod_California.ps ,width=9cm,height=7.5cm,angle=0} }
\centerline{ \psfig{file=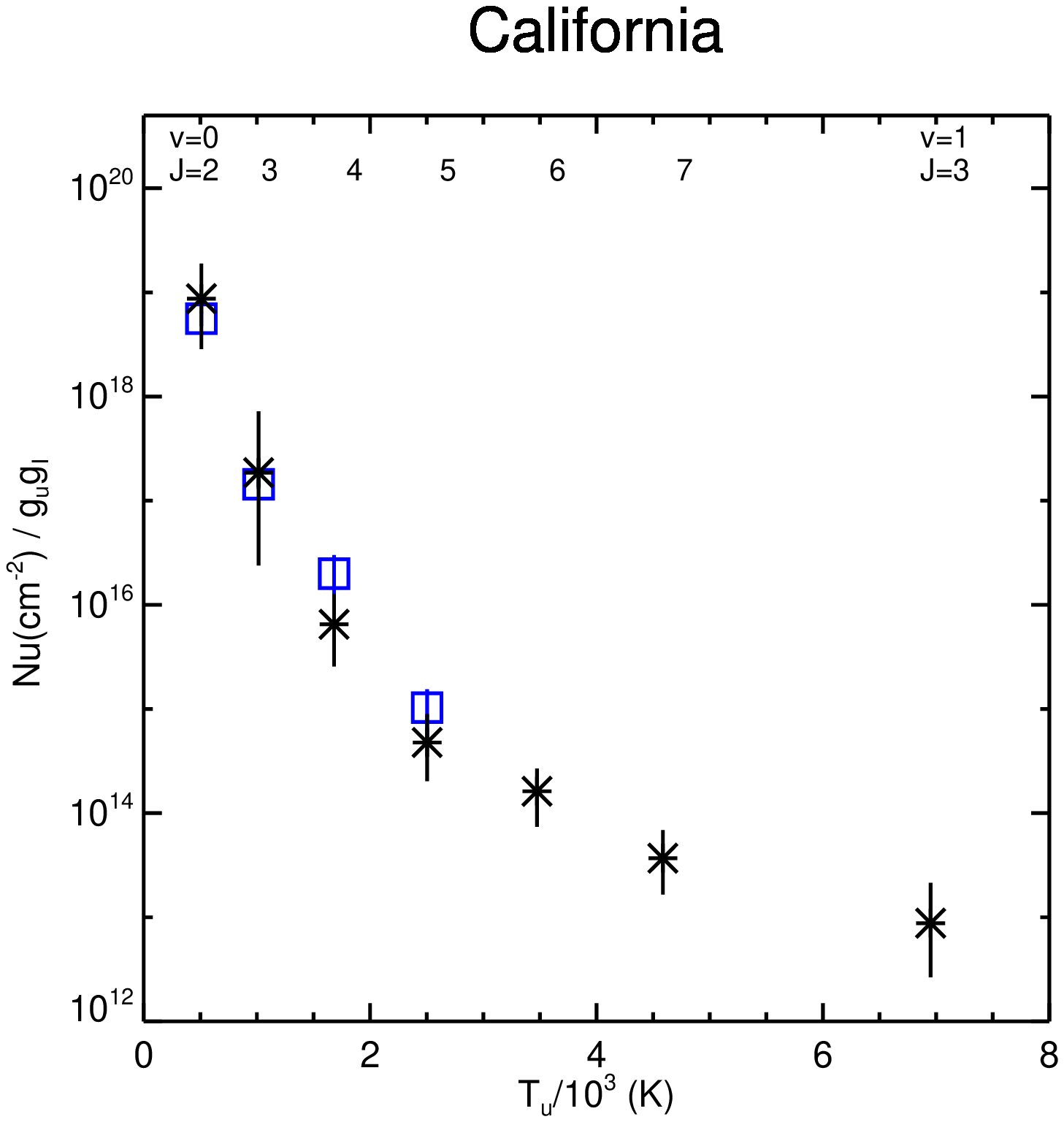 ,width=9cm,height=7.5cm,angle=0} }
\end{minipage} 
\begin{minipage}[c]{9cm} 
%\centerline{ \psfig{file=/Users/emiliehabart/PDR/PDR_plot/all_objets/figure_diagramme_obs_mod_N7023E.ps ,width=9cm,height=7.5cm,angle=0} }
\centerline{ \psfig{file=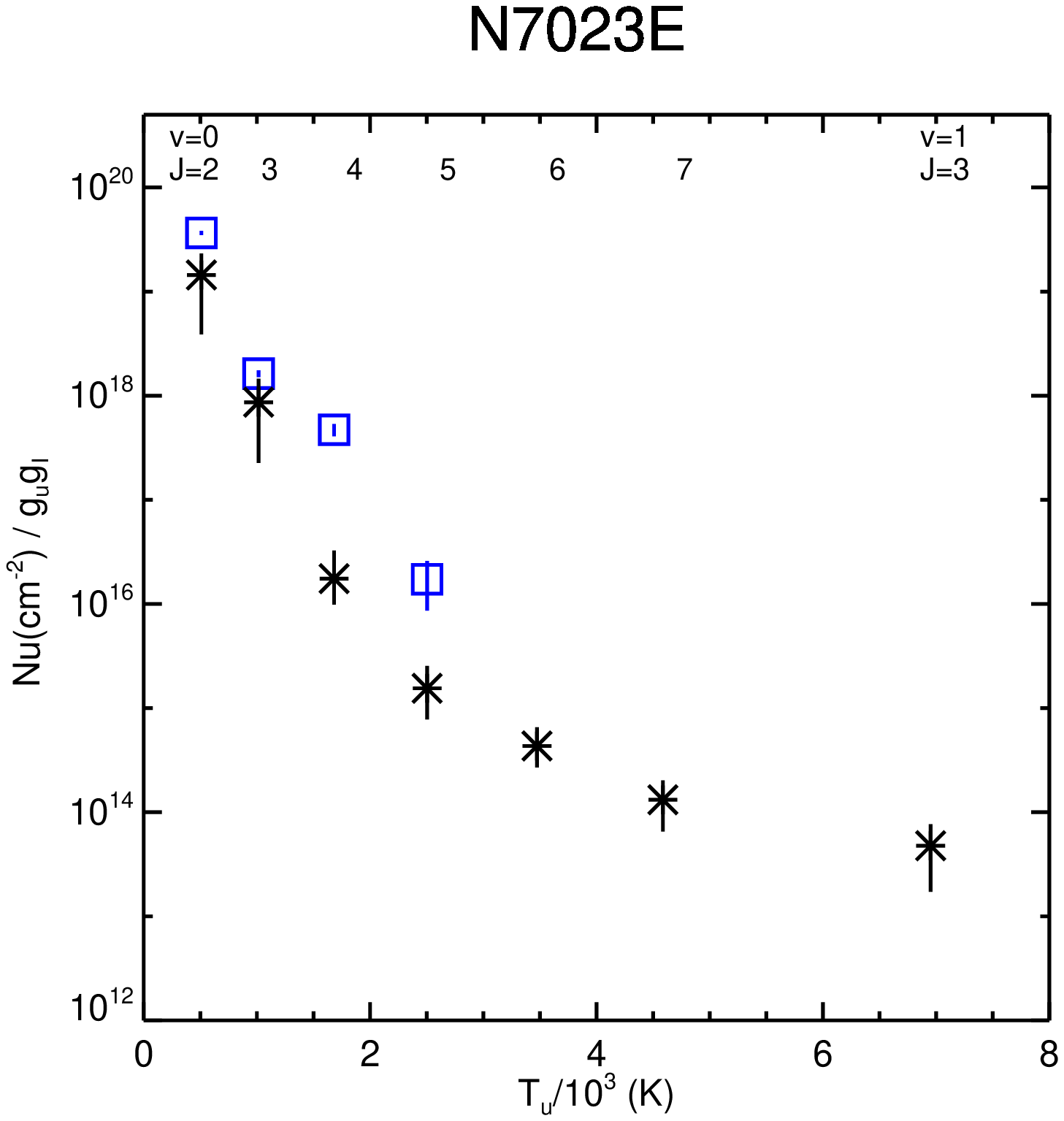 ,width=9cm,height=7.5cm,angle=0} }
\end{minipage} 
\begin{minipage}[c]{9cm} 
%\centerline{ \psfig{file=/Users/emiliehabart/PDR/PDR_plot/all_objets/figure_diagramme_obs_mod_Horsehead.ps ,width=9cm,height=7.5cm,angle=0} }
\centerline{ \psfig{file=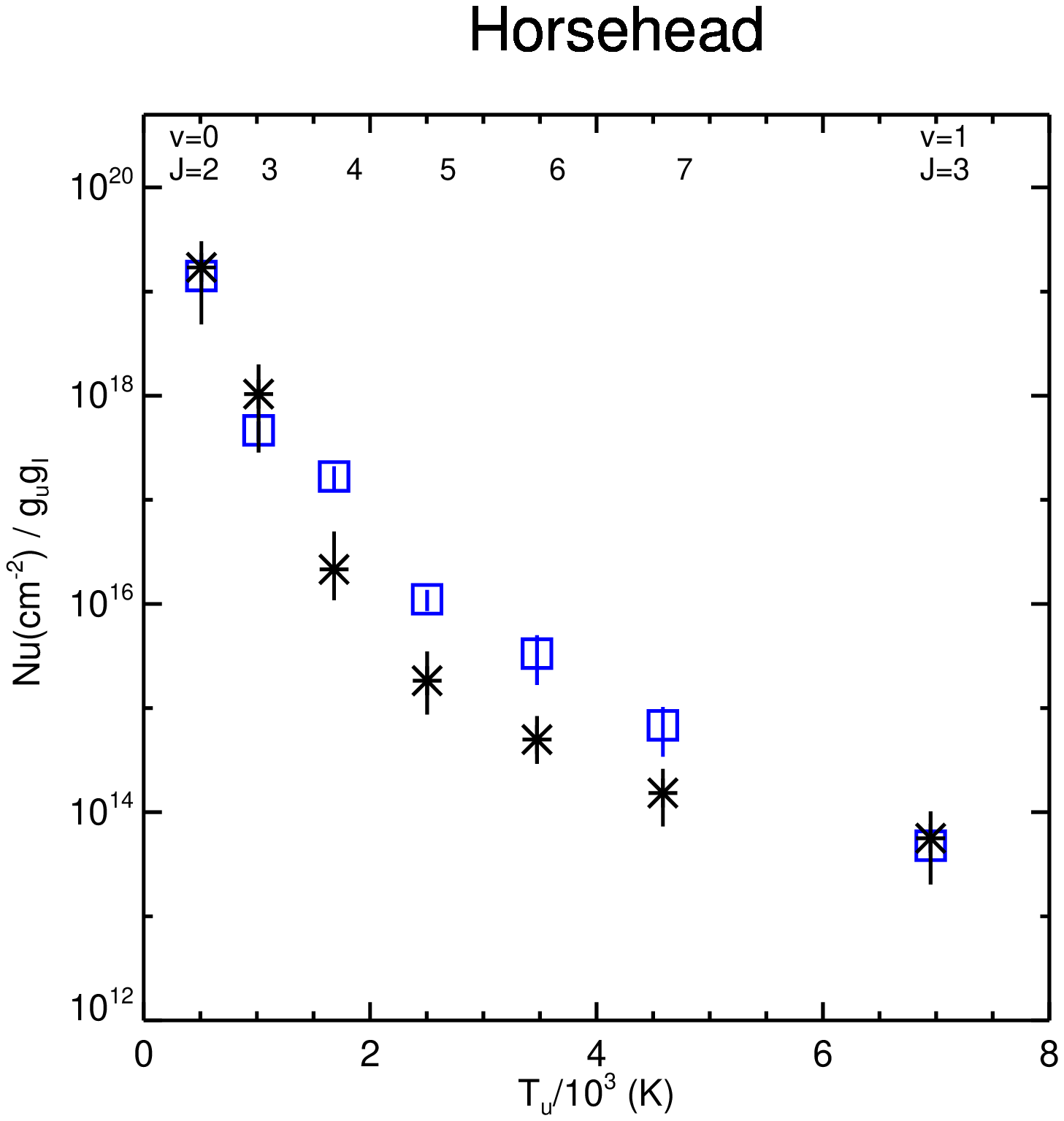 ,width=9cm,height=7.5cm,angle=0} }
\end{minipage}
\begin{minipage}[c]{9cm} 
%\centerline{ \psfig{file=/Users/emiliehabart/PDR/PDR_plot/all_objets/figure_diagramme_obs_mod_rho_Oph.ps ,width=9cm,height=7.5cm,angle=0} }
\centerline{ \psfig{file=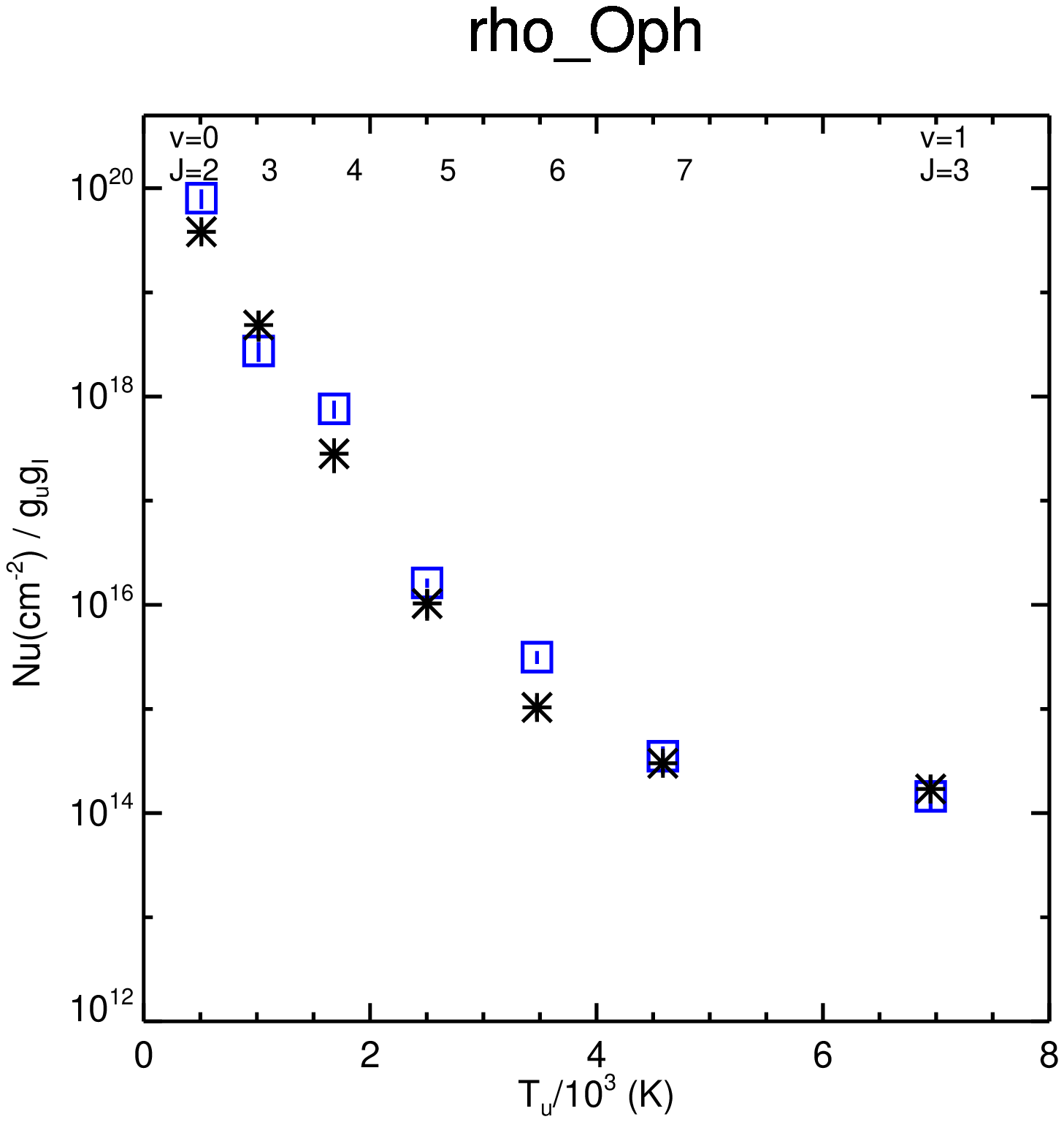 ,width=9cm,height=7.5cm,angle=0} }
\end{minipage} 
\begin{minipage}[c]{9cm} 
%\centerline{ \psfig{file=/Users/emiliehabart/PDR/PDR_plot/all_objets/figure_diagramme_obs_mod_N2023N.ps ,width=9cm,height=7.5cm,angle=0} }
\centerline{ \psfig{file=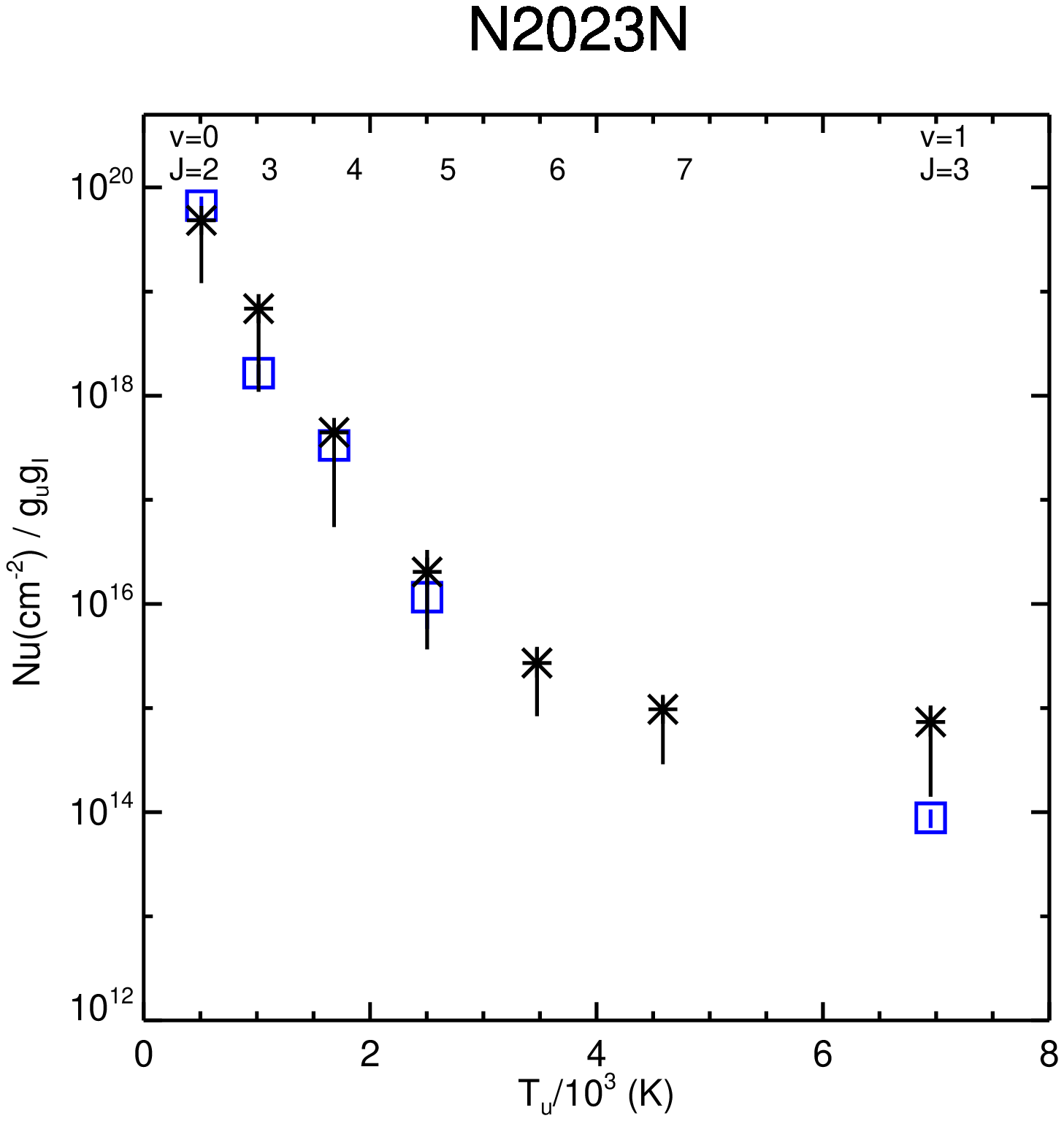 ,width=9cm,height=7.5cm,angle=0} }
\end{minipage} 
%--------------------------------------------------------------------------------------
\caption{\em 
Excitation diagram of H$_2$. $N_u$ is the column density of the transition upper level, $g_u$ is the degeneracy of the upper level and $T_u$ is the upper level energy in Kelvin. 
The squares show the data. The H$_2$ ortho-to-para ratio is taken equal to 3.
The error bars are small in size compared to the symbol size. The arrows indicate upper limits.
The stars show the PDR model predictions.
For each object, H$_2$ line intensities were computed from linear interpolation of our grid of models with the suitable values of the FUV incident radiation field and gas density (see Table \ref{table_sample}).
%For each object, isochoric models with the associated values of the gas density and strength of the FUV incident radiation field given in Table \ref{table_sample} were computed. 
We use $\chi$=4.5 and $n=10^3$ cm$^{-3}$ for L1721; $\chi$=19 and $n$=5 $10^3$ cm$^{-3}$ for the California; $\chi$=51 and $n=10^4$ cm$^{-3}$ for N7023E; $\chi$=60 and $n=10^4$ cm$^{-3}$ for the Horsehead; $\chi$=250 and $n=10^4$ cm$^{-3}$ for rho Oph; $\chi$=550 and $n=5~10^4$ cm$^{-3}$ for N2023N. 
Error bars due to uncertainties in the UV field are given. 
For comparison of the predicted and observed line intensities, in the model the PDRs have been tilted with respect to the line of sight with cos($\theta$)=0.15.}
\label{fig_diagram_mod_obs}
\end{figure*}

In this section, we compare the models predictions to the observations.
Figure \ref{fig_diagram_mod_obs} presents the excitation diagrams as observed and as predicted for each object.
For each object, H$_2$ line intensities were computed from linear interpolation of our grid of models with the suitable values of the FUV incident radiation field and gas density.
%For each object, specific models with the associated values of the FUV incident radiation field and gas density given in Table \ref{table_sample} were computed. 
Error bars due to uncertainties in the FUV field have been estimated.
In order to compare the predicted and observed line intensities, the PDR is tilted with respect to the line of sight.
For simplicity, we adopt the same viewing angle $\theta$ between the line-of-sight and the normal to the PDR for all the models.
We assume $\theta \simeq 80^{\circ}$, which is suitable for PDRs viewed nearly edge-on as most of our PDRs are.
The optically thin H$_2$ lines surface brightness are enhanced by a factor 1/cos($\theta$) $\sim$6 relative to the face-on surface brightness.

\begin{figure*}[htbp]
\leavevmode
%\centerline{ \psfig{file=/Users/emiliehabart/PDR/PDR_plot/all_objets/figure_article_IH2_mod_obs_Tg_PDRs.ps,width=15cm,angle=0} }
\centerline{ \psfig{file=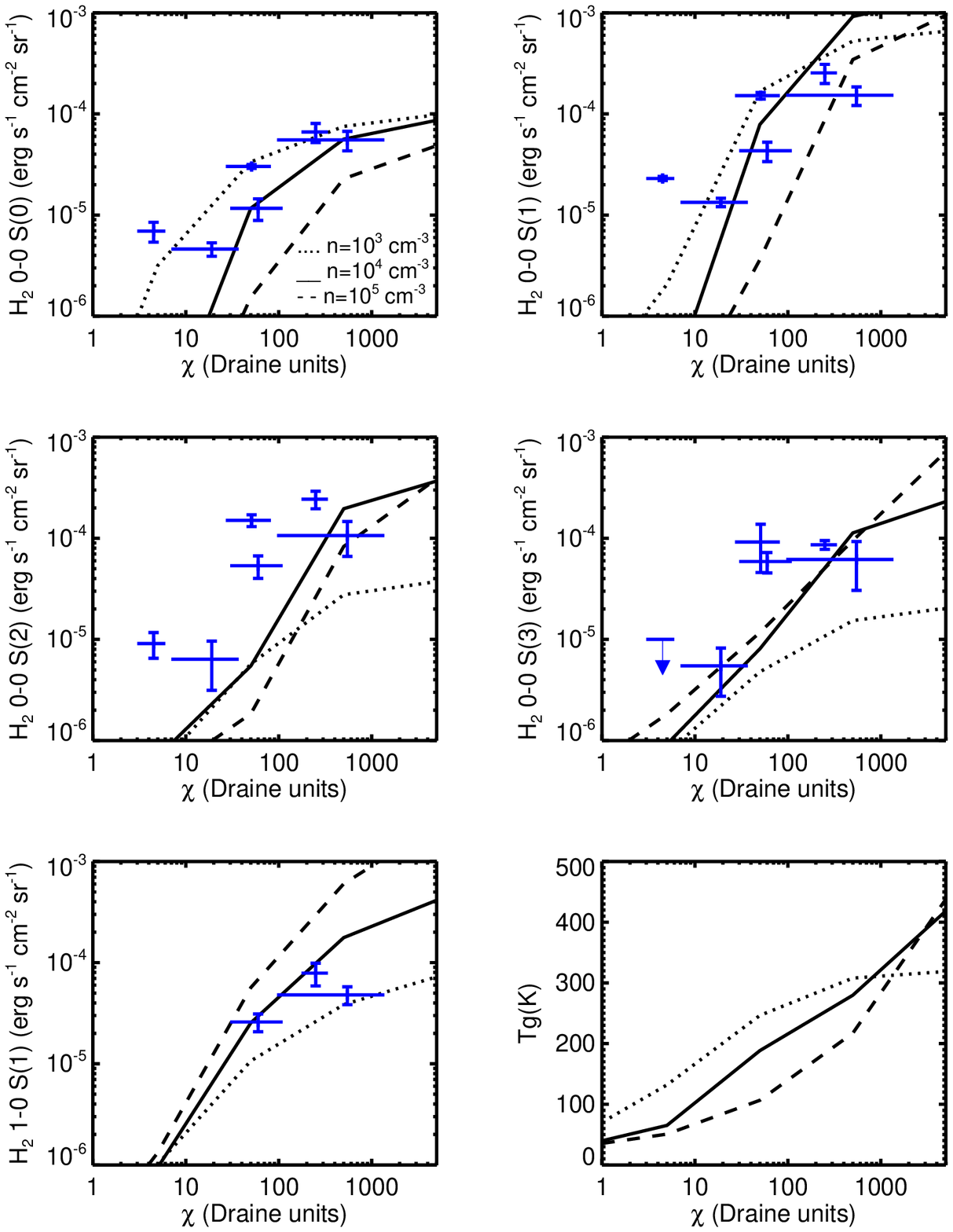,width=15cm,angle=0} }
\caption{\em 
Intensites of several pure rotational lines 0-0 S(0) through S(3) as well as for the 
ro-vibrational line 1-0 S(1) predicted by the PDR model for different gas densities - $n=10^5$ cm$^{-3}$ (dashed lines), $n=10^4$ cm$^{-3}$ (solid lines),
$n=10^3$ cm$^{-3}$ (dotted lines) - as a function of the strength of the FUV incident radiation field, $\chi$. The kinetic temperature at the photodissociation front $T_g$ obtained
in the grid of models is also presented in the lower right panel. 
The blue points with error bars show the observed intensity for our PDR sample. 
From left to right, we have L1721 ($\chi$=4.5), California ($\chi$=19), N7023E ($\chi$=51), Horsehead ($\chi$=60), rho Oph ($\chi$=250), and N2023N ($\chi$=550).
Horizontal error bars due to uncertainties in the FUV field are given.
Vertical error bars for uncertainties in the intensities are also given. %small of size comparable to the symbol size. 
The arrows indicate the upper limits. 
For comparison of the predicted and observed line intensities, in the model the PDRs have been tilted with respect to the line of sight 
with cos($\theta$)=0.15.
}
\label{fig_intensity_mod_obs}
\end{figure*}

Figure \ref{fig_intensity_mod_obs} shows the intensity of the rotational lines 0-0 S(0) through S(3), as well as for the ro-vibrational line 1-0 S(1), as observed and as predicted as a function of $\chi$ and for different $n$.  
Error bars for uncertainties in the FUV field are given. 
The same viewing angle $\theta$ between the line-of-sight and the normal to the PDR  is applied for all the models.
Figure \ref{fig_intensity_mod_obs} also presents the kinetic temperature $T_g$ at the H/H$_2$ photodissociation front (n(H$^0$)=n(H$_2$))  obtained in the grid of models. At the edge of the cloud, the main heating gas process is the photoelectric effect on dust. 
The FUV photons heat the gas by means of photoelectric emission from grain surfaces and PAHs.
As expected, $T_g$ increases with $\chi$ since  the heating photoelectric rate $\Gamma _{PE} \propto \chi$. %But for high $\chi$ and $n\le 10^4$, grain ionization can be so large at the edge that the photoelectric effect is less efficient and $T_g$ does not anymore increase with $\chi$. 
In the model, both the photoelectric effect heating and UV pumping increase with $\chi$, which implies strong enhance by several orders of magnitude of the H$_2$ rotational and ro-vibrational emission line intensities (see Fig. \ref{fig_intensity_mod_obs}).

Comparing the excitation diagrams as observed and as predicted (Fig. \ref{fig_diagram_mod_obs}),
we find that for all the objects - except for the less excited PDR L1721 - the models account for the first low rotational levels $J=2, 3$ (or the 0-0 S(0) and S(1) lines). 
These low levels probing the bulk of the molecular gas at moderate temperature (80$\lesssim T \lesssim $400 K) are in fact reproduced well by the PDR model calculations as shown in the upper panels of the Fig. \ref{fig_intensity_mod_obs}.
For the ro-vibrational $\rm v=1$, $J=3$ level (or the 1-0 S(1) line) probing the UV pumped gas, the model also fits the data well when available. In the Horsehead, rho Oph, and N2023N, the observed 1-0 S(1) line intensity is in fact in good agreement with the PDR model UV pumping predictions (Fig. \ref{fig_diagram_mod_obs} and lower left panel in Fig. \ref{fig_intensity_mod_obs}).
On the other hand, the PDR model underestimates the column densities of the excited rotational levels with $J=4, 5$ (or the 0-0 S(2)-S(3) lines) by large factors (factors $\ge$10) in many of our PDRs - L1721, California, N7023E, Horsehead - as shown in Fig. \ref{fig_diagram_mod_obs} and middle panels of Fig. \ref{fig_intensity_mod_obs}.
%For the moderate excited Horsehead PDR, the intensity ratio of the excited to low rotational lines (e.g., 0-0 S(2)/S(0) or 0-0 S(3)/S(1)), as well as the ratio  of the rotational to rovibrational lines (e.g., 0-0 S(2)/1-0 S(1) or 0-0 S(3)/1-0 S(1)), are underestimated by one order of magnitude (see Fig. \ref{fig_diagram_mod_obs}).
%For the N7023E PDR, the intensities of the 0-0 S(2)-S(3) lines are underestimated by similar factors ($\sim$10).
%For the California, the discrepancy starts from $J=4$
In the lowest excited  L1721 PDR, the discrepancy between model and data starts even from $J=3$ (i.e., the 0-0 S(1) line, see Fig. \ref{fig_diagram_mod_obs} or upper right panel in Fig. \ref{fig_intensity_mod_obs}).
From the intensity of the rotational lines 0-0 S(2) and S(3) as a function of $\chi$ (middle panels of Fig. \ref{fig_intensity_mod_obs}), 
one can indeed clearly see that the discrepancy between models and observations  mostly concern the low/moderate PDRs with $\chi \le 100$.
In these low/moderate UV environments, photon heating (by pumping and/or inelastic collisions via photoelectric effect) as described in the model cannot account for the observed H$_2$ excited rotational line intensities.
For the higher excited PDRs with $\chi > 100$ (such as rho Oph, N2023N), we find conversely that the model roughly reproduces most of the observed excited rotational H$_2$ lines. However, for N2023N, it is difficult to conclude about the uncertainty on $\chi$.

The discrepancy between the model and the observations cannot be due to uncertainties on the PDR inclination, 
since the limb brightening factor 1/cos($\theta$) can be at most a few only of what we assumed.
Furthermore, enhancement geometrical factors would increase all the H$_2$ lines intensities, producing too much emission of the low-J rotational and rovibrational lines and not fitting the H$_2$ line intensity ratios.
Considering the error bars for uncertainties in the FUV field shown in Figs. \ref{fig_diagram_mod_obs} and \ref{fig_intensity_mod_obs}, we also find that  the discrepancy between the model and the data cannot come from uncertainties on the strength of the FUV incident radiation field $\chi$. 
Similarly,  uncertainties on the local gas density $n$ cannot be invoked as the predicted intensities for excited rotational lines are comparable for models with $n=10^3-10^5$ cm$^{-3}$ (see Fig. \ref{fig_intensity_mod_obs} for $\chi < 100$). 
Density structure changes would also have minor effects, since modeling the H$_2$ excitation mainly depends upon the average gas density in the H$_2$ emission zone.  
Detailed calculations that take a steep gas density gradient into account as observed from the Horsehead PDR  edge
 from previous complementary data  \cite[]{habart2005} led in fact to similar results.
%Models with blackbody or Kurucz model spectrum with the suitable stellar parameters give also comparable results.
However, clumpiness that can lead to a substantial reduction of the effective optical depths could increase the portion of hot gas and explain a significant amount of rotationally excited H$_2$.

\section{Comparison with H$_2$ observations of the diffuse interstellar medium in the Solar Neighborhood}
\label{h2_diffus}

Early far-UV absorption studies with the Copernicus satellite, as well as far-UV spectroscopic data 
obtained more recently with FUSE show significant amounts of excited H$_2$ in diffuse clouds \cite[]{spitzer73,spitzer74,spitzer76,savage77,gry2002,browning2003,gillmon2006}.
In particular, \cite{gry2002} analyze FUSE H$_2$ spectra over three lines of sight in the Chamaeleon clouds
probing gas heated by the mean Solar Neighborhood radiation field.
These data, which include the populations of the lowest two $J$=0 and $J$=1 levels, clearly reveal a multi-temperature structure.
The $J\ge 3$ levels characterized by an excitation temperature of $T_{53}\sim$200-250 K lie far above the temperature derived from the lowest level transitions with $T_{10}\sim$60 K.
Using the Meudon PDR code with $\chi =0.4$ and $n\sim 80$ cm$^{-3}$, 
\cite{nehme2008} find a good fit to a number of observables, including the column density of H$_2$ in the 
$J$=0 and 1 levels, but  their model  fails to account  for the H$_2$ in levels $J\ge 3$. 
The discrepancy between the observed and predicted column densities of these levels is about one order of magnitude, 
a difference similar to that  we report in low-excitation PDRs. 
%Moreover, the fraction of the unexpected excited H$_2$ (about few percents) are roughly comparable in diffuse clouds and our PDRs.

The H$_2$ excitation in PDRs and  in the diffuse interstellar medium differ 
in the power radiated per hydrogen atom.  The power radiated from the first  rotational levels per nucleon,
i.e. $P_{H_2^*}\sim I_{H_2^*} / N_H \times 4\pi$ (in erg s$^{-1}$ H$^{-1}$), where $I_{H_2^*}$ is the intensity of the high rotational lines and $N_H$ the total hydrogen column density of the emission zone,
is much stronger in low-excitation PDRs than in diffuse clouds.
For the L1721 PDR with $I_{H_2^*} \sim I_{0-0~S(1)}+I_{0-0~S(2)} \sim 4.8~10^{-5}$ erg s$^{-1}$ cm$^{-2}$ sr$^{-1}$ and  $N_H \sim 2\times N_{H_2}\sim 10^{21}$ cm$^{-2}$ (see Table \ref{table_obs}), we find $P_{H_2^*}\sim 6~10^{-25}$ erg s$^{-1}$ H$^{-1}$. For the diffuse interstellar medium towards HD 102065, the observed H$_2$ column densities  translate into  $I_{H_2^*}\sim$10$^{-6}$ erg s$^{-1}$ cm$^{-2}$ sr$^{-1}$, adding emission in the S(1) and S(2) lines.
With $N_H\sim 10^{21}$ cm$^{-2}$, we find $P_{H_2^*}\sim 2~10^{-26}$ erg s$^{-1}$ H$^{-1}$.
This value is  more than one order of magnitude lower  than the L1721 value.
The difference is  even greater for other PDRs in our sample.
This result indicates that the energy deposited per hydrogen atom is much higher in PDRs than in diffuse interstellar clouds. This is not surprising since physical conditions in PDRs differ from those in the diffuse interstellar medium. In
particular, the gas kinetic energy per nucleon and its dissipation rate could be higher, because  PDRs in our sample are all 
in massive star-forming regions where turbulence may be powered by the mechanical action of luminous stars on their environment. 
Thus, the mechanical processes proposed to account for the H$_2$ excitation in diffuse interstellar  clouds \cite[]{flower98,falgarone2005}  could also contribute to the H$_2$ excitation in PDRs,  even if the power radiated per H atom in the rotational excited lines is much greater. This possibility
 is further discussed in Sect. \ref{origin_mechanical}.

\section{Origin of the rotationally excited H$_2$}
\label{origin}

Our Spitzer results show the presence of significant rotationally excited H$_2$ in the low/moderate excited PDRs that is not predicted by PDR models.
The cause of this could be that, in the models, the column density of H$_2$ is too low in the outer zones where H$_2$ is excited %(by inelastic collisions or UV pumping) 
or alternatively that a small fraction of the H$_2$ gas is hotter than predicted by models.
The effect should be selective by only significantly increasing the intensities of the excited  H$_2$ rotational lines (e.g., 0-0 S(2)-S(5)).
The intensities of the low rotational lines (e.g., 0-0 S(0)-S(1)) probing the bulk of the molecular H$_2$ gas at moderate temperature, % (80$<T<$400 K),
 as well as of the rovibrational line (e.g., 1-0 S(1)) probing the UV pumping excitation, should mostly not be affected in order to fit the data.
In the following, we examine some of the issues with this in mind.

\subsection{H$_2$ formation rate} 

The H$_2$ line intensities depend mainly on the H$_2$ formation rate $R_f$ since it controls the location of the H/H$_2$ transition zone in PDRs.
The line intensities change via the variation in both the gas temperature $T_g$ and the flux pumping H$_2$ in the H/H$_2$ transition zone.
Enhanced $R_f$ could result in a strong increase in the intensity of the excited rotational lines.
%Enhanced $R_f$ could result in a sharper increase in high rotational lines than in low rotational lines and rovibrational lines.
Previous ISO studies of moderately excited PDRs (such as rho Oph, S140, and IC 63) suggest that a high $R_f$ about five times the standard value is required to reproduce the H$_2$ rotational lines \cite[]{habart2004}.
However, we find here that, even if the H$_2$ formation rate is significantly enhanced, the model intensities fall short by an order of magnitude.
The same conclusion is reached by \cite{li2002} using ISO data of the low-excited PDR the S140 periphery regions (with $\chi\sim$10-30 and $n\sim 10^3$ cm$^{-3}$). 
This can be understood as follows.
In this low/moderately excited PDR regime with low $\chi/n$ ratios ($\chi/n\lesssim 0.01$), H$_2$ self-shields efficiently enough so that the  H$^0$/H$_2$ transition zone is closer to the edge ($N \le 10^{21}$ cm$^{-2}$), even without an enhanced H$_2$ formation rate. % \cite[see, e.g., ]{hollenbach99}. 
H$_2$ already competes effectively with dust in absorbing FUV photons % the fraction of the FUV photon flux pumping H$_2$ cannot be much higher
and the gas temperature in the  H$^0$/H$_2$ transition zone is comparable to the PDR edge temperature, so although the H$_2$ line intensities are increased in the high H$_2$ formation case, it does not increase much.
We conclude that a global enhancement in the H$_2$ formation rates improves the comparison with observations, but is not sufficient to match the excited rotational lines as observed in our low-excited PDRs.

\subsection{Advected H$_2$}

Our model results are based on the assumption that the PDR is in equilibrium.
In reality, advection of molecular gas from the shielded cloud interior to the warm surface - created by turbulent motions and/or fast progression of the photodissociation front into the cloud -
can create relatively high nonequilibrium abundances of H$_2$ in the outer layers. This could produce extra column densities of excited H$_2$.
%H$_2$ is the most likely species to go out of equilibrium, because of its slow destruction rate by (self-shielded) photodissociation and its relatively slow formation rate on grain surfaces.  
However, as discussed in \cite{stoerzer98}, there is a range of $\chi/n$ ($0.01 \lesssim \chi/n \lesssim 0.2 \rm v $ with $\rm v$ the gas velocity in km/s) for which the effects of advection will have the strongest influence  on line intensities compared with equilibrium values. 
For high $\chi/n$, the dissociation timescales are so short that advected H$_2$ will not survive in the outer layers.
For very low $\chi/n$, the differences between equilibrium and nonequilibrium line intensities are expected to be small, since the H/H$_2$ front is in the equilibrium model near the edge at $N \le 10^{21}$ cm$^{-2}$, where most of the excited H$_2$ comes from. 
For our low/moderately excited PDRs with $10^{-3} \lesssim \chi/n \lesssim 10^{-2}$,
the condition that advected H$_2$ will survive in the outer layers without being dissociated is verified, even for low velocity. %($\rm v \gtrsim$0.1 km/s).
However, the line intensities are expected not to increase significantly, since the H/H$_2$ front is near the edge, even without advection. 
Out-of-equilibrium PDR model calculations %made for $\chi/n=0.01-10$ 
by \cite{stoerzer98} 
predict that the integrated H$_2$ line intensities do not vary by more than factors of 3.
However,  their calculations concern the high-density, high-$\chi$ case, which does not apply here.
Moreover, local increase in the H$_2$ lines emissivity at the PDR edge can be important.
Further detailed calculations for the low-$\chi$ case would be needed to fully test the non-equilibrium effects.
%However, further calculations would be needed to test the non-equilibrium effects.

Another point that one can examine is the OTP ratio expected in  the advection case.
If there is advection and the dynamical timescale ($t_f$) is much shorter than the OTP equilibrium time ($t_{OTP}$), 
the OTP ratio must be out-of-equilibrium. 
Simple calculations of these different timescales for the conditions prevailing in our PDRs\footnote{
The advection time can be given as $t_f=N_0 / (n \times v)  \sim$ 3 $10^4$ yrs $\times$ [ (1km/s) / $\rm v$ ] $\times$ [ ($10^4$ cm$^{-3}$) / $n$] with $N_0\simeq 10^{21}$ cm$^{-2}$ the total hydrogen column density in the surface PDR zone.
The dominant mechanisms that control the OTP ratio conversion via gas-phase spin exchange reactions are collisions with H atoms and/or protons.
The rate coefficient for the proton reaction can be given by $3~10^{-10}$ cm$^3$ s$^{-1}$ \cite[]{gerlich90}. 
%The ortho-para conversion timescale corresponding is $t_{OTP} \sim$100/n(+) yrs, where n(+) represent the density of H$^+$, H$_3^+$ or H$_3$O$^+$ in cm$^{-3}$. 
The rate coefficient for the reactive collisions with H atoms can be given by  $8~10^{-11}$ e$^{(-3900/T_g)}$ cm$^3$ s$^{-1}$ \cite[e.g.,][]{Schofield67}. 
Due to the high activation barrier of the reaction with H atoms ($\sim$3900 K), the dominant process that controls the OTP ratio conversion is collisions with protons  in cool gas ($T_g \le$ 100-200 K).
Taking the gas temperature profiles as predicted by the equilibrium model, 
we  find that $T_g \le$200 K for PDR with $\chi \sim 5$, %(such as L1721) 
while  $T_g \le$350 K for PDR with $\chi \sim 50$. % (such as N7023E, Horsehead). 
%The differences between equilibrium and nonequilibrium model must be small since in the range of $\chi$  and $n$ studied here the gas temperature profiles as a function of depth in the PDR are fairly insensitive to the fraction of H$_2$ (heating is mainly due to photoelectric emission from dust grains and cooling to fine structure line emission).
Thus, for low-excited PDRs the proton reaction dominate all the PDR, while for moderatly excited PDRs, the collisions with H atoms are expected to be important at the PDR edge. 
Within the range of conditions prevailing in our PDRs, 
we estimate $t_{OTP} \sim$ 10$^4$-10$^7$ yrs for the proton reaction 
%using the rate coefficient given by \cite{gerlich90} ;
and $t_{OTP} \sim$ (3 $10^7$ - $10^{11}$)/$n$  yrs for reaction with H atoms.
% using the rate given by \cite{Schofield67} .
}
and assuming $\rm v \sim$1 km/s suggest that $t_f \le t_{OTP} $ for the low excited PDRs (such as L1721 with $\chi \sim 5$, $n\sim$10$^3$ cm$^{-3}$) while $t_f \ge t_{OTP}$  for the outer layers of the moderate excited PDRs (such as N7023E, Horsehead with $\chi \sim 50$, $n\sim$10$^4$ cm$^{-3}$).
Consequently, the OTP ratio expected in the advection case should be equal to 1 or 3 according to the PDR zone studied.
Unfortunately, for many of our PDRs, not enough H$_2$ lines have been measured to do a proper fit of the OTP ratio of the excited rotational levels (see Sect. \ref{observation}).
Nevertheless, for moderately excited PDRs (Horsehead, rho Oph), where several excited H$_2$ rotational lines have been measured, we derive the OTP ratio of about $\sim$1 for the $J$=3 to 5 levels  while $\sim$3 for the $J$=5 to 7 levels.  
The different OTP findings could result from the fact these rotational levels probe different PDR zones where the OTP conversion is more or less efficient.
In either case, we suggest that these findings should not eliminate advection as a solution to extra warm/hot H$_2$.
However, more sensitive searches to  better constrain the OTP ratio of the excited H$_2$ gas, as well as constraints on the kinematics in the H$_2$ emission zone, are needed for any conclusion.

\subsection{Local heating via the photoelectric effect}

One alternative solution is to locally increase the gas temperature in the PDR outer zone. 
At the edge of the cloud, the main heating gas process is the  photon heating via photoelectric effect on dust.
A local increase in the photoelectric heating rate can have various causes.
In particular, there is increasing observational evidence of evolution of the very small dust particles, including PAHs and very small grains (VSGs) in the illuminated parts of the cloud \cite[]{boulanger88b,boulanger90,bernard93,abergel2002,rapacioli2005,berne2007,compiegne2008}.
Increasing the abundance of the smaller grains that dominate the photoelectric effect would lead to higher gas temperatures \cite[e.g.,][]{bakes94,weingartner01}.
Previous ISO study of the low-excited PDR L1721 by \cite{habart2001a} shows that abundance variations of small dust grains should be taken into account to reproduce the cooling gas due to fine structure lines ([C$^+$] 158 $\mu$m and [O$^0$] 63 $\mu$m).
Also, they predict that the H$_2$ lines emission is increased by a factor of $\sim$5-10, in line with the present data. 
%Note that taking into account our Spitzer measurements of the H$_2$ lines, the heating efficiency $\epsilon$ is at most increased by a factor of 2.

Analysis of the full IRS Spitzer data spectrum obtained in the frame of the SPECPDR program permits us to characterize the small grain population. 
In particular, \cite{berne2007} and \cite{compiegne2008} observed significant variations in the relative emission between PAHs and VSGs in our PDRs sample\footnote{N7023E and rho Oph spectra were analyzed by \cite{berne2007}; Horsehead and N2023N spectra by \cite{compiegne2008}. L1721 and California do not present any detected spatial variation within the small field observed with IRS.}.
%This confirms previous ISO studies showing PAH/VSG emission ratio variations at dense illuminated ridge (e.g. Abergel et al. 2002; Rapacioli et al. 2005).
\cite{compiegne2008}, who focus on the Horsehedad and N2023N, shows that excitation effects cannot account for these variations and interprets it in term of PAH/VSG relative abundance variations.
\cite{berne2007} extracted the mid-IR spectra of PAH and VSG populations.
For N2023N, where spectral evolution is observed, \cite{compiegne2008} find that the PAH/VSG abundance  ratio is $\sim$5 times lower in the dense, cold zones of the PDR than in its diffuse illuminated part where dust properties seem to be the same as in cirrus.
For the Horsehead, where they did not spatially resolve any spectral variation, they found that this ratio is 2.4 less at the peak emission of the PDR than in Cirrus. %the reference diffuse (Cirrus) case. 
%The obtained PAH/VSG relative abundance for the Horsehead is almost the median between the diffuse and dense part of N2023N.
However, stronger variations on spatial scales below the available Spitzer resolution cannot be excluded.
To quantify this impact, the absolute grain abundance profile of each dust population as a function of depth should be implemented in the PDR model.
Towards our PDRs where spectral evolution is observed, more detailed studies of the spatial emission profiles of the dust components, in relation with the H$_2$ lines, are currently in progress \cite[]{joblin2010}.

\subsection{Heating by dissipation of mechanical energy} 
\label{origin_mechanical}
Here, we consider H$_2$ heating associated with the dissipation of the gas kinetic energy.
PDRs in our sample are all located in massive star-forming regions,  where the expansion of HII gas, stellar winds and/or supernovae explosions are significant sources of mechanical energy. 
The energy released in shocks or by the dissipation of turbulence could increase the H$_2$($J$) populations in comparison to PDR model predictions.
We find that the MHD shock model of \cite{flower98}\footnote{The MHD shock models have been updated, in particular with the new cross sections of collisional excitation of H$_2$ \cite{lebourlot99}.}
 fits the excitation well as do as the absolute intensities of the H$_2$ lines observed in the low-excited PDR L1721, 
 for a gas density $n\sim 10^3$ cm$^{-3}$ and a shock velocity $\rm v \sim$ 15 km/s. 

The dissipation rate of turbulence ($P_{diss}$) is often related to the amplitude of turbulent motions (the 
gas velocity dispersion $v_{rms}$) and the size ($l_i$) at which the energy is injected.
Quantitatively we have $P_{diss}/M_H  \sim 3/2 \times  v_{rms}^3 / l_i = 6 \times  10^{-25}  \times (v_{rms}/20~km\, s^{-1})^3 / (l_i/10\, pc) ~ {\rm erg~s^{-1}~H^{-1}}$. 
This estimate quantifies the magnitude of the turbulent gas velocity, % ($\sim$10 to 15 km/s), 
which is necessary to account for the H$_2$  S(1) and S(2) line emission in L1721. % (see Sect. \ref{h2_diffus}).
High spectral resolution observations of H$_2$ ($R=\lambda/\Delta\lambda$ greater than 20,000 achievable with 
the CRIRES spectrograph on the VLT) are needed in order to test this possibility. High spectral resolution 
observations of the C$^+$ far-IR line emission from PDRs,  to be performed with the Herschel Space Telescope, 
will also quantify turbulence in the outer PDR layers where CO is photodissociated.

\section{A diffuse contribution to the H$_2$ emission from galaxies}
\label{galaxies}

Spitzer spectroscopic observations have provided detections of the H$_2$
rotational lines for many galaxies, in particular a large
fraction of galaxies in the SINGS sample \cite[]{roussel2007}.
The H$_2$ emission of star-forming galaxies is tightly correlated with
the PAH emission. It is particularly striking that the H$_2$/PAH emission ratio is observed to be
independent of the intensity of the dust heating, as measured by the 24 $\mu$m to far-IR dust emission ratio.
\cite{roussel2007} argue that the H$_2$ and dust emission are powered by a common energy source,
the UV light from young stars.  The detection of H$_2$ line emission in low-excitation PDRs, such as
L1721, supports this interpretation.  Low-excitation PDRs, such as L1721, may add up to make a
diffuse component that may be a dominant contribution to the H$_2$ and PAH emissions from star-forming galaxies.

Rotationally excited H$_2$ has been detected
along Galactic lines of sight away from luminous star
forming regions with ISO and Spitzer observations \cite[]{falgarone2005,hewitt2009}.
Thus, Galactic observations do indicate the presence of a diffuse component of H$_2$ emission in galaxies.
However, it is still unclear whether this emission comes from low-excitation PDRs or if it arises from
the diffuse interstellar medium.  Like us, \cite{falgarone2005} used the Meudon code
to estimate the contribution of low-excitation PDRs to the observed line emission. They concludes that
the S(0) line emission may be accounted for by PDRs but that the model estimates fall short
of the observed S(1) and S(2) lines by a large factor ($\sim$5-10). 
This discrepancy is similar to the one we faced in modeling our observations of low-excitation PDRs. 
The absolute line intensities measured along their galactic line of sight are comparable to the intensities measured in our L1721 PDR, but the S(0)/S(1) line intensity ratio is significantly higher in the Galactic lines of sight.  
% which suggests that the rotationally excited H$_2$ detected along this long line of sight could  have the same origin to that seen in our PDRs.
\cite{falgarone2005} concludes that the UV emission cannot be the unique energy source and that much of the H$_2$ line emission
must come from the diffuse interstellar medium where H$_2$ may be heated by the localized dissipation
of the gas turbulent kinetic energy. 
This conclusion needs to be revisited in the light of PDR observations, but it is not necessarily contradicted by the correlation between
H$_2$ and PAH emissions, since the diffuse interstellar medium
is also thought to provide a significant contribution to the PAH emission \cite[]{draine2007}.

\section{Conclusion}
\label{conclusion}

Thanks to Spitzer spectroscopic observations, we were able to detect H$_2$ pure rotational lines emission in PDRs with modest FUV fields and densities.
The low/moderate $\chi$ regions studied here are very widespread  in galaxies and may contain a large fraction of the molecular gas.
However, this intermediate regime between the diffuse cloud and the bright PDRs has been poorly studied.
To analyze the H$_2$ line emission observations, we used an updated version of the Meudon PDR code.
Our results allow strong constraints to be placed upon the H$_2$ excitation in the interstellar medium of galaxies.
The main results from this work can be summarized as follows.

%\item
The IRS wavelength coverage allows us to detect several strong H$_2$ pure rotational lines from 0-0 S(0) to S(3) at 28.2, 17.03, 12.29, and 9.66 $\mu$m, the aromatic band features at 6.2, 7.7, 8.6, 11.3 $\mu$m, the dust mid-IR continuum emission, 
and the fine structure lines of ionized gas [NeII] at 12.8 $\mu$m, [SIII] at 18.7 and 33.4 $\mu$m and [SiII] at 34.9 $\mu$m. 
The observed mid-IR spectra are typical of PDRs.

%\item
A single temperature cannot describe the full set of observed H$_2$ line intensities.
A combination of at least two H$_2$ gas components, with one cool/warm ($\sim$100-300K) and another warm/hot ($\sim$300-700K) with much lower column densities (a few percent of the first component) is required.
The ortho-to-para ratios derived are about $\sim$1 for the first component and about $\sim$3 for the second warmer component.
The non-equilibrium behavior has already been noted in previous ISO observations of PDRs \cite[]{moutou99,fuente99,habart2003a}.

%\item
By comparing the observations with the PDR model predictions, we find that the model can account for the first low H$_2$ rotational line (e.g., 0-0 S(0)-S(1)) probing the bulk of the gas at moderate temperature, as well as the ro-vibrational  line (e.g., 1-0 S(1) observed with ground-based telescopes) probing the UV-pumped gas.
In contrast, the model underestimates the excited rotational lines (e.g., 0-0 S(2)-S(3)) by large factors ($\ge$10 in some objects).
In the lowest excited PDR, the discrepancy between the model and the data starts even from the rotational $J=3$ level (e.g., 0-0 S(1)).
This discrepancy between observations and the PDR model predictions has the 
same  order of magnitude as reported for diffuse interstellar clouds. 
However, in PDRs the power radiated per H atom in the rotational excited levels is more than one order of magnitude
larger than in diffuse clouds.

%\item
The discrepancy between the data and the PDR model  could indicate that, in the models, the column density of H$_2$ is too low in the outer zones where H$_2$ is excited
(by inelastic collisions or UV pumping), or, alternatively, that a small fraction of the H$_2$ gas is hotter than predicted by models.
We discuss whether  an enhancement in the H$_2$ formation rate, or a  local increase in photoelectric heating, as proposed for brighter PDRs in former ISO studies, may also work in low-excitation PDRs.
An enhancement in the H$_2$ formation rate improves the comparison with observations, but the models still fall short. 
Further work is needed to quantify the impact of the evolution of very small dust particles across PDRs on the gas energetics.
Out-of-equilibrium processes or mechanical heating (by weak shocks or dissipation of turbulence as proposed 
for diffuse interstellar clouds) 
are alternative possibilities. % to explain the unexpectedly large amount of rotationally excited H$_2$.

Although we cannot decide at this point, we emphasize the need for further development of the models. Progress
is also expected from PDR spectroscopy of additional cooling lines and data on the 
gas kinematics,  to be obtained from the ground and Herschel.  
In particular, observations of gas cooling lines of species such as C$^+$ at high spectral resolution with the Heterodyne Instrument for the Far Infrared (HIFI)  will provide missing information  on  
the gas velocity within the PDR layer where CO is photodissociated.
%More sensitive searches for the excited rotational lines of H$_2$ are needed to  better constrain the ortho-to-para ratio of the excited H$_2$ gas observed.
The combination of these data with models %will be very powerful and 
should help constrain the relative radiative and dynamical influence of stars on the physical conditions within PDRs.
In the longer term, the Mid-InfraRed Instrument (MIRI) on James Webb Space Telescope (5-28 $\mu$m,  diffraction-limited resolution of $\sim$0.2'' at 5 $\mu$m) will provide 
maps of the H$_2$ rotational line emission, as well as of the small dust emission with unprecedented angular resolution.
%Such high-sensitivity and high resolution space-based spectrographs would be required in order to better understand the physics of H$_2$.

%Sensitive ground based studies already possible %with CRIRES 
%at the VLT %(wavelength range 1-5 $\mu$m, $R\sim$20,000- 100,000) 
%will allow the observation of high $J$ H$_2$ rotational transitions as well as a variety of ro-vibrational lines with both high spectral and spatial resolution.
%VISIR Not sensitive enough : In addition, VISIR will permit at longer wavelengths observation of the 0-0 S(1) and S(3) H2 transitions with a spatial resolution of  0.3Ã and R  20,000 at 10 ?m. 
%Of course, the sensitivity of the ground based instruments is limited by the atmosphere and thus the insight gained by such measurements will be for the relative bright H$_2$ emission sources.Such ground based instruments are hopefully the forerunners of high class instrumentation in space or from airborne platforms.
% and thus the insight gained by such measurements will be greatest for compact sources of H$_2$ emission. 
%SOFIA ?? %Such ground based instruments are hopefully the forerunners of high class instrumentation in space or from airborne platforms. The spectrometers on SOFIA (covering 0.3-1600 ?m) will soon observe several H2 lines including the 28 ?m S(0) transition with diffraction limited resolution of 1.5Ã and R   104 ? 105 at 20 ?m. I

%\newpage

\begin{appendix}

\section{Map in the H$_2$ rotational emission lines}

Figure A.1 shows maps in the H$_2$ 0-0 S(0) line emission at 28.2~$\mu$m obtained with the IRS long-low module towards N7023E, the Horsehead, rho Oph, and N2023N.
Maps in the H$_2$ 0-0 S(1) line emission at 17.03~$\mu$m are presented in Fig. \ref{fig_h2_map_obs}.
Figure A.2 shows maps in the H$_2$ 0-0 S(2) and S(3) lines emission at 12.29 and 9.66~$\mu$m obtained with the IRS short-low module towards the Horsehead.
Projection onto the same spatial grid (2.5''/pixel) and convolution into the same beam (Gaussian with a full width high maximum of 10.7'') were performed to compare the different line maps obtained.
There is a good agreement in the intensity distribution between the H$_2$ lines.
The same substructures are present in the maps.
The emission in the lowest rotational line 0-0 S(0) is somewhat more extended and peaks more inside the nebula. This is easily to understand since the upper state of this line is relatively low-lying. The signal-to-noise ratio in the 0-0 S(0) line maps is lower than in the other line maps since this line is fainter. 
Towards the Horsehead, the spatial distribution of the excited 0-0 S(2) and S(3) lines is comparable. The 0-0 S(3) line is slighlty shifted nearer the edge.
%Not S(2) S(3) show N2023N since very small area corresponding to average area.
%Not show for 7023 and rho Oph as not obtained since CVF data

%----------------------------------------------------
\begin{figure*}[htbp]
\leavevmode
\begin{minipage}[c]{9cm} 
%\centerline{ \psfig{file=/Users/emiliehabart/PDR/PDR_plot/all_objets/figure_article_carte_h2_28mic_convolved_N7023E.ps,width=8.0cm,angle=0} }
\centerline{ \psfig{file=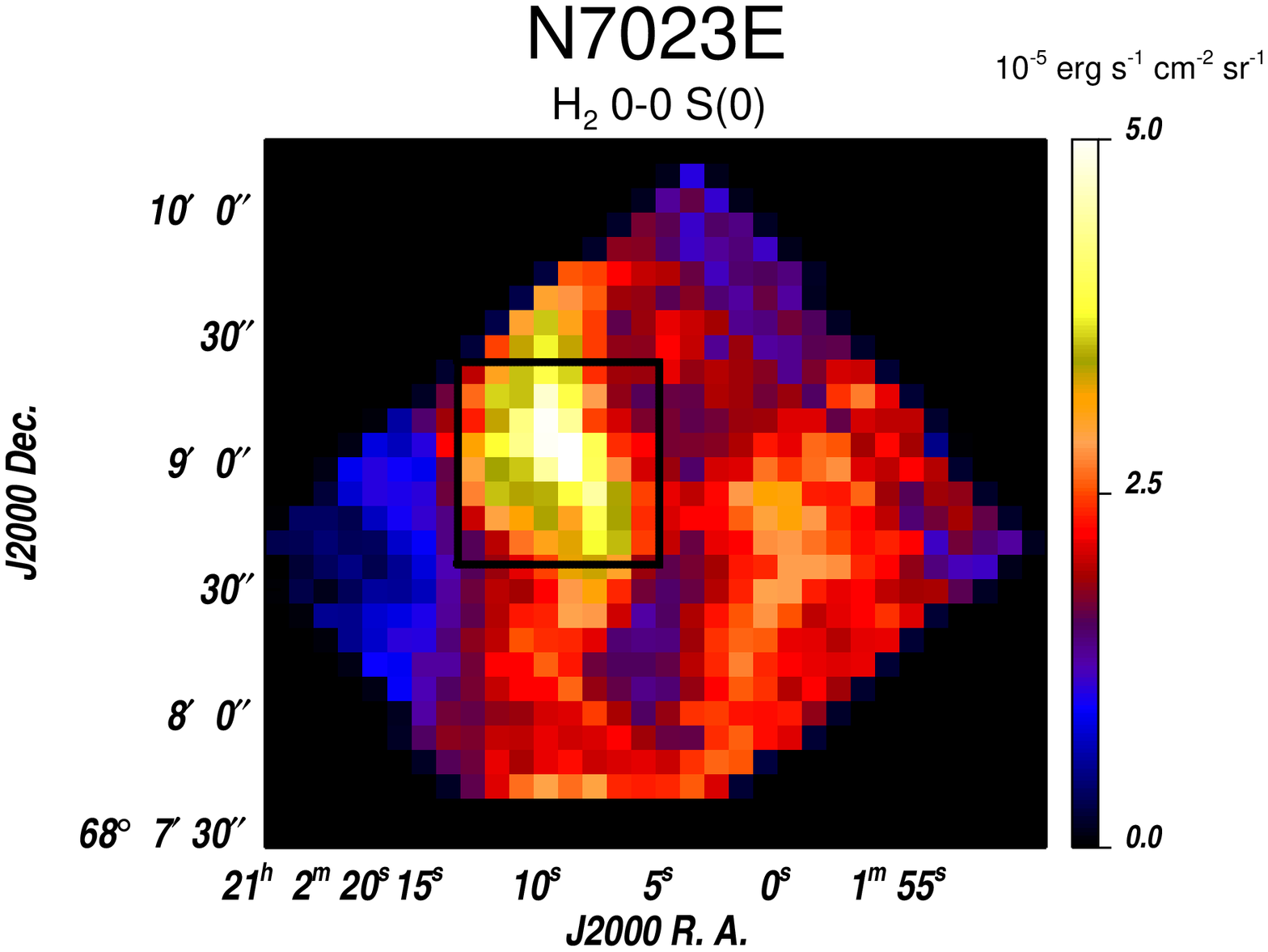,width=8.0cm,angle=0} }
\end{minipage}
\begin{minipage}[c]{9cm} 
%\centerline{ \psfig{file=/Users/emiliehabart/PDR/PDR_plot/all_objets/figure_article_carte_h2_28mic_convolved_Horsehead.ps,width=8.0cm,angle=0} }
\centerline{ \psfig{file=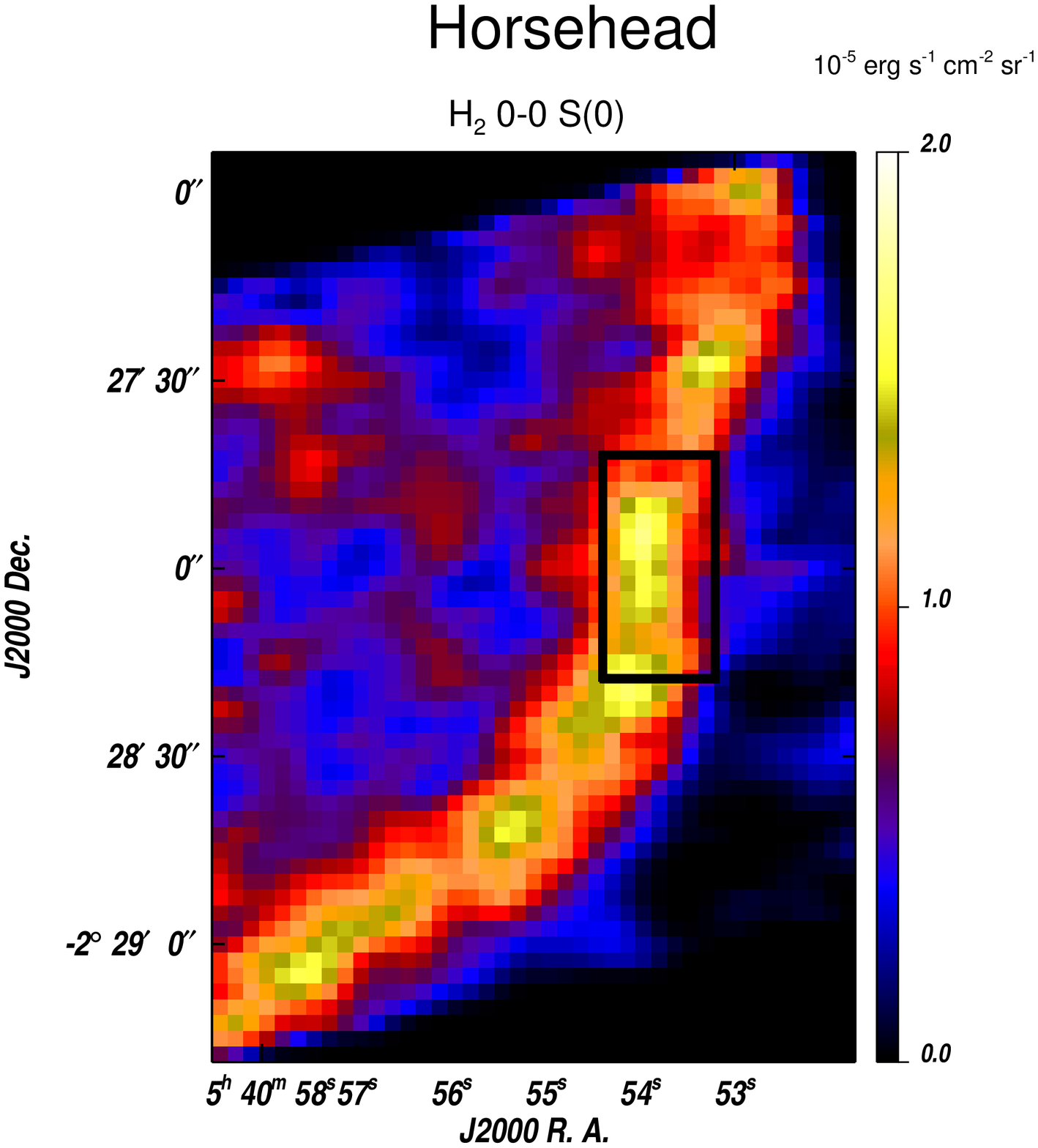,width=8.0cm,angle=0} }
\end{minipage}
\begin{minipage}[c]{9cm} 
%\centerline{ \psfig{file=/Users/emiliehabart/PDR/PDR_plot/all_objets/figure_article_carte_h2_28mic_convolved_rho_Oph.ps,width=8.0cm,angle=0} }
\centerline{ \psfig{file=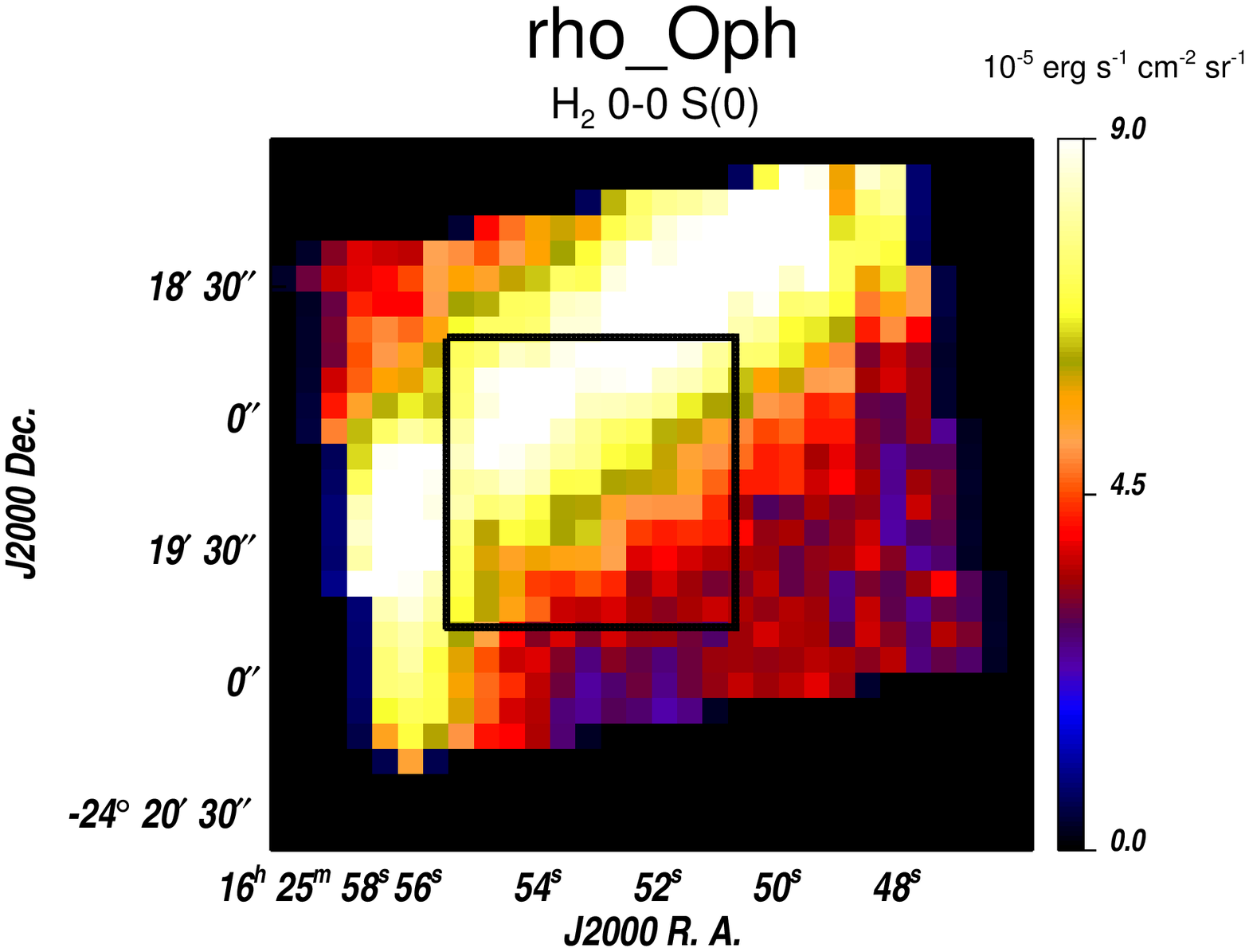,width=8.0cm,angle=0} }
\end{minipage}
\begin{minipage}[c]{9cm} 
%\centerline{ \psfig{file=/Users/emiliehabart/PDR/PDR_plot/all_objets/figure_article_carte_h2_28mic_convolved_N2023N.ps,width=8.0cm,angle=0} }
\centerline{ \psfig{file=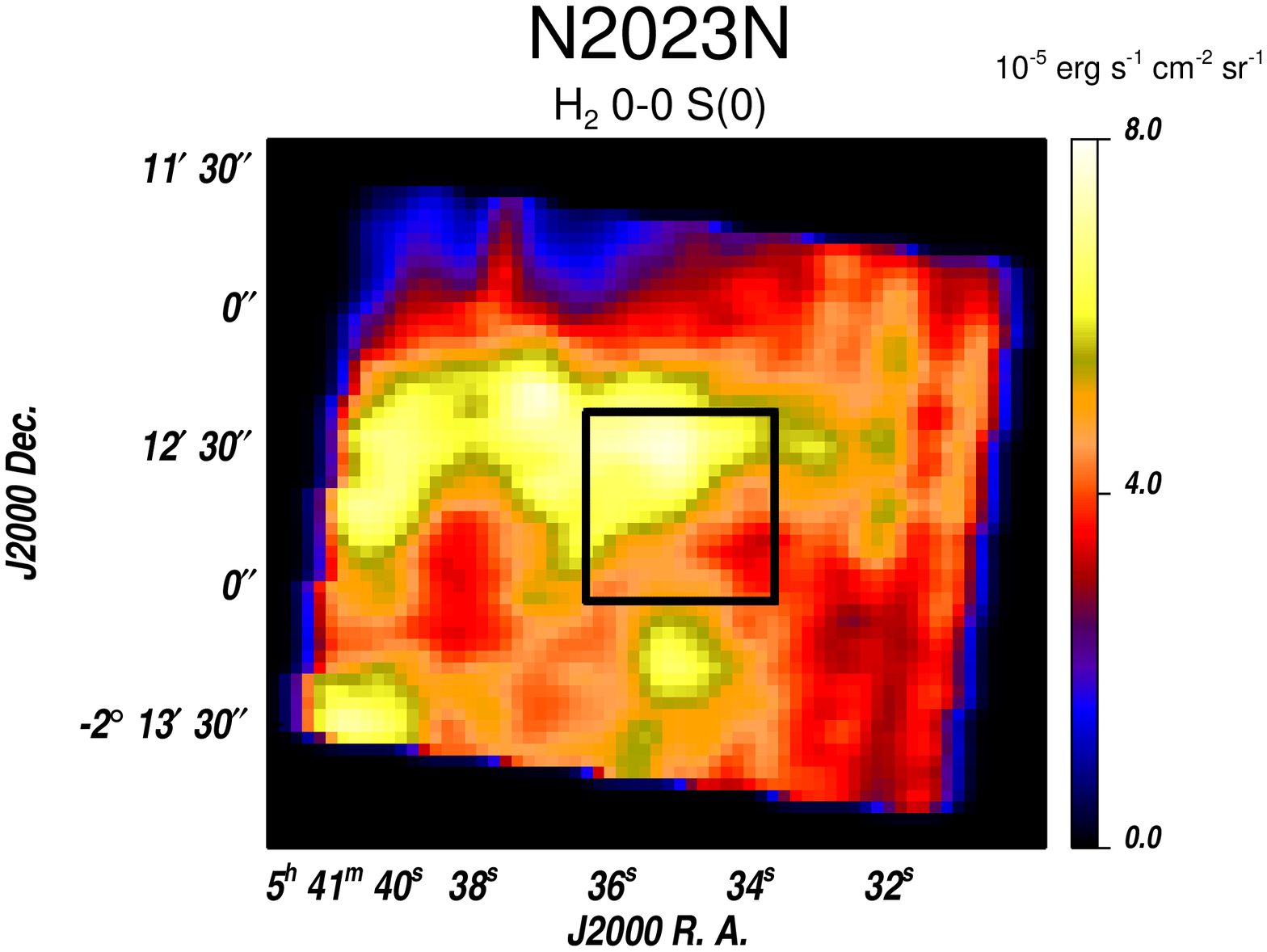,width=8.0cm,angle=0} }
\end{minipage}
\caption{\em 
Map of the H$_2$ 0-0 S(0) emission line obtained with the IRS low spectral resolution mode. The black boxes show the region where the spectra of Fig. \ref{fig_spectrum_low_resolution} have been averaged.
}
\label{App_fig_h2_map_obs}
\end{figure*}

%----------------------------------------------------
\begin{figure*}[htbp]
\leavevmode
\begin{minipage}[c]{8cm} 
%\centerline{ \psfig{file=/Users/emiliehabart/PDR/PDR_plot/all_objets/figure_article_carte_h2_12mic_convolved_Horsehead.ps,width=7.0cm,angle=0} }
\centerline{ \psfig{file=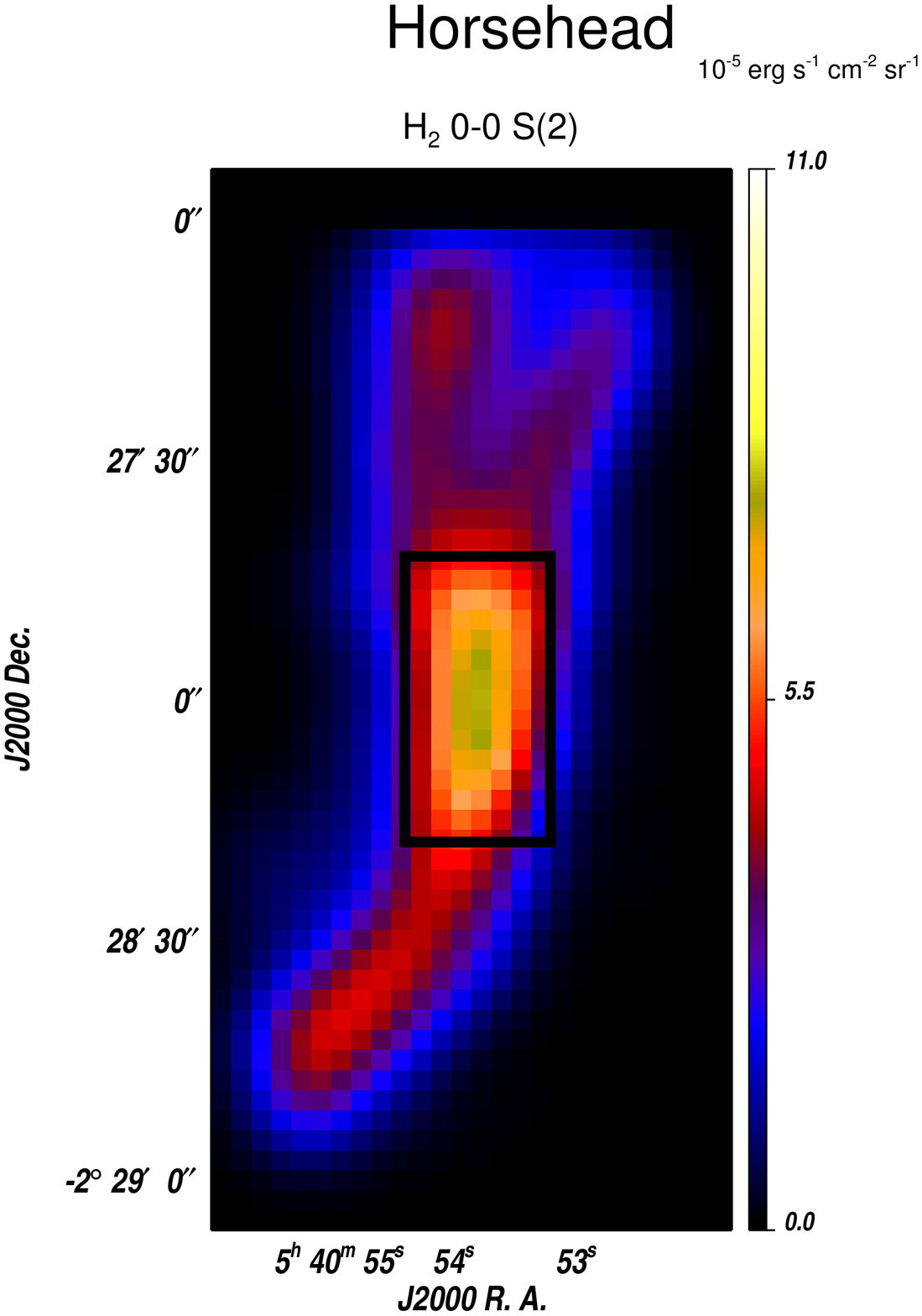,width=7.0cm,angle=0} }
\end{minipage}
\begin{minipage}[c]{8cm} 
%\centerline{ \psfig{file=/Users/emiliehabart/PDR/PDR_plot/all_objets/figure_article_carte_h2_9mic_convolved_Horsehead.ps,width=7.0cm,angle=0} }
\centerline{ \psfig{file=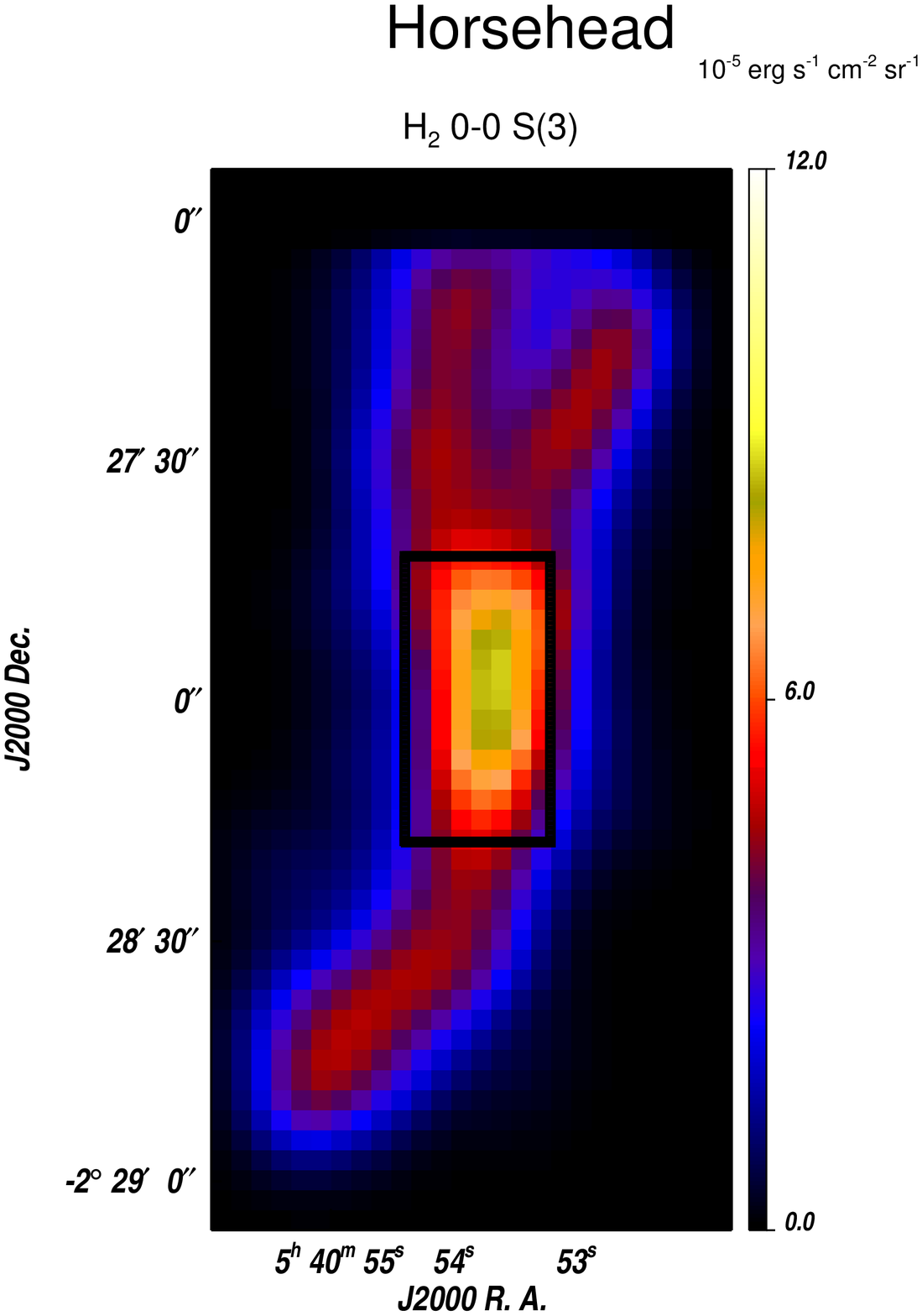,width=7.0cm,angle=0} }
\end{minipage}
\caption{\em 
Map of the H$_2$ 0-0 S(2) and S(3) emission line obtained with the IRS low spectral resolution mode towards the Horsehead nebula. The black boxes show the region where the spectra of Fig. \ref{fig_spectrum_low_resolution} have been averaged. %The white boxes show the region observed by the IRS Short-Low (SL) module. 
}
\label{App_fig_h2_map_obs_bis}
\end{figure*}

\end{appendix}

%TO HAVE BIBLIOGRAPHY

\bibliographystyle{natbib}

%\bibliographystyle{/Users/emiliehabart/gabi/natbib}
%\bibliography{/Users/emiliehabart/biblio/all}

\begin{thebibliography}{{Flower} \& {Pineau Des Forets}(1998)}

\bibitem[{Abergel} et~al.(1999)]{abergel99}
{Abergel}, A, {Andre}, P, {Bacmann}, A, {Bernard}, J.~P, {Bontemps}, S,
  {Boulanger}, F, {Coulais}, A, {Desert}, F.~X, {Falgarone}, E, {Kaas}, A.~A,
  {Huldtgren}, M, {Miville-Deschenes}, M.~A, {Nordh}, L, {Olofsson}, G,
  {Perault}, M, \& {Puget}, J.~L.
\newblock Spatial distribution of dust from cirrus to dense clouds.
\newblock In {\em The Universe as seen by ISO}, page 615, 1999.


\bibitem[{Abergel} et~al.(2002)]{abergel2002}
{Abergel}, A, {Bernard}, J.~P, {Boulanger}, F, {Cesarsky}, D, {Falgarone}, E,
  {Jones}, A, {Miville-Deschenes}, M.~A, {Perault}, M, {Puget}, J.~L,
  {Huldtgren}, M, {Kaas}, A.~A, {Nordh}, L, {Olofsson}, G, {Andr\'e}, P,
  {Bontemps}, S, {Casali}, M.~M, {Cesarsky}, C.~J, {Copet}, M.~E, {Davies}, J,
  {Montmerle}, T, {Persi}, P, \& {Sibille}, F.
\newblock 2002, {\em \AaA}, 389:239.

\bibitem[{Abergel} et~al.(2003)]{abergel2003}
{Abergel}, A, {Teyssier}, D, {Bernard}, J.~P, {Boulanger}, F, {Coulais}, A,
  {Fosse}, D, {Falgarone}, E, {Gerin}, M, {Perault}, M, {Puget}, J.~L, {Nordh},
  L, {Olofsson}, G, {Huldtgren}, M, {Kaas}, A.~A, {Andr\'E}, P, {Bontemps}, S,
  {Casali}, M.~M, {Cesarsky}, C.~J, {Copet}, E, {Davies}, J, {Montmerle}, T,
  {Persi}, P, \& {Sibille}, F.
\newblock 2003, {\em \AaA}, 410:577.

\bibitem[{Allen} et~al.(2004)]{allen2004}
{Allen}, R.~J, {Heaton}, H.~I, \& {Kaufman}, M.~J.
\newblock 2004, {\em \ApJ}, 608:314.

\bibitem[{Bakes} \& {Tielens}(1994)]{bakes94}
{Bakes}, E. L.~O \& {Tielens}, A. G. G.~M.
\newblock 1994, {\em \ApJ}, 427:822.

\bibitem[{Bernard} et~al.(1993)]{bernard93}
{Bernard}, J.~P, {Boulanger}, F, \& {Puget}, J.~L.
\newblock 1993, {\em \AaA}, 277:609.

\bibitem[{Bern\'e} et~al.(2007)]{berne2007}
{Bern\'e}, O, {Joblin}, C, {Deville}, Y, {Smith}, J.~D, {Rapacioli}, M,
  {Bernard}, J.~P, {Thomas}, J, {Reach}, W, \& {Abergel}, A.
\newblock 2007, {\em \AaA}, 469:575.

\bibitem[{Bertoldi}(1997)]{bertoldi97}
{Bertoldi}, F.
\newblock Iso: A novel look at the photodissociated surfaces of molecular
  clouds (invited paper).
\newblock In {\em First ISO Workshop on Analytical Spectroscopy}, page~67,
  1997.

\bibitem[{Boulanger} et~al.(1988)]{boulanger88b}
{Boulanger}, F, {Beichman}, C, {Desert}, F.~X, {Helou}, G, {Perault}, M, \&
  {Ryter}, C.
\newblock 1988, {\em \ApJ}, 332:328.

\bibitem[{Boulanger} et~al.(1990)]{boulanger90}
{Boulanger}, F, {Falgarone}, E, {Puget}, J.~L, \& {Helou}, G.
\newblock 1990, {\em \ApJ}, 364:136.

\bibitem[{Browning} et~al.(2003)]{browning2003}
{Browning}, M.~K, {Tumlinson}, J, \& {Shull}, J.~M.
\newblock 2003, {\em \ApJ}, 582:810.

\bibitem[{Cazaux} \& {Tielens}(2004)]{cazaux2004}
{Cazaux}, S \& {Tielens}, A. G. G.~M.
\newblock 2004, {\em \ApJ}, 604:222.

\bibitem[{Compi\`egne} et~al.(2007)]{compiegne2007}
{Compi\`egne}, M, {Abergel}, A, {Verstraete}, L, {Reach}, W.~T, {Habart}, E,
  {Smith}, J.~D, {Boulanger}, F, \& {Joblin}, C.
\newblock 2007, {\em \AaA}, 471:205.

\bibitem[{Compi\`egne} et~al.(2008)]{compiegne2008}
{Compi\`egne}, M, {Abergel}, A, {Verstraete}, L, \& {Habart}, E.
\newblock 2008, {\em \AaA}, 491:797.

\bibitem[{Draine} \& {Bertoldi}(1999)]{draine99a}
{Draine}, B \& {Bertoldi}, F.
\newblock Theoretical models of photodissociation fronts.
\newblock In {\em H2 in Space}, page 131, 1999.

\bibitem[{Draine} \& {Lee}(1984)]{draine84}
{Draine}, B.~T \& {Lee}, H.~M.
\newblock 1984, {\em \ApJ}, 285:89.

\bibitem[{Draine} et~al.(2007)]{draine2007}
{Draine}, B.~T, {Dale}, D.~A, {Bendo}, G, {Gordon}, K.~D, {Smith}, J. D.~T,
  {Armus}, L, {Engelbracht}, C.~W, {Helou}, G, jr~{Kennicutt}, R.~C, {Li}, A,
  {Roussel}, H, {Walter}, F, {Calzetti}, D, {Moustakas}, J, {Murphy}, E.~J,
  {Rieke}, G.~H, {Bot}, C, {Hollenbach}, D.~J, {Sheth}, K, \& {Teplitz}, H.~I.
\newblock 2007, {\em \ApJ}, 663:866.

\bibitem[{Draine}(1978)]{draine78}
{Draine}, B.~T.
\newblock 1978, {\em \ApJS}, 36:595.

\bibitem[{Duley} \& {Williams}(1984)]{duley84}
{Duley}, W.~W \& {Williams}, D.~A.
\newblock Interstellar chemistry.
\newblock In {\em Interstellar Chemistry}, 1984.

\bibitem[{Falgarone} et~al.(2005)]{falgarone2005}
{Falgarone}, E, {Verstraete}, L, {Pineau Des For√™ts}, G, \& {Hily-Blant}, P.
\newblock 2005, {\em \AaA}, 433:997.

\bibitem[{Field} et~al.(1998)]{field98}
{Field}, D, {Lemaire}, J.~L, {Pineau Des For\^ets}, G, {Gerin}, M, {Leach}, S,
  {Rostas}, F, \& {Rouan}, D.
\newblock 1998, {\em \AaA}, 333:280.

\bibitem[{Fitzpatrick} \& {Massa}(1990)]{fitzpatrick90}
{Fitzpatrick}, E.~L \& {Massa}, D.
\newblock 1990, {\em \ApJS}, 72:163.

\bibitem[{Flower} \& {Pineau Des Forets}(1998)]{flower98}
{Flower}, D.~R \& {Pineau Des Forets}, G.
\newblock 1998, {\em \MNRAS}, 297:1182.

\bibitem[{Fuente} et~al.(1999)]{fuente99}
{Fuente}, A, {Martin-Pintado}, J, \& {Rodriguez-Fernandez}, N.
\newblock 1999, {\em \ApJ}, 518:L45.

\bibitem[{Gerlich}(1990)]{gerlich90}
{Gerlich}, D.
\newblock 1990, {\em \JChPh}, 92:2377.

\bibitem[{Gillmon} et~al.(2006)]{gillmon2006}
{Gillmon}, K, {Shull}, J.~M, {Danforth}, C, \& {Tumlinson}, J.
\newblock 2006, {\em ASPC}, page 439.

\bibitem[{Goicoechea} et~al.(2006)]{goicoechea2006}
{Goicoechea}, J.~R, {Pety}, J, {Gerin}, M, {Teyssier}, D, {Roueff}, E,
  {Hily-Blant}, P, \& {Baek}, S.
\newblock 2006, {\em \AaA}, 456:565.

\bibitem[{Gould} \& {Salpeter}(1963)]{gould63}
{Gould}, R.~J \& {Salpeter}, E.~E.
\newblock 1963, {\em \ApJ}, 138:393.

\bibitem[{Gry} et~al.(2002)]{gry2002}
{Gry}, C, {Boulanger}, F, {Nehm\'e}, C, {Pineau Des For\^ets}, G, {Habart}, E,
  \& {Falgarone}, E.
\newblock 2002, {\em \AaA}, 391:675.

\bibitem[{Habart} et~al.(2001)]{habart2001a}
{Habart}, E, {Verstraete}, L, {Boulanger}, F, {Pineau Des For\^ets}, G, {Le
  Peintre}, F, \& {Bernard}, J.~P.
\newblock 2001, {\em \AaA}, 373:702.

\bibitem[{Habart} et~al.(2003)]{habart2003a}
{Habart}, E, {Boulanger}, F, {Verstraete}, L, {Pineau Des For\^ets}, G,
  {Falgarone}, E, \& {Abergel}, A.
\newblock 2003, {\em \AaA}, 397:623.

\bibitem[{Habart} et~al.(2004)]{habart2004}
{Habart}, E, {Natta}, A, \& {Kr\"ugel}, E.
\newblock 2004, {\em \AaA}, 427:179.

\bibitem[{Habart} et~al.(2005)]{habart2005}
{Habart}, E, {Abergel}, A, {Walmsley}, C.~M, {Teyssier}, D, \& {Pety}, J.
\newblock 2005, {\em \AaA}, 437:177.

\bibitem[{Hewitt} et~al.(2009)]{hewitt2009}
{Hewitt}, J.~W, {Rho}, J, {Andersen}, M, \& {Reach}, W.~T.
\newblock 2009, {\em \ApJ}, 694:1266.

\bibitem[{Hollenbach} \& {Salpeter}(1971)]{hollenbach71}
{Hollenbach}, D \& {Salpeter}, E.~E.
\newblock 1971, {\em \ApJ}, 163:155.

\bibitem[{Hollenbach} \& {Tielens}(1999)]{hollenbach99}
{Hollenbach}, D.~J \& {Tielens}, A. G. G.~M.
\newblock 1999, {\em \RvMP}, 71:173.

\bibitem[{Hoogerwerf} et~al.(2000)]{hoogerwerf2000}
{Hoogerwerf}, R, {De Bruijne}, J. H.~J, \& {De Zeeuw}, P.~T.
\newblock 2000, {\em \ApJ}, 544:L133.

\bibitem[{Joblin} et~al.(2005)]{joblin2005}
{Joblin}, C, {Abergel}, A, {Bernard}, J.~P, {Berne}, O, {Boulanger}, F,
  {Cernicharo}, J, {Compi\`egne}, M, {Deville}, Y, {Gerin}, M, {Goicoechea}, J,
  {Habart}, E, {Le Bourlot}, J, {Maillard}, J.~P, {Rapacioli}, M, {Reach}, W,
  {Simon}, A, {Smith}, J.~D, {Teyssier}, D, \& {Verstraete}, L.
\newblock 2005, {\em \IAUS}, 231:153.

\bibitem[{Joblin et al.}(2011)]{joblin2010}
{Joblin et al.}
\newblock 2011, {\em in preparation}.

\bibitem[{Jura}(1975)]{jura75}
{Jura}, M.
\newblock 1975, {\em \ApJ}, 197:575.

\bibitem[{Kaufman} et~al.(2006)]{kaufman2006}
{Kaufman}, M.~J, {Wolfire}, M.~G, \& {Hollenbach}, D.~J.
\newblock 2006, {\em \ApJ}, 644:283.

\bibitem[{Kemper} et~al.(1999)]{kemper99}
{Kemper}, C, {Spaans}, M, {Jansen}, D.~J, {Hogerheijde}, M.~R, {van Dishoeck},
  E.~F, \& {Tielens}, A. G. G.~M.
\newblock 1999, {\em \ApJ}, 515:649.

\bibitem[{Laor} \& {Draine}(1993)]{laor93}
{Laor}, A \& {Draine}, B.~T.
\newblock 1993, {\em \ApJ}, 402:441.

\bibitem[{Le Bourlot} et~al.(1999)]{lebourlot99}
{Le Bourlot}, J, {Pineau des For\^ets}, G, \& {Flower}, D.~R.
\newblock 1999, {\em \MNRAS}, 305:L802.

\bibitem[{Le Petit} et~al.(2006)]{lepetit2006}
{Le Petit}, F, {Nehm\'e}, C, {Le Bourlot}, J, \& {Roueff}, E.
\newblock 2006, {\em \ApJS}, 164:506.

\bibitem[{Li} et~al.(2002)]{li2002}
{Li}, W, {Evans}, N. J.~I, {Jaffe}, D.~T, {van Dishoeck}, E.~F, \& {Thi}, W.~F.
\newblock 2002, {\em \ApJ}, 568:242.

\bibitem[{Lynds}(1962)]{lynds62}
{Lynds}, B.~T.
\newblock 1962, {\em \ApJS}, 7:L1.

\bibitem[{Meyer} et~al.(1997)]{meyer97}
{Meyer}, M.~R, {Calvet}, N, \& {Hillenbrand}, L.~A.
\newblock 1997, {\em \AJ}, 114:288.

\bibitem[{Meyer} et~al.(1998)]{meyer98}
{Meyer}, D.~M, {Jura}, M, \& {Cardelli}, J.~A.
\newblock 1998, {\em \ApJ}, 493:222.

\bibitem[{Motte} et~al.(1998)]{motte98}
{Motte}, F, {Andre}, P, \& {Neri}, R.
\newblock 1998, {\em \AaA}, 336:150.

\bibitem[{Moutou} et~al.(1999)]{moutou99}
{Moutou}, C, {Verstraete}, L, {Sellgren}, K, \& {L\'eger}, A.
\newblock The rich spectroscopy of reflection nebulae.
\newblock In {\em The Universe as Seen by ISO}, page 727, 1999.


\bibitem[{Nehm\'e} et~al.(2008)]{nehme2008}
{Nehm\'e}, C, {Le Bourlot}, J, {Boulanger}, F, {Pineau Des For\^ets}, G, \&
  {Gry}, C.
\newblock 2008, {\em \AaA}, 483:485.

\bibitem[{Perryman} et~al.(1997)]{perryman97}
{Perryman}, M. A.~C, {Lindegren}, L, {Kovalevsky}, J, {Hoeg}, E, {Bastian}, U,
  {Bernacca}, P.~L, {Cr\'ez}, M, {Donati}, F, {Grenon}, M, {Van Leeuwen}, F,
  {Van Der Marel}, H, {Mignard}, F, {Murray}, C.~A, {Le Poole}, R.~S,
  {Schrijver}, H, {Turon}, C, {Arenou}, F, {Froeschl\'E}, M, \& {Petersen},
  C.~S.
\newblock 1997, {\em \AaA}, 323:L49.

\bibitem[{Pety} et~al.(2004)]{pety2004}
{Pety}, J, {Teyssier}, D, {Fosse}, D, {Gerin}, M, {Roueff}, E, {Abergel}, A, \&
  {Habart}, E.
\newblock 2004, {\em \AaA, in press}.

\bibitem[{Pety} et~al.(2005)]{pety2005}
{Pety}, J, {Teyssier}, D, {Foss√©}, D, {Gerin}, M, {Roueff}, E, {Abergel}, A,
  {Habart}, E, \& {Cernicharo}, J.
\newblock 2005, {\em \AaA}, 435:885.

\bibitem[{Pety} et~al.(2007)]{pety2007}
{Pety}, J, {Goicoechea}, J.~R, {Hily-Blant}, P, {Gerin}, M, \& {Teyssier}, D.
\newblock 2007, {\em \AaA}, 464:L41.

\bibitem[{Philipp} et~al.(2006)]{philipp2006}
{Philipp}, S.~D, {Lis}, D.~C, {Gusten}, R, {Kasemann}, C, {Klein}, T, \&
  {Phillips}, T.~G.
\newblock 2006, {\em \AaA}, 454:213.


\bibitem[{Pound} et~al.(2003)]{pound2003}
{Pound}, M.~W, {Reipurth}, B, \& {Bally}, J.
\newblock 2003, {\em \AJ}, 125:2108.

\bibitem[{Rapacioli} et~al.(2005)]{rapacioli2005}
{Rapacioli}, M, {Joblin}, C, \& {Boissel}, P.
\newblock 2005, {\em \AaA}, 429:193.

\bibitem[{Rapacioli} et~al.(2006)]{rapacioli2006}
{Rapacioli}, M, {Calvo}, F, {Joblin}, C, {Parneix}, P, {Toublanc}, D, \&
  {Spiegelman}, F.
\newblock 2006, {\em \AaA}, 460:519.

\bibitem[{Roussel} et~al.(2007)]{roussel2007}
{Roussel}, H, {Helou}, G, {Hollenbach}, D.~J, {Draine}, B.~T, {Smith}, J.~D,
  {Armus}, L, {Schinnerer}, E, {Walter}, F, {Engelbracht}, C.~W, {Thornley},
  M.~D, {Kennicutt}, R.~C, {Calzetti}, D, {Dale}, D.~A, {Murphy}, E.~J, \&
  {Bot}, C.
\newblock 2007, {\em \ApJ}, 669:959.

\bibitem[{Savage} \& {Sembach}(1996)]{savage96}
{Savage}, B.~D \& {Sembach}, K.~R.
\newblock 1996, {\em \ARAA}, 34:279.

\bibitem[{Savage} et~al.(1977)]{savage77}
{Savage}, B.~D, {Bohlin}, R.~C, {Drake}, J.~F, \& {Budich}, W.
\newblock 1977, {\em \ApJ}, 216:291.

\bibitem[{Schofield}(1967)]{Schofield67}
{Schofield}, K.
\newblock 1967, {\em Pl. \& Spa. Sc.}, 643:15.

\bibitem[{Shaw} et~al.(2005)]{shaw2005}
{Shaw}, G, {Ferland}, G.~J, {Abel}, N.~P, {Stancil}, P.~C, \& {van Hoof}, P.
  A.~M.
\newblock 2005, {\em \ApJ}, 624:794.

\bibitem[{Spitzer} \& {Cochran}(1973)]{spitzer73}
{Spitzer}, L \& {Cochran}, W.~D.
\newblock 1973, {\em \ApJ}, 186:L23.

\bibitem[{Spitzer} \& {Morton}(1976)]{spitzer76}
{Spitzer}, L \& {Morton}, W.~A.
\newblock 1976, {\em \ApJ}, 204:731.

\bibitem[{Spitzer} et~al.(1974)]{spitzer74}
{Spitzer}, L, {Cochran}, W.~D, \& {Hirshfeld}, A.
\newblock 1974, {\em \ApJS}, 28:373.

\bibitem[{St\"{o}rzer} \& {Hollenbach}(1998)]{stoerzer98}
{St\"{o}rzer}, H \& {Hollenbach}, D.
\newblock 1998, {\em \ApJ}, 495:853.

\bibitem[{Teyssier} et~al.(2004)]{teyssier2004}
{Teyssier}, D, {FossÈ}, D, {Gerin}, M, {Pety}, J, {Abergel}, A, \& {Roueff}, E.
\newblock 2004, {\em \AaA}, 417:135.

\bibitem[{Thi} et~al.(1999)]{thi99}
{Thi}, W.~F, {van Dishoeck}, E.~F, {Black}, J.~H, {Jansen}, D.~J, {Evans},
  N.~J, \& {Jaffe}, D.~T.
\newblock Weak molecular hydrogen emission from diffuse and translucent clouds.
\newblock In {\em The Universe as Seen by ISO}, page 529, 1999.

\bibitem[{Walmsley} et~al.(2000)]{walmsley2000}
{Walmsley}, C.~M, {Natta}, A, {Oliva}, E, \& {Testi}, L.
\newblock 2000, {\em \AaA}, 364:301.

\bibitem[{Ward-Thompson} et~al.(2006)]{ward-thompson2006}
{Ward-Thompson}, D, {Nutter}, D, {Bontemps}, S, {Whitworth}, A, \& {Attwood},
  R.
\newblock 2006, {\em \MNRAS}, 369:1201.

\bibitem[{Weingartner} \& {Draine}(2001)]{weingartner01}
{Weingartner}, J.~C \& {Draine}, B.~T.
\newblock 2001, {\em \ApJS}, 134:263.



\end{thebibliography}

\end{document}